# Mergers and acquisitions transactions strategies in diffusion - type financial systems in highly volatile global capital markets with nonlinearities

Dimitri O. Ledenyov and Viktor O. Ledenyov

*Abstract* – The M&A transactions represent a wide range of unique business optimization opportunities in the corporate transformation deals, which are usually characterized by the high level of total risk. The M&A transactions can be successfully implemented by taking to an account the size of investments, purchase price, direction of transaction, type of transaction, and using the modern comparable transactions analysis and the business valuation techniques in the diffusion – type financial systems in the finances. We developed the *MicroM&A software program* with the *embedded optimized near-real-time artificial intelligence algorithm* to create the *winning virtuous M&A strategies,* using the financial performance characteristics of the involved firms, and to estimate the *probability of the M&A transaction completion success*. We believe that the fluctuating dependence of M&A transactions number over the certain time period is quasi-periodic. We think that there are many factors, which can generate the quasi periodic oscillations of the M&A transactions number in the time domain, for example: the stock market bubble effects. We performed the research of the nonlinearities in the M&A transactions number quasi-periodic oscillations in Matlab, including the ideal, linear, quadratic, and exponential dependences. We discovered that the average of a sum of random numbers in the M&A transactions time series represents a time series with the quasi periodic systematic oscillations, which can be finely approximated by the polynomial numbers. We think that, in the course of the M&A transaction implementation, the ability by the companies to absorb the newly acquired knowledge and to create the new innovative knowledge bases, is a key pre-determinant of the M&A deal completion success as in Switzerland. We would like to state that the winning virtuous mergers and acquisitions transactions strategies in the diffusion - type financial systems in the highly volatile global capital markets with the nonlinearities can only be selected through the decision making process on the various available M&A choices, applying the econophysical econometrical analysis with the use of the inductive, deductive and abductive logics.

# Introduction

The *globalization* results in a strong necessity to originate and implement the *new corporate strategies* towards the businesses re-structurizations through the various types of the **Mergers and Acquisitions transactions** in order to optimize the *organizational structures, management capabilities, financial indicators*, aiming to establish the *fully optimized profitable corporations* at the various *business operation scales and scopes* within the different product and services line in the various markets. Thus, let us begin the discussion on the topic of our research interest by exploring all the possible impacts by the *globalization* on the *M&A transaction strategies selection*, reviewing the research opinions by various authors in the *M&A* literature and trying to understand the main ideas behind the M&A transactions.

*Lall (2002)* writes: "International *mergers and acquisitions (M&As)*, particularly those with giant *transnational companies (TNCs)* spending vast sums of money to take over firms in other countries, are one of the most visible aspects of *globalization*. Such *M&As* are now the most important form of **Foreign Direct Investments (FDI)**, far outstripping investment in new facilities ('greenfield' investments) in terms of value (see various issues of *UNCTAD World Investment Report*)."

*Sakai (2002)* explains: "Global industrial restructuring in the current era is characterized by an increase in cross-border strategic alliances, *mergers and acquisitions (M&As)* and other types of business networking."

*Hussinger (2005)* states: "Due to increasing *globalization* of markets, it is often hypothesized that firms engage in *merger and acquisition (M&A)* activities in order to secure their **international competitiveness**. While the former *merger waves* in the *1960s* and *1980s* were characterized by **diversification endeavors**, *M&A* activities in the *1990s* are said to be driven by firm strategies that aimed at strengthening competitiveness and market power within their field of core competencies. *M&As* provide firms the possibility to grow, lower sector or technological competition, and to benefit from economies of scale and scope, which has been important for increasing or maintaining market power in opened, international markets."

*Neto, Brandão, Cerqueira (2008)* note: "When a company decides to **invest abroad**, it can do it in two different ways: i) through the establishment of a *greenfield investment* in new asset in a foreign country, ii) or through an *investment by acquiring a pre-existent foreign firm or merging with a foreign firm*. Therefore, the two main components of *Foreign Direct Investment (FDI)* are *greenfield investments* and *mergers and acquisitions (M&A)*."

*King, Slotegraaf, Kesner (2008)* comment: "The value of worldwide *merger and acquisition (M&A)* activity set a new record in *2006* with *$3.79 trillion* worth of transactions—a *38%* increase over *2005* (*Berman (2007a)*). The dominant rationale used to explain acquisition activity is that acquiring firms seek **higher performance** (*Bergh (1997), Hoskisson and Hitt (1990), Sirower (1997)*)."

*Hussinger (2012)* emphasizes: "In times of increasing technological competition the **access to technological knowledge** is one of the major objectives for *mergers and acquisitions (M&As)* (*Chakrabarti et al. (1994); Capron et al. (1998); Puranam et al. (2003); Graebner (2004)*)."

In other words, in the *diffusion - type financial system*, ***the M&A transactions represent a wide range of unique business optimization opportunities and techniques in the corporate transformation deals, which are mainly aimed to facilitate the value creation process within the corporation, increase the operating efficiency of corporation and raise the competitiveness of corporation in the selected markets, however it makes sense to note that the M&A transactions are usually characterized by the high total risk factors.***

Let us review the extensive list of literature on the *M&A transactions* in the *diffusion-type financial system*, which has been created by the world renowned scientists in the *USA and Canada, Eastern and Western Europe, Asia and Australia* regions over the recent decades.

The **North American academicians** are commonly regarded as the pioneers in the research on the *M&A transactions* in the *World*. We would like to make a chronological literature review on the *M&A transactions* in the chronological order. Among the early research works on the *M&A transactions*, it is necessary to highlight the research on the *merger movements* in the *American* industry in *1895 – 1956* in *Nelson (1959)*. The forces generating and limiting concentration under the *Schumpeterian competition* in *Schumpeter (1934)* have been investigated in *Nelson, Winter (1978)*. The *theory of the growth of the firm* has been researched in *Penrose (1959)*. The *mergers* and the *market for the corporate control* have been studied in *Manne (1965)*. The financial motivation for the *conglomerate mergers* has been analyzed in *Levy, Sarnat (1970), Lewellen (1971)*. The efficiency performance of *conglomerate firms* has been researched in *Weston, Mansinghka (1971)*. Some issues on the *corporate bankruptcy* and the *conglomerate merger* have been uncovered in *Higgins, Schall (1975)*. The cycle of highly innovative research articles *by Jensen* and his co-authors can be definitely considered as a significant contribution to the theory on the *M&A transactions*, namely the *new theory of the firm* has been proposed in *Jensen, Meckling (1976)* and the *market for the corporate control* has been described in *Jensen, Ruback (1983)*. The agency costs of the free cash flow, corporate finance and *takeovers* have been researched in *Jensen (1986)*. The *takeover controversy* has been analyzed in *Jensen (1987)*. The causes and consequences of *takeovers* have been fully explained in *Jensen (1988)*. The performance pay and top-management incentives have been placed at the center of research in *Jensen, Murphy (1990)*. The problems on the *corporate control* and the *politics of finance* have been considered in *Jensen (1991)*. The modern industrial revolution has been selected as a main topic of research in *Jensen (1993)*. The *mergers*, including their motives, effects and policies, have been also researched in *Steiner (1976)*. The *corporate mergers* and the *co-insurance of corporate debt* have been studied in *Kim, McConnell (1977)*. The *diversification through acquisition* has been described in *Salter, Weinhold (1979)*. The *takeover bids*, the free rider problem, and the theory of the corporation have been considered in *Grossman, Hart (1980)*. The *determinants and effects of mergers* have been discovered in *Mueller (1980)*. The problems of the *mergers* and the *market share* have been researched in *Mueller (1984)*. Some issues on the corporation growth, diversification and *mergers* have been studied in *Mueller (1987)*. The *mergers*, their causes, effects and policies have been investigated in *Mueller (1989)*. The risk reduction as a managerial motive for the *conglomerate mergers* has been discussed in *Amihud, Baruch (1981)*. The impact of *merger bids* on the participating firms' security holders has been extensively considered in *Asquith, Kim (1982)*. The *merger bids*, uncertainty, and stock-holder returns have been described in *Asquith (1983)*. The gains to the bidding firms from the *mergers* have been calculated in *Asquith, Bruner, Mullins (1983)*. The strategies of *Japanese* investors in the *United States* have been defined in *Hennart, Park (1993)*. The investment choices between the *mergers/acquisitions* and the *joint ventures* by the *Japanese* investors in the *United States* have been discussed in *Hennart, Reddy (1997)*. The *mergers*, debt capacity, and the valuation of corporate loans have been estimated in *Stapleton (1982)*. The *horizontal mergers* and *stockholder wealth* have been considered to certain degree in *Eckbo (1983)*. The *bidding strategies* and *takeover premiums* have been reviewed in *Eckbo (2009)*. An empirical test of the *redistribution effect* in the *pure exchange mergers* has been conducted in *Eger (1983)*. The valuable research contributions in the *M&A transactions* science by *Lubatkin* resulted in a better understanding of a number of complex issues during the *M&A transactions* implementation in various organizations. The *mergers* and the *performance of the acquiring firm* have been investigated in *Lubatkin (1983)*. The *merger strategies* and *stockholder value* have been researched in *Lubatkin (1987)*. The study towards the reconciliation of market performance measures to the strategic management research has been done in *Lubatkin, Shrieves (1986)*. The *merger strategies* and *shareholder value* in the case of *large mergers* during the *1980s* in the conditions of relaxed antitrust enforcement rules have been studied in *Lubatkin, Srinivasan,*

*Merchant (1997)*. The top management turnover in related *M&As* has been investigated in *Lubatkin, Schweiger, Weber (1998)*. The ecological investigation of firm effects in the *horizontal mergers* has been conducted in *Lubatkin, Schulze, Mainkar, Cotterill (2001)*. The wealth effect of *merger activity* and the objective functions of *merging firms* have been considered in details in *Malatesta (1983)*. The possible losses from the *horizontal merger*, including the effects of an exogenous change in the industry structure on the *Cournot-Nash equilibrium*, have been studied in *Salant, Switzer, Reynolds (1983)*. The abnormal returns to the *acquired firms* by the *type of acquisition* and the method of payment have been researched in *Wansley, Lane, Yang (1983)*. The important problem on how to achieve the integration on the human side of the *merger* has been discussed in *Blake, Monton (1984)*. The organizational performance measurement in the absence of objective measures in the case of the *privately held firm* and *conglomerate business unit* has been investigated in *Dess, Robinson (1984)*. The relationship between the *aggregate merger* activity and the *stock market* has been established in *Geroski (1984)*. The categorical *bank acquisitions* have been considered in *Lobue (1984)*. The multiple cultures integration issues in the *acquisitions* during the corporate transitions processes have been discussed in *Sales, Mirvis (1984)*. The role of market structure in the *merger behavior* has been highlighted in *Stewart, Harris, Carleton (1984)*. The anatomy of a *merger*, including the multicultural differences problems, has been studied in *Buono, Bowditch, Lewis (1985)*. The conjectures on the cognitive simplification in the *acquisition and divestment decision making processes* have been suggested in *Duhaime (1985)*. The certain characteristic of *takeover targets* have been researched in *Hasbrouch (1985)*. An exploratory study of *strategic acquisition factors*, relating to the organization performance, has been completed in *Kusewitt (1985)*. A model of stock price reactions with an application to the *corporate acquisitions* has been developed in *Malatesta, Thompson (1985)*. The oligopoly and the incentive for the *horizontal merger* have been considered in *Perry, Porter (1985)*. The determinants of tender offer premiums have been described in *Walkling, Edmister (1985)*. *Chatterjee* wrote a cycle of research articles on the *M&A transactions*, namely the impacts of *acquisitions* on the *merging and rival firms* have been investigated in *Chatterjee (1986)*. The *corporate mergers stock holder diversification* and the *changes in the systematic risk* have been analyzed in *Chatterjee, Lubatkin (1990)*. The gains in the *vertical acquisitions* and the market power have been researched in *Chatterjee (1991)*. The sources of value in the *takeovers*, synergy or restructuring, including the implications for the *target and bidder firms*, have been investigated in *Chatterjee (1992)*. The cultural differences and the shareholder value in the *related mergers* have been considered in *Chatterjee, Lubatkin, Schweiger, Weber (1992)*. The *corporate mergers* and the *security returns* have been studied in *Dennis, McConnell (1986)*. Some issues in the *corporate acquisitions* have been researched in *Jemison, Sitkin (1986a, b)*. The *mergers* that last in the frames of a predictable pattern have been researched in *Montgomery, Wilson (1986)*. A methodological and empirical analysis *toward the* prediction of *takeover targets* has been made in *Palepu (1986)*. The *merger* and the *bankruptcy alternative* have been discussed in *Pastena, Ruland (1986)*. The hubris hypothesis of *corporate takeovers* has been uncovered in *Roll (1986)*. The interesting cycle of research works has been written by *Shleifer* and co-authors. The large shareholders and corporate control topics have been discussed in *Shleifer, Vishny (1986)*. The breach of trust in the *hostile takeovers* has been selected as a main research topic in *Shleifer, Summers (1988)*. The *takeover wave* of the *1980s* has been found to exist in *Shleiper, Vishny (1990)*. The *takeovers* in the *1960s* and the *1980s* have been researched in *Shleifer, Vishny (1991)*. The *inefficient markets* have been researched in *Shleifer (2001)*. The *stock market driven acquisitions* have been considered in *Shleifer, Vishny (2003)*. The *post-merger integration* has been explored in *Shrivastava (1986)*. The effects of the *interstate bank mergers* on the shareholder wealth have been considered in *De, Duplichan (1987)*. A theory for the choice of exchange medium in the mergers and acquisitions has been created in *Hansen (1987)*. A contingent framework for the *acquisition integration process* has been proposed in *Haspeslagh, Farquhar (1987)*. The *target abnormal returns*, associated with the *acquisition announcements* have been researched in *Huang, Walkling (1987)*. The returns to the *acquirers* and the competition in the *acquisition market* in the case of banking industry have been considered in *James, Wier (1987)*. The *banking acquisitions*, including the *acquirer* and *target shareholder returns*, have been investigated in *Neely (1987)*. The life after the *takeover* has been described in *Ravenscraft, Scherer (1987a)*. The *mergers*, selloffs, and economy efficiency have been analyzed in *Ravenscraft, Scherer (1987b)*. The *acquisition of divested assets* and the *shareholder wealth* have been researched in *Sicherman, Pettway (1987)*. The *corporate acquisition strategies* and the economic performance have been studied in *Singh, Montgomery (1987)*. The *corporate takeover bids*, methods of payment, and *bidding firms' stock returns* have been investigated in *Travlos (1987)*. The executive compensation, method of payment and abnormal returns to the *bidding firms* at the *takeover announcements* have been analyzed in *Travlos, Waegelein (1992)*. The *interstate bank mergers* have been researched in *Trifts, Scanlon (1987)*. The overpaying in the *corporate takeovers* has been discussed in *Varaiya, Ferris (1987)*. Some aspects of the *mergers and acquisitions* have been discussed in *Auerbach (1988)*. The causes and consequences of *corporate takeovers* have been explained in *Auerbach (1989)*. The returns to the bidding firms in the *mergers and acquisitions* have been estimated in *Barney (1988)*. The firm resources and sustained competitive advantage have been analyzed in *Barney (1991)*. The synergistic gain from the *corporate acquisitions*, including its division between the *stockholders of target* and the *acquiring firms,* have been researched in *Bradley, Desai, Kim (1988)*. The impact of *foreign acquisition* on the labour has been studied in *Brown, Medoff (1988)*. The effect of *takeover activity* on the corporate research and development has been considered in *Hall (1988)*. The impact by the corporate restructuring on the industrial research and development has been described in *Hall (1990)*. The *mergers* and the *R&D* have been revisited in *Hall (1999)*. The characteristics of the *hostile and friendly takeovers* have been summarized in *Morck, Shleifer, Vishny (1988)*. The supposition that the unwise managerial objectives can drive the bad *acquisitions* has been analyzed in *Morck, Shleifer, Vishny (1990)*. The *hostile takeovers* in the *1980s* have been reviewed in *Bhagat, Shleifer, Vishny (1990)*. The acculturation in the *mergers and acquisition* has been researched in *Nahavandi, Malekzadeh (1988)*. The impact by the *merger-related regulations* on the shareholders of bidding firms has been evaluated in *Schipper, Thompson (1983)*. The strategic business fits in the *corporate acquisition* have been studied in *Shelton (1988)*. The top management turnover following the *mergers and acquisitions* has been investigated in *Walsh (1988)*. The *merger and acquisition negotiations* and their impact upon the target company top management turnover have been extensively studied in *Walsh (1989)*. The failed *bank acquisitions* and the successful bidders' returns have been studied in *Bertin, Ghazanfari, Torabzadeh (1989)*. The preemptive bidding and the role of the medium of exchange in the *acquisitions* have been researched in *Fishman (1989)*. The determinants of the tender offer and the *post-acquisition financial performance* have been investigated in *Fowler, Schmidt (1989)*. A time-series analysis of the *mergers and acquisitions* in the *US* economy has been completed in *Golbe, White (1989)*. The returns to the bidders and targets in the *acquisition process* in the banking industry have been evaluated in *Hannan, Wolken (1989)*. The tax attributes as the main determinants of shareholder gains in the *corporate acquisitions* have been investigated in *Hayn (1989)*. The cartels, collusion and *horizontal merger* have been considered *in Jacquemin, Slade (1989)*. The *New Hampshire bank mergers* have been reviewed in *Kaen, Tehranian (1989)*. The profitability of *mergers* has been studied in *Ravenscraft, Scherer (1989)*. The market valuation effects of *bank acquisitions* have been researched in *Wall, Gup (1989)*. The *hostile bank takeover offers* have been analyzed in *Baradwaj, Fraser, Furtado (1990)*. The bidder returns in the interstate and intrastate bank acquisitions have been forecasted in *Baradwaj, Dubofsky, Fraser (1992)*. The *airline mergers* airport dominance and the market power have been discussed in *Borenstein (1990)*. The cycle of research papers by *Data* and co-authors attracted the considerable attention from the side of scientists. The relationships between the *type of acquisition*, the autonomy given to the *acquired* *firm*, and the *acquisition success* have been empirically analyzed in *Datta, Grant (1990)*. The organizational fit and the *acquisition performance effects* during the *post-acquisition integration* have been considered in *Datta (1991)*. The executive compensation and the *corporate acquisition* decisions have been studied in *Datta, Iskander-Datta, Raman (1992)*. The *corporate partial acquisitions*, total firm valuation and the effect of financing method have been researched in *Datta, Iskandar-Datta (1995)*. The executive compensation and the *corporate acquisition* decisions have been researched in *Datta, Iskandar-Datta, Raman (2001)*. The *horizontal mergers* have been analyzed in *Farrell, Shapiro (1990)*. The *mergers and acquisitions* in the *US* banking industry have been researched in *Hawawini, Swary (1990)*. The corporate performance after the *mergers* has been researched in *Healy, Palepu, Ruback (1990)*. The changing pattern of acquisition behavior in the *takeovers* and the consequences for the *acquisition processes* have been discussed in *Hunt (1990)*. The *horizontal mergers* have been analyzed in *Farrell, Shapiro (1990)*. *Hitt* wrote a number of valuable research articles on the *M&A transactions*, for example, the *mergers and acquisitions* and the managerial commitment to the innovation have been considered in *Hitt, Hoskisson, Ireland (1990)*. The effects of the *acquisition* on the *R&D* inputs and outputs have been considered in *Hitt, Hoskisson, Ireland, Harrison (1991)*. The market for the corporate control and firm innovation has been studied in *Hitt, Hoskisson, Johnson, Moesel (1996)*. The attributes of the *successful and unsuccessful acquisitions* of *US* firms have been listed in *Hitt, Harrison, Ireland, Best (1998)*. The *mergers and acquisitions* as a value generation opportunity for the stakeholders have been considered in *Hitt, Harrison, Ireland (2001)*. The limits of monopolization through the *acquisition* have been considered in *Kamien, Zang (1990)*. The *competitively cost advantageous mergers* and the monopolization have been studied in *Kamien, Zang (1991)*. The monopolization by the *sequential acquisition* has been researched in *Kamien, Zang (1993)*. The sensitivity of the *acquiring firms' returns* to the alternative model specification and disaggregation have been discussed in *Lahey, Conn (1990)*. The *corporate acquisitions* by listed firms have been summarized in *Loderer, Martin (1990)*. The question: Do union wealth concessions explain takeover premiums?, has been answered in *Rosett (1990)*. The sources of value creation in the *acquisitions* have been empirically investigated in *Seth (1990)*. The value creation and destruction in the *cross-border acquisitions* have been discussed in *Seth, Song, Pettit (2002)*. The differential effects of the *mergers* on the corporate security values have been described in *Shastri (1990)*. Some thoughts on the corporate ownership structure and performance have been presented in *Smith (1990)*. The common stock returns in the corporate takeover bids with the evidence from interstate bank mergers have been estimated in *Cornett, De (1991)*. The *postmerger share-price performance* of *acquiring firms* has been evaluated in *Franks, Harris, Titman (1991)*. The value creation through the corporate renewal during the *acquisition* has been described in *Haspeslagh, Jemison (1991)*. The role of *acquisitions* in the *foreign direct investment* with the evidence from the *US* stock market has been explained in *Harris, Ravenscraft (1991)*. The synergies and *post-acquisition performance* of corporation have been considered in *Harrison, Hitt, Hoskisson, Ireland (1991)*. The *acquisition activity* and equity issues have been discussed in *Mann, Sicherman (1991)*. A longitudinal field experiment through the communication with the employees, following a *merger*, has been conducted in *Schweiger, Denisi (1991)*. The *Tobin's Q* and the gain from the *takeovers* have been discussed in *Servaes (1991)*. The synergy, agency, and the determinants of premia paid in the *mergers* have been uncovered in *Slusky, Caves (1991)*. The *merger analysis*, industrial organization theory, and merger guidelines have been presented in *Willig (1991)*. The *post-merger performance* of *acquiring firms* has been evaluated in *Agrawal, Jaffe, Mandelker (1992), Agrawal, Jaffe (2000, 2002)*. The effect of a set of comparable firms on the accuracy of the price-earnings valuation method has been described in *Alford (1992)*. It is noteworthy to mention that the cycle of research articles by *Berger* and *co-authors* added to the better understanding of the *M&A transactions techniques* in the banking industry. The *megamergers* in the banking and the use of cost efficiency as an antitrust defense have been discussed in *Berger, Humphrey (1992)*. The differences in the efficiencies of financial institutions have been described in *Berger, Mester (1997)*. The effect of *bank mergers and acquisitions* on the small business lending has been discussed in *Berger, Saunders, Scalise, Udell (1997, 1998)*. The efficiency effects on the *bank mergers and acquisition*, using the *1990s data*, have been discussed in *Berger (1998)*. The globalization of financial institutions, evaluating the cross-border banking performance, has been studied in *Berger, DeYoung, Hesna, Udell (1999)*. The consolidation of the financial services industry, including its causes, consequences, and implications for the future, has been researched in *Berger, Demsetz, Strahan (1999)*. A comparison of methods and sources for obtaining the estimates of new venture performance has been made in *Brush, Vanderwerf (1992)*. The predicted change in the operational synergy and *post-acquisition performance* of *acquired businesses* has been discussed in *Brush (1996)*. The limits of monopolization through the *acquisition* have been set in *Gaudet, Salant (1992)*. The *external technology acquisition* in the large multi-technology corporations has been explained in *Grandstrand, Bohlin, Oskarsson, Sjoberg (1992)*. The question: Does corporate performance improve after the *mergers*?, has been answered in *Healey, Palepu, Ruback (1992)*. The *mergers and profitability problems* have been studied in *Ingham, Kran, Lovestam (1992)*. The *post-acquisition performance* of *acquiring firms* has been revealed in *Loderer, Martin (1992)*. The agreement between the top management teams and the expectations for the *post acquisition performance* have been studied in *Shanley, Correa (1992)*. The question: Are there cost savings from the *bank mergers*?, has been answered in *Srnivasan (1992)*. The *federal merger guidelines* have been formulated in *United States Department of Justice (1992)*. The effects of executive departures on the performance of the *acquired firms* have been researched in *Cannella, Hambrick (1993)*. The *mergers*, leveraged buyouts, and performance in food retailing industry have been studied in *Cotterill (1993)*. An antitrust economic analysis of stop & shop's proposed *acquisition* of the big *V* shop retail supermarket chain has been investigated in *Cotterill (2002)*. The determinants of corporate restructuring, including the relative importance of corporate governance, *takeover threat*, and free cash flow, have been discussed in *Gibbs (1993)*. A framework for understanding departures of *acquired executives* has been proposed in *Hambrick, Cannella (1993)*. The *mergers* and market power issues with particular focus on the airline industry have been described in *Kim, Singal (1993)*. The *bank mergers* with an accent on the integration and profitability have been highlighted in *Linder, Crane (1993)*. The efficiency effects of *horizontal bank mergers* have been described in *Rhoades (1993)*. A summary of *merger performance* studies in banking in *1980 - 1993*, and an assessment of the "operating performance" and "event study" methodologies have been presented in *Rhoades (1994a, b)*. The *takeovers performance improvement* in the banking industry have been researched in *Schrantz (1993)*. The impacts of managerial ownership on the *acquisition attempts* and *target shareholder wealth* have been discussed in *Song, Walking (1993)*. The performance of *acquisitions* of distressed firms has been evaluated in *Bruton, Oviatt, White (1994)*. Some issues in the process to *acquire* the technological firms have been discussed in *Chakrabarti, Hauschildt, Suverkup (1994)*. The *mergers* as a means of restructuring distressed firms have been empirical investigated in *Clark, Ofek (1994)*. The overall gains from the *large bank mergers* have been computed in *Houston, Ryngaert (1994)*. The *bank mergers* from the perspective of insiders and outsiders have been considered in *Houston, James, Ryngaert (2001)*. The *innovation through the acquisition* corporate development model has been suggested in *Hudson (1994)*. An agency theory perspective on the role of representatives in the *brokering mergers* has been created in *Kesner, Shapiro, Sharma (1994)*. The long-term valuation effects of *bank acquisitions* have been observed in *Madura, Wiant (1994)*. The shareholder benefits from the corporate international diversification on the base of evidence from *US* international acquisitions have been researched in *Markides, Ittner (1994)*. The determinants of *acquisition integration level* and their impacts on the decision-making perspective have been studied in *Pablo (1994)*. The combined effects of the fee cash flow and financial slack on the bidder and target stock returns have been estimated in *Smith, Kim (1994)*. The effects of *mergers* in the differentiated products industries have been measured in *Werden, Froeb (1994)*. The attacker's advantage in the *acquisition process*, including the technological paradigms, organizational dynamics, and value network problems, have been reviewed in *Christensen, Rosenbloom (1995)*. A microanalysis of both the tax reform and the *foreign acquisitions* has been completed in *Collins, Kemsley, Shackelford (1995)*. An exploratory empirical study on the successful integration of *R&D* functions after the *acquisition* has been conducted in *Gerpott (1995)*. The incorporating dynamic efficiency concern in the *merger* in the innovation markets has been researched in *Gilbert, Sunshine (1995)*. The intellectual property guidelines have been created in *Gilbert, Tom (2001)*. The competition and innovation have been researched in *Gilbert (2006)*. The valuation of cash flow forecasts has been considered in *Kaplan, Ruback (1995)*. The effects of the trade liberalization on the *cost-reducing horizontal mergers* have been researched in *Long, Vousden (1995)*. The *cross-border acquisitions* have been investigated in *Sudarsanam (1995)*. The wealth effects in the *US bank takeovers* have been considered in *Zhang (1995)*. The predicted change in the operational synergy and the *post-acquisition performance* in the *acquired businesses* have been discussed in *Brush (1996)*. The interest-rate exposure and *bank mergers* have been studied in *Esty, Narasimhan, Tufano (1996)*. The *hostile takeovers* and the correction of managerial failure have been considered in *Franks, Mayer (1996)*. The *M&A transactions strategies* have been privately discussed in *Brighton* in the *UK* in *Gerstein (1996)*. The impact of industry shocks on the *takeover* and the restructuring activity have been analyzed in *Mitchell, Mulherin (1996)*. The performance changes and the shareholder wealth creation, associated with the *mergers of publicly traded banking institutions* have been selected to discuss in *Pilloff (1996)*. The markup pricing in the *mergers and acquisitions* has been chosen as a research theme in *Schwert (1996)*. The *hostility* in the *takeovers* has been described in *Schwert (2000)*. The value of diversification during the *conglomerate merger wave* has been estimated in *Servaes (1996)*. The *mergers* with the differentiated products have been described in *Shapiro (1996)*. Some issues on the organizational learning through the *acquisitions* have been considered in *Vermeulen, Barkema (1996)*. The corporate cultural fit and performance in the *mergers and acquisitions* topics have been discussed in *Weber (1996)*. The effects of *megamergers* on the efficiency and prices with the evidence from a bank profit function have been presented in *Akhavein, Berger, Humphrey (1997)*. The asset redeployment, acquisitions and corporate strategy in declining industries have been researched in *Anand, Singh (1997)*. An integrative model for the prediction of divestiture of the *unrelated acquisitions* has been proposed in *Bergh (1997)*. The interesting cycle of research articles on the *M&A transactions* has been written by *Capron*. The outcomes of international *telecommunications acquisitions*, including the analysis of the four cases with the implications for the acquisitions theory, have been discussed in *Capron, Mitchell (1997)*. The bilateral resource redeployment and the capabilities improvement, following the *horizontal acquisitions*, have been considered in *Capron, Mitchell (1998)*. The resource re-deployment, following the *horizontal acquisitions* in *Europe* and *North America* in *1988 – 1992*, have been researched in *Capron, Dussauge, Mitchell (1998)*. The long-term performance of *horizontal acquisitions* with the multiple empirical evidences of the *US* and *European firms*, has been estimated in *Capron (1999a, b)*. The re-deployment of brands, sales forces, and general marketing management expertise, following the *horizontal acquisitions*, from a resource-based perspective have been researched in *Capron, Hulland (1999)*. The cases, when the *acquirers* earn the abnormal returns, have been analyzed in *Capron, Pistre (2002)*. The changes in the value-relevance of earnings and book values over the past forty years have been analyzed in *Collins, Maydew, Weiss (1997)*. The leadership style and the *post-merger satisfaction* have been discussed in *Covin, Kolenko, Sightler, Tudor (1997)*. The *mergers* related problems, including the leadership, performance and corporate health, have been thoughtfully discussed in *Fubini, Price, Zollo (1997)*. The one of possible explanations of the premium paid for the *large acquisitions* has been proposed in *Hayward, Hambrick (1997)*. The problem: When do firms learn from their *acquisition experience*?, has been accurately considered, using the evidences from *1990 – 1995*, in *Hayward (2002)*. The *international mergers* and the welfare under the decentralized competition policy have been studied in *Head, Ries (1997)*. The "wallet game" and its applications in the *auctions* with the almost common values have been discussed in *Klemperer (1997)*. A critical determinant of the *acquisition performance* and the *CEO rewards* have been presented in *Kroll, Wright, Toombs, Leavell (1997)*. An analysis of the effects of the *foreign acquisitions* vs the *domestic acquisitions* of the *US targets,* including the problem of *post-acquisition turnover* among the *US* top management teams, has been done in *Krug, Hegarty (1997)*. A study of the top managers in the multinationals, including the prediction on the problem: Who does stay and leave after an *acquisition*, has been made in *Krug, Hegarty (2001)*. The question: Do the long term shareholders benefit from the *corporate acquisitions*?, has been answered in *Loughran, Vijh (1997)*. The theory and evidence on the *corporate acquisitions* has been presented in *Megginson, Morgan, Nail (1997)*. The modeling of *takeover likelihood* has been completed in *Powell (1997)*. The performance impact of strategic similarity in the *horizontal mergers* in the *US* banking industry have been fully explained in *Ramaswamy (1997)*. The *telecommunications mergers and acquisitions* in the *USA* have been studied in *Rosenberg (1997)*. The cycle of research works on the *M&A transactions* by *Siegel* and his *co-authors* has been written with the particular research focus on the various *M&A effects assessments*. The human resource management implications in the process of the adoption of advanced manufacturing technologies have been described in *Siegel, Waldman, Youngdahl (1997)*. The skill-biased technological change has been researched, using the evidence from a firm-level survey, in *Siegel (1999)*. The problems of the ownership change, productivity, and human capital have been considered, applying the new evidences from the matched employer-employee data in Swedish manufacturing, in *Siegel, Simons, Lindstrom (2005)*. The assessment of the effects of the *mergers and acquisitions* on the firm's performance, plant productivity, and workers has been conducted by matching the employer - employee data in *Siegel, Simons (2006)*. The assessment of the effects of the *mergers and acquisitions* on the women and minority employees with the application of the new evidences from the matched employer-employee data has been presented in *Marsh, Siegel, Simons (2007)*. The evaluation of the effects of the *mergers and acquisitions* on the employees has been completed, using the evidence from the matched employer-employee data, in *Siegel, Simons (2008)*. The assessment of the effects of the *mergers and acquisitions* on the firm's performance, plant productivity, and workers, using the new evidences from the matched employer-employee data, has been realized in *Siegel, Simons (2010)*. The problems, concerning the companies, which lose the acquisition games, because of the synergy trap, have been researched in *Sirower (1997)*. The question: Did the mergers improve the *X*-efficiency and scale efficiency of *US* banks in the *1980s*?, has been answered in *Stavros (1997)*. The small business lending by the banks, involved in the *mergers*, has been investigated in *Walraven (1997)*. A learning perspective on the international expansion through the start-up or *acquisitions* has been presented in *Barkema, Vermeulen (1998)*. The consolidation in the *US* banking industry by means of the *banks mergers and acquisitions*, including the possible implications for the efficiency and risk, has been studied in *Boyd, Graham (1998)*. The toeholds and takeovers have been researched in *Bulow, Huang, Klemperer (1998, 1999)*. The *takeovers* of privately held *targets*, methods of payments, and bidder returns have been investigated in *Chang (1998)*. An interesting hypothesis that the *bank mergers* reduce the availability of credit to small businesses has been tested in *Cole, Walraven (1998)*. It was found that the *bank takeover activity* is associated with the *greater* rather than reduced availability of credit to the small businesses in *Cole, Walraven (1998)*. In the next research article, *Cole, Fatemi, Vu (2006)* greatly contributed to the *M&A science*: "*1)* by investigating whether a *takeover attempt* signals the investors about the quality of firm management as well as the quality of the specific firm investment under the consideration; *2)* by examining how the *merger bids and terminations* affect the *relative values of bidders*, attempting, diversifying and focusing on the takeovers; 3) by using data from the *1991 - 2000* period to re-examine the important topic of who wins and who loses, when the *mergers* are terminated. The *M&A transactions* have been researched in *Jones T (2009)*, where it was found that an every *M&A transaction* is different and it requires a special attention to the details and a contingency plan for the unknown variables. The *mergers* in the open economies have been studied in *Falvey (1998)*. The undervaluation hypothesis in the *cross-border mergers and acquisitions* has been researched in *Gonzalez, Vasconcellos, Kish (1998)*. The role of managerial incentives in the *bank acquisitions* during the consolidation of the financial services industry has been discussed in *Hadlock, Houston, Ryngaert (1998)*. The emerging patterns in the global telecommunications alliances and *mergers* have been discovered in *Jamison (1998)*. A case survey of synergy realization, including the strategic, organizational, and human resource perspectives on the *mergers and acquisitions*, has been described in *Larsson, Finkelstein (1999)*. A co-competence and motivational approach to the synergy realization toward the *merger and acquisition success* has been researched in *Larsson, Brousseau, Driver, Sweet (2004)*. The first-mover (dis)advantages in the *M&A process* have been studied in *Lieberman, Montgomery (1998)*. The wealth creation versus the wealth redistribution in the *pure stock-for-stock mergers* have been compared in *Maquieria, Megginson, Nail (1998)*. The national cultural distance and the *cross-border acquisition* performance have been investigated in *Morosini, Shane, Singh (1998)*. The value effects of the *bank mergers and acquisitions* in the banking industry have been determined in *Piloff, Santomero (1998)*. The difficult complicated question: Do the *substantial horizontal mergers* generate the significant price effects?, has been answered, using the evidence from the banking industry, in *Prager, Hannan (1998)*. The *post-acquisition performance of acquiring firms* has been measured in *Rau, Vermaelen (1998)*. An overview of case studies of the *nine mergers* with the particular focus on the *efficiency effects of bank mergers* has been presented in *Rhoades (1998)*. The economic role of *mergers* in the *USA* has been investigated in *Andrade, Stafford (1999)*. The new perspectives on the *mergers* have been provided in *Andrade, Mitchell, Stafford (2001)*. An economic role of *mergers* has been revealed in *Andrade, Stafford (2004)*. Some topics on the knowledge transfer in the *international acquisitions* have been presented in *Bresman, Birkinshaw, Nobel (1999)*. The use of $R^2$ technique in the accounting research in application to the measurements of changes in the value relevance over the last four decades has been reviewed in *Brown, Lo, Lys (1999)*. An empirical investigation on the method of comparables and the tax court valuations of the private firms has been conducted in *Beatty, Riffe, Thompson (1999)*. The knowledge transfer in the *international acquisitions* has been described in *Bresman, Birkinshaw, Nobel (1999)*. The possible ways to capture the real value in the *high-tech acquisitions* have been described in *Chadhuri, Tabrizi (1999)*. An empirical assessment of the residual income valuation model has been done in *Dechow, Hutton, Sloan (1999)*. The straightforward research question: Are you paying too much for that *acquisition*?, has been discussed in *Eccles, Lanes, Wilson (1999)*. A behavioral learning perspective on the influence of *organizational acquisition experience* on the *acquisition performance* has been proposed in *Haleblian, Finkelstein (1999)*. The certain issues on the corporate cash reserves and *acquisitions* have been analyzed in *Harford (1999)*. The driving origins of the merger waves have been found in *Harford (2005)*. The trade-offs between the buyers and the sellers in the *merger and acquisitions* have been described in *Rappaport, Sirower (1999)*. The problems of the leverage and corporate performance, applying the evidences from the unsuccessful takeovers, have been researched in *Safieddine, Titman (1999)*. The effects of *banking mergers* on the loan contracts has been investigated in *Sapienza (1999)*. The topics the consolidation of banking industry and the universal banking have been considered in *Saunders (1999)*. The valuation effects of *bank mergers* have been considered in *Becher (2000)*. The *first international merger wave* (and the fifth and last *US* wave) have been placed in the center of research interest in *Black (2000)*. The *R&D* intensity and *acquisitions* in the high technology industries, considering the evidence from the *US* electronic and electrical equipment industries, have been researched in *Blonigen, Taylor (2000)*. The institutional, cultural and transaction cost influences on the decision to make the *acquisition* or establish the *greenfield start-up* have been considered in *Brouthers K D, Brouthers L E (2000)*. A *CEO* roundtable discussion on the *successful mergers* has been conducted in *Carey (2000)*. The valuation accuracies of the price – earnings and price-book benchmark valuation methods have been discussed in *Cheng, McNamara (2000)*. The industrial restructuring through the *mergers and acquisitions* in the case of *Argentina* in the *1990s* has been described in *Chudnovsky (2000)*. An empirical investigation on the equity undervaluation and the decisions, related to the repurchase tender offers, has been conducted in *DMello, Shroff (2000)*. The advances in the *mergers and acquisitions* between the professional service firms, which explored the undirected process of the integration, have been researched in *Empson (2000)*. The influence of the corporate *acquisitions* on the behavior of key inventors has been researched in *Ernst, Vitt (2000)*. The *cross-border acquisitions* of *US* technology assets have been described in *Inkpen, Sundaram, Rockwood (2000)*. The incentives for the *banking megamergers* have been discussed in *Kane (2000)*. The *cross-border mergers and acquisitions* with the particular focus on their role in the industrial globalization have been discussed in *Kang, Johansson (2000)*. The role of international strategic alliances in the industrial globalization has been described in *Kang, Sakai (2000)*. The comparison of the *acquisitions* and the *divestitures* has been done in *Mulherin, Boone (2000)*. The role of retaining human capital in the *acquisitions* of high-tech firms has been discussed in *Ranft, Lord (2000)*. A review on the influence of cultural compatibility within the *cross-border acquisitions* has been discussed in *Schoenberg (2000)*. The *hostility* in the *takeovers* has been uncovered in *Schwert (2000)*. An empirical examination of motives for the *foreign acquisitions* of *US* firms has been completed in *Seth, Song, Pettit (2000)*. The characteristics of *merging firms* have been described in *Sorensen (2000)*. The *corporate takeovers*, strategic objectives, and *acquiring-firm shareholder wealth* have been reviewed in *Walker (2000)*. The determinants of *US* bank failures and *acquisitions* have been provided in *Wheelock, Wilson (2000)*. The *cross border mergers* and *acquisitions* analysis has been presented in *World Investment Report (2000)*. A longitudinal study on the *technological acquisitions* and the innovation performance of *acquiring firms* has been conducted in *Ahuja, Katila (2001)*. The probability of failure, survival and *acquisition of firms* in the financial distress has been discussed in *Astebro, Winter (2001)*. The executive retention and *acquisition* outcomes have been considered in *Bergh (2001)*. The *CEO* compensation and *bank mergers* have been reviewed in *Bliss, Rosen (2001)*. The *cross-border bank mergers* have been analyzed in *Buch, Delong (2001)*. The asymmetric information, bargaining, and *international mergers* have been studied in *Das, Sengupta (2001)*. The *mergers and acquisitions* through an intellectual capital perspective have been considered in *Gupta, Ross (2001)*. The *merger policies* and trade liberalization have been researched in *Horn, Levinsohn (2001)*. The *merger* and technological change in *1885 - 1998* have been selected as a subject of research in *Jovanovic, Rousseau (2001)*. The *Q-theory of mergers* has been proposed in *Jovanovic, Rousseau (2002a)*. The proposition on the *mergers* as a reallocation has been placed at the center of discussion in *Jovanovic, Rousseau (2002b)*. The question: Are the cash acquisitions associated with the better post-combination operating performance than the *stock acquisitions*?, has been answered in *Linn, Switzer (2001)*. Considering the market for the corporate assets, the following two questions have been raised: *1)* Who engages in the *mergers and asset sales*; and *2)* Are there the efficiency gains?, in *Maksimovic, Phillips (2001)*. The role of managerial incentives in the *corporate acquisitions* in the *1990s* has been discussed in *North (2001)*. The evidence on the *mergers and acquisitions* has been presented in *Paulter (2001)*. The business valuation discounts and premiums have been estimated in *Pratt (2001)*. The learning through the *acquisitions* process has been described in *Vermeulen, Barkema (2001)*. The absolute and relative resources as the determinants of international acquisitions have been investigated in *Anand, Delios (2002)*. The effects of the partners' heterogeneity of experience on the *corporate acquisitions* have been researched in *Beckman, Haunschild (2002)*. The main problems on the sales force optimization after the *merger* have been researched in *Bekier, Shelton (2002)*. A valuation-based approach to the selection of comparable firms has been explored in *Bhojraj, Lee (2002)*. The corporate governance in the case of the cross-*border mergers* has been researched in *Bris, Cabolis (2002)*. The problem: What do the returns to the *acquiring firms* tell us?, has been researched, using the evidence from the firms that make many acquisitions in *Fuller, Netter, Stegemoller (2002)*. The *mergers, acquisitions* and *corporate restructurings* have been researched in *Gaughan (2002)*. The effect of the *mergers and acquisitions* on the technological performance of companies in a high-tech environment has been researched in *Hagedoorn, Duysters (2002)*. The comparison of international strategies such as the *acquisitions* strategy versus the *greenfield investments* strategy has been conducted in *Harzing (2002)*. The question: When do the firms learn from their *acquisition experience*?, has been answered, using the evidences from *1990 – 1995* in *Hayward (2002)*. The operating performance and the method of payments in the *takeovers* have been discussed in *Heron, Lie (2002)*. A beginner's guide on the implications of *cross-border mergers* and *acquisitions* by *TNC*s in developing countries: has been published in *Lall (2002)*. A grounded model of *acquisition implementation* has been proposed in *Ranft, Lord (2002)*. The implications for the small firms of global industrial restructuring have been studied in *Sakai (2002)*. The predictions of the *successful takeovers* and the *risk arbitrage* have been investigated in *Branch, Taewon (2003)*. The bidding wars over the *R&D* intensive firms to obtain the corporate control have been documented in *Coff (2003)*. The market valuations in the new economy have been investigated in *Core, Guay, Van Buskirk (2003)*. The *cross-border mergers and acquisitions wave* of the late *1990s* has been registered in *Evenett (2003)*. The question: What can go wrong and how to prevent it in the process of the *mergers?*, has been at the focus of research in *Gaughan (2003)*. An international comparison of the effects of *mergers* has been completed in *Gugler, Mueller, Yurtoglu, Zulehner (2003)*. The effect of *mergers* on the company employment in the *USA* and *Europe* has been researched in *Gugler, Yurtoglu (2004)*. *Officer* with his co-authors wrote a cycle of research articles on the *M&A transactions*. For instance, the termination fees in *the mergers and acquisitions* have been determined in *Officer (2003)*. The collars and renegotiation in the *mergers and acquisitions* have been researched in *Officer (2004)*. The market pricing of implicit options in the merger collars has been described in *Officer (2006)*. The research on the *acquisition discounts* for the unlisted targets with the determination of the price of corporate liquidity has been completed in *Officer (2007)*. *King* conducted an advanced research program on the *M&A transactions* at the *Wright Peterson Air Force Base* in *Ohio* in the *USA*. For example, the investigation on the integration of *acquired firms* in the high-technology industries has been completed in *King, Driessnack (2003)*. Some issues on the complementary resources and the exploitation of technological innovations have been considered in *King, Covin, Hegarty*

*(2003)*. The meta-analyses of *post-acquisition performance* in *King, Dalton, Daily, Covin (2004)*. The bondholder wealth effects in the *mergers and acquisitions* have been characterized, using the new evidence from the *1980s* and *1990s* in *Billett, Tao-Hsien, King, Mauer (2004)*. The performance implications of firm resource interactions in the *acquisition* of *R&D*-intensive firms have been analyzed in *King, Slotegraaf, Kesner (2008)*. The integration trade-offs in the *technology-grafting acquisitions* have been identified in *Puranam, Singh, Zollo (2003)*. The *mergers* trends have also been researched in *Schonfeld, Malik (2003)*. An explorative study on the relationship between the acquisitions, divestitures and innovations has been conducted in *Van Beers, Sadowski (2003)*. The *mergers and acquisitions* in the telecommunications industry have been studied in *Warf (2003)*. The *applied mergers and acquisitions* have been researched in *Bruner (2004)*. The mandatory bids, squeeze-out, sell-out and the dynamics of tender offer process have been researched in *Burkart, Panunzi (2004)*. The secrets of *successful mergers* have been revealed in *Camara, Renjen (2004)*. An empirical investigation of early mover advantages in acquisitions has been completed in *Carow, Heron, Saxton (2004)*. The sources of gains in the *horizontal mergers*, have been identified, using the evidences from the customers, supplier, and rival firms, in *Feea, Thomas (2004)*. The *CEOs* compensations and incentives, have been researched, using the evidences from the *M&A* bonuses, in *Grinstein (2004)*. The personal benefits, obtained by *CEOs*, whose firms are acquired, have been researched in *Hartzell, Ofek, Yermack (2004)*. The *merger profitability* and the trade policy have been evaluated in *Huck, Konrad (2004)*. The *bank mergers* have been researched in *Humphrey, Vale (2004)*. The progress in the *mergers and acquisitions* has been documented in *Javidan, Pablo, Singh, Hitt, Jemison (2004)*. The *merger policy* and its impact on the innovation has been researched in *Katz, Shelanski (2004)*. The nature of discipline by the *corporate takeovers* has been investigated in *Kini, Kracaw, Mian (2004)*. The price pressure around the *mergers* has been researched in *Mitchell, Pulvino, Stafford (2004)*. The firm size and the gains from the *acquisitions* have been estimated in *Moeller, Schlingemann, Stultz (2004)*. A comparison between the cross-border acquisitions and the domestic acquisitions has been conducted in *Moeller, Schlingemann (2005)*. The interesting problem: How do the diversity of opinion and the information asymmetry affect the *acquirer returns*?, has been researched in *Moeller, Schlingemann, Stulz (2007)*. A survey by the *CFOs* on the *merger motives* and *target valuation* in *Mukherjee, Kiymaz, Baker (2004)*. The gains in the *bank mergers* with the evidences from the bond markets have been researched in *Penas, Unal (2004)*. The *merger programs* and compensation have been studied in *Rosen (2004)*. The cross-country determinants of the *mergers and acquisitions* have been found in *Rossi, Volpin (2004)*. Some research topics on the resources allocation in the *acquisitions* have been explored in *Saxton, Dollinger (2004)*. A review of research and recommendations on the theoretical foundations of the *cross-border mergers and acquisitions* has been conducted in *Shimizu, Hitt, Vaidyanath, Pisano (2004)*. The *post-acquisition strategies* and the *integration capability* in the *US bank mergers* in *Zollo, Singh (2004)*. The *M&A performance* has been considered in *Zollo, Meier (2008)*. The important issues on the corporate business valuation for the *mergers and acquisitions* have been studied in *Aluko, Amidu (2005)*. The valuation for the *mergers, buyout,* and *restructuring* has been completed in *Arzac (2005)*. The generation of the *merger waves* by the *cross-border mergers and acquisitions* has been considered in *Brakman, Garretsen, Van Marrewijk (2005)*. The bank consolidation through the *merger* of *Fleet* and *BankBoston* has been studied in *Calomiris, Pornrojnangkool (2005)*. An empirical analysis of the possible impacts of the *M&A*s on the *R&D* process in the companies has been conducted in *Cassiman, Colombo, Garrone, Veugelers (2005)*. The influence of the *mergers and acquisitions* on the ability of companies to innovate has been researched in *Cassiman, Colombo (2006)*. A conceptual framework on the *M&A* and innovation has been created in *Cassiman, Ueda (2006)*. The consolidation in the wireless phone industry in the *USA* has been described in *Fox (2005)*. A *theory of preemptive mergers* has been proposed in *Fridolfsson, Stennek (2005)*. An exploratory analysis of *reverse takeovers* has been completed in *Gleason, Rosenthal, Wiggins (2005)*. The assessment of the *international mergers and acquisitions* as a mode of foreign direct investment has been conducted in *Globerman, Shapiro (2005)*. Some topics on the measurement and management of the companies values have been researched in *Goedhart, Koller, Wessels (2005)*. A new *theory of mergers and merger waves* has been formulated in *Gorton, Khal, Rosen (2005)*. The multinationals and the global capitalism from the nineteenth century to the twenty-first century have been accurately described in *Jones (2005)*. The *post-merger performance* of bank-holding companies in *1987-1998* has been discussed in *Knapp, Gart, Becher (2005)*. The problem of emotions management in the *mergers and acquisitions* has been discussed in *Kusstatscher, Cooper (2005)*. The various impacts of shareholder control on the *merger payoffs* have been explained in *Moeller (2005)*. The effect of the *food industry mergers and acquisitions* on the employment and wages has been determined in *Ollinger, Nguyen, Blayney, Chambers, Nelson (2005)*. A review of recent research on the banking consolidation and the small business lending in the *USA* has been done in *Ou (2005)*. The effect of *mergers and acquisitions* on the small business lending by the large banks has been revealed in *PM KeyPoint LLC (2005)*. The impact of the acquisitions on the innovations has been analyzed in *Prabhu, Chandy, Ellis (2005)*. The need for a hybrid approach in the organizational integration of the *acquired biotechnology companies* into the pharmaceutical companies has been evidently presented in *Schweizer (2005)*.

The industry structure and the *horizontal takeovers* through the prism of an analysis of the wealth effects on the rivals, suppliers, and corporate customers have been researched in *Shahrur (2005)*. The *cross-border mergers and acquisitions* in the conditions of various international economics have been considered in *Brakman, Garretsen, Van Marrewijk (2006)*. A review and the detailed research agenda, which answers the question: What have we *acquired* and what should we *acquire* in the divestiture research?, has been outlined in *Brauer (2006)*. The problem: How have borrowers fared in the *banking megamergers*?, has been investigated in *Carow, Kane, Narayanan (2006)*. The *mergers and acquisitions* with their effect on the innovative performance of companies in the high-tech industries have been discussed in *Clodt, Hagedoorn, Van Kranenburg (2006)*. The questions: Does the investors misevaluation drive the takeover market?, has been considered in *Dong, Hirschleifer, Richardson, Teoh (2006)*. The *cross-border merger waves* have been detected in *Fumagalli, Vasconcelos (2006)*. The *R&D* investment level and the business environment as the predictors of the *firm's acquisition* have been at the center of consideration in *Heeley, King, Covin (2006)*. The outsourcing of the *R&D* through the *acquisitions* in the pharmaceutical industry has been discussed in *Higgins, Rodriguez (2006)*. The *national mergers* versus the *international mergers* in the conditions of unionized oligopoly has been raised in *Lommerud, Straume, Sorgard (2006)*. The problem on the disruption of inventors in the *acquired companies*, namely the *acquisition integration* and the productivity losses in the technical core, has been formulated in *Paruchuri, Nerkar, Hambrick (2006)*. The management of the coordination-autonomy dilemma in the *technology acquisitions*, when building the organization with the particular focus on the innovation, has been researched in *Puranam, Singh, Zollo (2006)*. The *international mergers* have been considered with the particular attention on the incentives and welfare in *Qiu, Zhou (2006)*. The international linkages between the trade and the *merger policies* have been established in *Saggi, Yildiz (2006)*. The comparable transactions analysis and data manipulation during the *M&A processes* tools have been developed in *Schnoor (2006a, b)*. The literature review on *M&A* and *R&D* with the particular stress on the innovation impacts has been done in *Veugelers (2006)*. The foreign currency exposure and hedging processes by means of the *foreign acquisitions* have been studied in *Bartram, Burns, Helwege (2007)*. The *mergers* in the multidimensional competition have been deeply analyzed in *Davidson, Ferrett (2007)*. The information spillovers in the execution and valuation of the *commercial banks M&As*, using the principles of learning by observing, have been researched in *DeLong, DeYoung (2007)*. The multi-sided platform businesses have been studied by creating an empirical framework with an application to the *Google's* purchase of the *DoubleClick* in *Evans, Noel (2007)*. The *mergers and acquisitions* in conditions of the globalization have been researched in *Hutson (2007)*. The trends and determinants of the *mergers and acquisitions* in the developing countries in *1990s* have been reviewed in *Kamaly (2007)*. The impact of the *acquisition* on the innovation performance by the inventors as well as the productivity of company at the semiconductor companies have been researched in *Kapoor, Lim (2007a, b)*. The accounting for the distress in the *bank mergers* has been analyzed in *Koetter, Bos, Heid, Kolari, Kool, Porath (2007)*. The *acquisition premiums*, subsequent workforce reductions and *post-acquisition performance* have been investigated in *Krishnan, Hitt, Park (2007)*. The customer information sharing in the result of the completed *M&A deal* has been researched in *Kim, Choi (2007)*. The *corporate governance* and the *acquirer returns* have been considered in *Masulis, Wang, Xie (2007)*. The *cross-border mergers* as the instruments of comparative advantage have been studied in *Neary (2007)*. The *cross-border mergers and acquisitions* vs *greenfield foreign direct investment* have been investigated in *Nocke, Yeaple (2007)*. A new evidence from the *corporate takeover market* has been presented in *Boone, Mulherin (2008)*. The *mergers*, corporate control and governance issues, going from the principles of corporate finance, have been described in *Brealy (2008a, b)*. The trade liberalization and industrial restructuring through the *mergers and acquisitions* have been discussed in *Breinlich (2008)*. The *mergers*, *acquisitions*, and other restructuring activities have been studied in *DePamphilis (2008)*. The market valuations of start-up ventures around the technology bubble have been investigated in *Gavious, Schwartz (2008)*. The *post-merger restructuring* and the *boundaries of the firm* have been researched in *Maksimovic, Phillips, Prabhala (2008)*. The globalization and profitability of the *cross-border mergers and acquisitions* have been considered in *Norbäck, Persson (2008a)*. The *cross-border mergers & acquisitions policy* in service markets has been studied in *Norbäck, Persson (2008b)*. The efficiency and tax revenues issues at the *cross-border mergers & acquisitions* have been researched in *Norbäck, Persson, Vlachos (2009)*. The *strategic merger waves* have been researched in *Toxvaerd (2008)*. The financial distress and the firm's exit, including the determinants of involuntary exits, voluntary liquidations and restructuring exits, have been researched in *Balcaen, Buyze, Ooghe (2009)*. A review of literature on the *mergers and acquisitions of financial institutions* after *2000* has been created in *DeYoung, Evanoff, Molyneux (2009)*. The problem on the added value during the firm valuation by the financial experts has been considered in *Elnathan, Gavious, Hauser (2009)*. The *mergers and acquisitions review* has been published in *Jones K (2009)*. The *mergers* and innovation in the big pharma have been researched in *Ornaghi (2009)*. The question: Do the *mergers* improve the information?, using the research evidences from the loan market, has been clarified in *Panetta, Schivardi, Shum (2009)*.

The notes on the *post-merger integration* have been compiled in *Patel, Bourgeois (2009)*. The basics of the *mergers & acquisitions* have been described in *Wong (2009)*. The stock market bubble effects on the *mergers and acquisitions* have been researched in *Aharon, Gavious, Yosef (2010)*. The *acquisitions* as a response to the deregulation, going from the evidences in the cable television industry in *Canada*, have been researched in *Byrne (2010)*. The bidders' strategic timing of *acquisition announcements* and the effects of payment method on the target returns and competing bids have been researched in *Chen, Chou, Lee (2011)*. The anticipation, *acquisitions*, and bidder returns topics have been considered in details in *Jie, Song, Walkling (2011)*. A large sample study of the *mergers and acquisitions* from *1992* to *2009*, including the implications of data screens on the *merger and acquisition analysis*, has been conducted in *Netter, Stegemoller, Wintoki (2011)*. The practical aspects, regarding the stages of *mergers* through the *company acquisition* have been discussed in *Rus (2012)*. The possible reasons: Why do some *targets* accept the *very low and even negative takeover premiums*?, have been identified in *Weitzel, Kling (2012)*.

The **Asian and Australian M&A subject experts** have been worked hard to make the advanced research on the *M&A transactions* in *Asia* during the recent years. The cross border *M&As* in the crisis-affected *Asia* have been analyzed in *UNCTAD (2000)*. An analysis of *mergers* in the private corporate sector in *India* has been done in *Beena (2000)*. The impact of market cycle on the performance of the *Singapore acquirers* has been investigated in *Pangarkar (2004)*. A comparative perspective toward the understanding of the *merger-wave* in the *Indian* corporate sector has been suggested in *Beena (2004)*. An exploratory analysis on the *mergers and acquisitions* in the *Indian* pharmaceutical industry in *Beena (2006a)*. The *mergers and acquisitions* in the *Indian* pharmaceutical industry, including their nature, structure and performance, have been researched in *Beena (2006b)*. The *overseas mergers and acquisitions* by the Indian enterprises, including their patterns and motivations, have been researched in *Pradhan, Abraham (2005)*. The most complicated problem: Which is the best internationalization strategy for the *Indian* pharmaceutical enterprises: The *overseas acquisition* versus the *green-field foreign investment*?, has been researched in *Pradhan, Alakshendra (2006)*. The growth of *Indian* multinationals in the *World economy*, including some implications for their development, has been considered in *Pradhan (2007a)*. The modern trends and patterns of *overseas acquisitions* by the *Indian* multinationals have been discovered in *Pradhan (2007b)*. The causes and consequences of *cross-border acquisitions* in a transition economy, using the *1998 - 2006* deal data for targeted *Chinese* and *Indian* firms and *foreign acquirers*, have been examined in *Nagano, Yuan (2007)*, confirming the fact that that a recent increase in the *cross-* *border acquisition* contributed to the *Gross Domestic Product* (*GDP)* growth in *P.R. China* and *India*. The bank consolidation and the *soft information acquisition* in the small business lending in *Japan* has been researched in *Ogura, Uchida (2007)*. The *mergers and acquisitions* in *Japan, Germany, France, the UK* and *USA* have been explained in *Jackson, Miyajima (2007)*. The *M&A* management from the strategic intent to the integration, including the consideration on the *IOC's acquisition* of the *IBP* in *India*, has been analyzed in *Venkiteswaran (2008)*. The international venturing emerging paradigms consideration in the frames of a study of the *Indian IT industry* has been performed in *Varma (2008)*. The effect of the *mergers and acquisitions* on the market concentration and interest spread in *Pakistan* has been researched in *Mehwish, Kayani, Javid (2012)*.

The **European M&A transactions trends** have been analyzed by the *European* scientists in their various research articles. The economics and politics of *European merger control* have been selected as the main research topics in *Neven, Nuttall, Seabright (1993)*. The evolutionary characteristics of *acquisitions* in *France* in *1959 – 1992* have been identified in *Derhy (1995)*. The effect of the *mergers and acquisitions* on the efficiency and profitability of *EC* credit institutions has been evaluated in *Vennet (1996)*. The wealth creation and the bid resistance in the *UK takeover bids* have been researched in *Holl, Kyriazis (1997)*. The research problem such as: Can the *mergers* foster the efficiency?, in the conditions, when the introduced regulation fosters the *mergers* in the *Italian* researched case, has been analyzed in *Resti (1998)*. The *European* lessons on the consolidation in the banking industry have been provided in *Boot (1999)*. The study on the economics of *bank mergers* in the *European Union* has been completed in *Dermine (1999)*. The problem: Why do banks *merge* in the conditions of financial crisis?, has been clearly answered and further researched with the particular interest in the full spectrum of possible implications for the banking industry and the policy regulation in *Italy* in *Focarelli, Fabio, Salleo (1999), Focarelli, Panetta, Salleo (2002), Focarelli, Panetta, Salleo (2003)*. The *mergers and shareholder wealth* in the *European* banking has been evaluated in *Cybo-Ottone, Murgia (2000)*. The privatization and foreign direct investment in *Central* and *Eastern Europe* have been researched in *Hunya, Kalotay (2000)*. The toeholds, bid-jumps and expected pay-offs in the *takeovers* have been considered in *Betton, Eckbo (2000)*. The *corporate takeovers* have been considered in *Betton, Eckbo, Thorburn (2008)*. The *merger negotiations* and the toehold puzzle have been studied in *Betton, Eckbo, Thorburn (2009)*. The efficiency gains from the *mergers* in the *European economy* have been evaluated in *Roeller, Stenneck, Verboven (2001)*. The banking culture in *Italian financial system* has been discussed in *Carretta (2001)*. The banking culture in the *Italian financial system* has been further analyzed in *Carretta, Farina,*

*Schwizer (2005a)*. The "culture of compliance" between the banks and the supervisory authorities in the *Italian financial system*, including the management of the reputation, investor relation, corporate governance and leadership, has been discussed in *Carretta, Farina, Schwizer (2005b)*. The change management in the process of *post merger integration* of banks in the *Italian financial system* has been described in *Carretta (2007)*. The *M&As* and *post merger integrations* in the *Italian banking industry* have been considered in *Carretta, Farina, Schwizer (2008)*. The market access strategies in the *EU banking sector* have been investigated in *Beckmann, Eppendorfer, Neimke (2002)*. The cycle of research articles by *Conyon* and his co-authors greatly contributed to the *M&A transactions understanding* in the *UK*. The impact of the *mergers and acquisitions* on the company employment in the *United Kingdom* has been investigated in *Conyon, Girma, Thompson, Wright (2002a)*. The question: Do the *hostile mergers* destroy jobs?, has been discussed in *Conyon, Girma, Thompson, Wright (2002b)*. The productivity and wage effects of *foreign acquisitions* in the *United Kingdom* have been investigated in *Conyon, Girma, Thompson, Wright (2002c)*. The question: Do the wages rise or fall after the *merger*?, has been answered in *Conyon, Girma, Thompson, Wright (2004)*. The effect of *foreign acquisitions* on the total factor productivity in the manufacturing industry in the *UK* in *1987 – 1992* has been outlined in *Harris, Robinson (2002)*. An analysis of the *mergers and acquisitions* in the *Turkish* banking sector has been conducted in *Mumcu, Zenginobuz (2002)*. The *mergers and acquisition* in *Germany* have been studied in *Schmid, Wahrenburg (2002)*. An economic perspective on the *corporate governance* in *Germany* has been presented in *Schmidt (2003)*. The people, culture and politics impacts on the *cross-border mergers* have been characterized by the *Danish* scientists in *Soderberg, Vaara (2003)*. A qualitative review on the impact by the cultural differences on the *merger and acquisition performance* has been reported in *Denmark* in *Stahl, Voigt (2003)*. The *UK* evidences on the glamour acquirers, method of payment and *post-acquisition performance* have been presented in *Sudarsanam, Mahate (2003)*. The effect of *mergers and acquisitions* on the bank efficiency in *Greece* has been analyzed in *Athanasoglou, Brissimis (2004)*. The *M&A* success in the *European bank mergers and acquisitions* has been determined in *Beitel, Schiereck, Wahrenburg (2004)*. The economic integration and the profitability of *cross-border mergers and acquisitions* have been investigated in *Bjorvatn (2004)*. The domestic plants' employment and survival prospects after the *foreign acquisition* have been discussed in *Girma, Görg (2004)*. The shareholder wealth effects of the *European domestic and cross-border takeover bids* have been described in *Goergen, Renneboog (2004)*. The corporate governance convergence, has been investigated, applying the evidences from the *takeover regulation reforms* in *Europe* in *Goergen, Martynova, Renneboog (2005)*. The triggers, performance and motives of the *takeover waves* have been described in *Martynova, Renneboog (2005)*. The *mergers and acquisitions* in *Europe* have been accurately characterized in *Martynova, Renneboog (2006)*. The *Finish* perspective on the *acquiring and acquired companies* has been proposed in *Lehto, Lehtoranta (2004)*. The employment effects of *mergers* have been analyzed in *Lehto, Böckerman (2006)*. An event study of *M&As* in the *European banking industry* with the particular interest in the diversification problem versus the specialization problem has been completed in *Lepetit, Patry, Rous (2004)*. The economics of the proposed *European takeover directive* has been studied in *McCahery, Renneboog, Ritter, Haller (2004)*. The macroeconomic determinants of the *cross-border mergers and acquisitions* have been studied, using the *European* and *Asian* evidences, in *Aminian, Campart (2005)*. The question: Did the concentration on the core competencies drive the *merger and acquisition* activities in the *1990s*?, has been researched, using the empirical evidences in *Germany*, in *Hussinger (2005)*. The absorptive capacity and the *post-acquisition inventor productivity* after the *M&A deals* in *Germany* have been investigated in *Hussinger (2010, 2012)*. The modelling of the *European mergers*, including the theory, competition and case studies, has been performed in *Röller (2005)*. The impact on the *UK acquirers* of the *domestic cross-border public and private acquisitions* has been investigated in *Conn, Cosh, Guest, Hughes (2005)*. The case of cross-border *M&A activity* has been researched in *Di Giovanni (2005)*. The choice of payment method in *European mergers and acquisitions* has been considered in *Faccio, Masulis (2005)*. Some problems on the development of the country multinationals have been found to exist in *France* in *Aykut, Goldstein (2006)*. The determinants of the *mergers and bankruptcies* in *Switzerland* in *1995 - 2000* have been found in *Buehler, Kaiser, Jaeger (2006)*. The *M&A performance* in the *European financial industry* has been selected as a subject of main research interest in *Campa, Hernando (2006)*. The corporate governance and the agency costs of the debt and outside equity have been researched in *Szilagyi (2007)*. The corporate finance in *Norway*, including the capital structure and hybrid capital, has been studied in *Mjøs (2007)*. The taxation of the foreign profits in the cases of the *international mergers and acquisitions*, considering the source versus the residence based taxation, has been evidently described in *Becker, Fuest (2007a, b)*. The change management during the *post merger integration* in the case of the *Unicredit* group in *Italy* has been researched in *Farina (2007)*. The *mergers & acquisitions* and the innovation performance in the telecommunications equipment industry in *Germany* have been discussed in *Gantumur, Stephan (2007)*. The macroeconomic determinants of the *cross border mergers and acquisitions* and *greenfield investments* have been defined in *Neto, Brandão, Cerqueira (2008a)*. The impact of *FDI*, *cross border mergers* and *acquisitions* and *greenfield investments* on economic growth has been evaluated in *Neto, Brandão, Cerqueira (2008b)*. The *mergers, acquisitions* and *technological regimes*, going from the *European* experience over the period of *2002 – 2005*, have been presented in *Damiani, Pompei (2008)*. The *acquisitions*, *divestitures* and *innovation performance* in the *Netherlands* have been extensively researched in *Van Beers, Dekker (2009)*. The two central problems in the *M&A transactions* completion process: *1)* the reasons why *acquisitions* and *divestitures* occur, and *2)* the impact of *acquisitions* and *divestitures* on firms' innovation performance, have been considered in details in *Van Beers, Dekker (2009)*. The business problem: Do *financial conglomerates* create or destroy value?, going from the *EU* evidences, has been discussed in *Van Lelyveld, Knot (2009)*. The stock price reaction to the *M&A announcements* at the *London Stock Exchange* has been thoughtfully researched in *Spyrou, Siougle (2010)*. The new logic of the *merger and acquisition pricing*, using the accumulated knowledge base in *Germany*, has been explained in *Lenz (2010)*. The impact by the *European bank mergers* on the bidder default risk has been estimated in *Vallascas, Hegendorff (2011)*. The *cross-border mergers* and market segmentation have been described in *Chaudhuri (2011)*. The dynamics of *M&A transactions* during the recent crisis in the *European banking sector* has been investigated in *Beltratti, Paladino (2011)*. The *merger and acquisition performance* of *European banks* has been researched in *Asimakopoulos, Athanasoglou (2011)*. The *bank mergers and acquisitions* and the *lending relationships* in *Norway* have been considered in *Hetland, Mjøs (2011)*. The competition protection and the *Philip Kotler's* strategic recommendations, including some issues on the *M&A transactions* in *Poland*, have been researched in *Fornalczyk (2012)*. The overview of *the mergers & acquisitions* in *Switzerland* in *2012* and outlook for *the mergers & acquisitions* in *2013* have been presented in *Pfister, Kerler, Valk, Prien, Peyer, Kuhn, Hintermann, Ljaskowsky, Arnet (2013)*. The global trends in the *international mergers and acquisitions* in the financial sector and theirs features in *Ukraine* have been explored in *Shumska, Stepanenko-Lypovyk (2013)*. The next big luxury *M&A targets* in *2014* have been analyzed in *Friedman (2013)*.

We would like to explain that, in this research article, the authors prefer to limit our research considerations by an assumption that the selection of the *mergers and acquisitions transactions strategies* takes place in the *diffusion-type financial systems* in the *imperfect highly volatile global capital markets* with the *nonlinearities*. Since the time, when the first *financial systems* were established to govern the money markets in *Bagehot (1873, 1897), Fisher (1892)*, the *diffusion theory* has been frequently applied to accurately characterize the *diffusion - type financial systems* in the *finances*. The multiple evidences of the fact that the *diffusion processes* have the considerable influences on the various *econophysical* and *econometrical parameters* of the *diffusion-type financial systems* have been described in *Bachelier (1900), Volterra (1906), Slutsky (1910, 1912, 1913, 1914, 1915, 1922a, b, 1923a, b c, 1925a, b, 1926, 1927a, b, 1929, 1935, 1937a, b, 1942), Osborne (1959), Alexander (1961), Shiryaev (1961, 1963, 1964, 1965, 1967, 1978, 1998a, b, 2002, 2008a, b, 2010), Grigelionis, Shiryaev (1966), Graversen, Peskir, Shiryaev (2001), Kallsen, Shiryaev (2001, 2002), Jacod, Shiryaev (2003), Peskir, Shiryaev (2006), Feinberg, Shiryaev (2006), du Toit, Peskir, Shiryaev (2007), Eberlein, Papapantoleon, Shiryaev (2008, 2009), Shiryaev, Zryumov (2009), Shiryaev, Novikov (2009), Gapeev, Shiryaev (2010), Karatzas, Shiryaev, Shkolnikov (2011), Shiryaev, Zhitlukhin (2012), Zhitlukhin, Shiryaev (2012), Feinberg, Mandava, Shiryaev (2013)*, *Akerlof, Stiglitz (1966), Rothschild, Stiglitz (1976), Stiglitz, Weiss (1981), Richiardi, Gallegati, Greenwald, Stiglitz (2007), Jaffee, Russell (1976)*, *Leland, Pyle (1977), Bernanke (1979, 2002, 2004, 2007, 2009a, b, c, d, e, 2010a, b, 2012a, b, 2013a, b, c, d, e, f, g, h), Bernanke, Blinder (1992), Bernanke, Gertler (1995), Bernanke, Reinhart (2004), Bernanke, Reinhart, Sack (2004), Bernanke, Blanchard, Summers, Weber (2013)*, *Shiller, Pound (1989), Conley, Hansen, Luttmer, Scheinkman (1997)*, *Stock, Watson (2002), Xiaohong Chen, Hansen, Carrasco (2009), Ledenyov D O, Ledenyov V O (2013f, g, h, i).*

Finally, we would like to say that this short condensed research article on the selection and implementation of the *mergers and acquisitions transactions strategies* in the *diffusion - type financial systems* in *the highly volatile global capital markets* with the *nonlinearities,* has the *three main research goals*:

*1.* to present the extensive review on the current state of research on the *mergers and acquisitions transactions strategies* selection and *implementation* in the *diffusion - type financial systems* in *the highly volatile global capital markets* with the *nonlinearities*;

*2.* to report the innovative research proposals and cutting-edge research results on the *mergers and acquisitions transactions implementation strategies* in the *diffusion - type financial systems* in *the highly volatile global capital markets* with the *nonlinearities*;

*3.* to enhance our general understanding on *the nature of the nonlinearities in the finances* in *Ledenyov V O, Ledenyov D O (2012a, b), Ledenyov D O, Ledenyov V O (2012c, d)*, *Ledenyov D O, Ledenyov V O (2013a, b, c, d, e, f, g, h, i)*.

As always we would like to comment that this innovative research article explores the subject of our research interest, using the extended knowledge base on *the nonlinearities in the microwave superconductivity* in *Ledenyov D O, Ledenyov V O (2012e)*.

# Mergers and Acquisitions transactions: Review on business valuation methodologies and comparable transactions analysis techniques

We would like to continue this research article by writing a clear definition of the *Merger and Acquisition (M&A) transaction* in *Wikipedia (2013)*: "***Mergers and acquisitions (M&A) is an aspect of corporate strategy, corporate finance and management dealing with the buying, selling, dividing and combining of different companies and similar entities that can help an enterprise grow rapidly in its sector or location of origin, or a new field or new location, without creating a subsidiary, other child entity or using a joint venture.***"

Let us provide a few more possible definitions of the *M&A transaction* given by other researchers.

*Jones K (2009)* defines the *M&A transaction* as: "*Mergers and acquisitions* refers to the *corporate strategy, finance,* and *management* that deals with the buying, selling, and combining of companies; mergers and acquisitions can finance or help a growing company in a given industry expand rapidly without the necessity of creating another business entity *Wikipedia (2009)*. The ultimate goal of a merger is to create *value*. *Value* can only be created when the value of *Company A + Company B* is greater than the value of *Company A* and *Company B* separately."

*Fornalczyk (2012)* writes: "*Mergers and acquisitions (M&A)* offer an alternative to strategic alliances in order to strengthen market position. Mergers result in higher concentration of assets in the hands of a single company, the outcome of acquisitions is the creation and expansion of capital groups. Competition law treats a capital group as one economic entity because subsidiaries are coordinated by the dominant company within the group."

*Brakman, Garretsen, Van Marrewijk (2005)* explain the general motivation behind the *M&A deals*: Following *Neary (2004a)* various motives for *M&As* can be distinguished in general. In the *Industrial Organization (IO)* literature two basic motives stand out: an *efficiency motive* and a *strategic motive*. *Efficiency gains* arise because takeovers increase synergy between firms that increase economies of scale or scope. Furthermore, from a *strategic perspective*, *M&As* might change the market structure and as such have an impact on firm profits, which might even be reduced to zero (this is the so-called '*merger paradox*', *Salant et al. (1983)*)."

It makes sense to add that the **M&A main purposes** and business transformation goals may include in *Wikipedia (2013)*:

1. *Economy of scale*: the optimization of combined company by removing duplicate departments or operations, lowering the costs of the company relative to the same revenue stream, thus increasing profit margins.
2. *Economy of scope*: the optimization of combined company by increasing or decreasing the scope of marketing and distribution of different types of products.
3. *Increase of revenue or market share*: the combined company increase its market power (by capturing increased market share) to set prices.
4. *Cross-selling*: the combined company will increase the customer base.
5. *Synergy*: the combined company will have the opportunities of managerial specialization, increased order size and associated bulk-buying discounts.
6. *Taxation*: the combined profitable company can buy the loss making companies to reduce its tax liability.
7. *Geographical or other diversification*: the combined company will smooth the earnings results of a company, which over the long term smoothens the stock price of a company, giving conservative investors more confidence in investing in the company.
8. *Resource transfer*: the combined company resources can create value through either overcoming information asymmetry or by combining scarce resources.
9. *Vertical integration*: Vertical integration occurs when an upstream and downstream firm merge (or one acquires the other). There are several reasons for this to occur. One reason is to internalize an externality problem. A common example of such an externality is double marginalization. Double marginalization occurs when both the upstream and downstream firms have monopoly power and each firm reduces output from the competitive level to the monopoly level, creating two deadweight losses. Following a merger, the vertically integrated firm can collect one deadweight loss by setting the downstream firm's output to the competitive level. This increases profits and consumer surplus. A *merger* that creates a vertically integrated firm can be profitable.
10. *Hiring*: the combined company will use the acquisitions as an alternative to the normal hiring process
11. *Absorption*: the combined company will absorb the similar businesses under single management.
12. *Diversification*: the combined company may expect to hedge against a downturn in an individual industry.

13. *Manager's compensation*: the senior management team at the combined company may expect to increase their executive compensation, based on the total amount of profit of the company, instead of the profit per share, which would give the team a perverse incentive to buy companies to increase the total profit while decreasing the profit per share (which hurts the owners of the company, the shareholders).

There are the three **M&A transactions main categories** as explained in *Jones T (2009):* "Typically, there are three categories of mergers:

1. *Horizontal,*
2. *Vertical*, and
3. *Conglomerate/Diversification.*

Additional subcategories include product-extension merger, purchase merger, consolidation merger, accretive merger, and dilutive merger."

*Jones T (2009)* provides the following definitions for the ***three main M&A transactions categories:***

1. "A *horizontal merger* is a merger of two companies in the same line of business.
2. *Vertical mergers* involve companies that cater to different stages of production. The buyer expands back toward the source of raw materials or forward in the direction of the customer.
3. A *conglomerate or diversification merger* is the union of two companies that are not competitors, not in the same industry, and not a part of the same supply chain. A conglomerate merger is the acquisition of a company in a totally separate line of business. Typically, the acquired company is operated as a separate business unit or a wholly owned subsidiary of the parent company."

In the time of initiation of the *M&A* transaction, the special attention has to be focused on the ***Due Diligence*** process by which the *M&A* buyers inspect a target company. There are the two *Due Diligence* possibilities in *M&A* transaction:

1. *Friendly Bid*, when the acquisition is supported by the management of the target business. Frequently, the *material adverse change provisions (MAC)* are used to renegotiate the price of the transaction in this case.
2. *Hostile Bid*, when the acquisition is opposed by the management of the target business.

The ***Business Valuation*** in the *M&A* deal includes the following types of valuation in *Wikipedia (2013)*:

1. The asset valuation,

2. The historical earnings valuation,
3. The future maintainable earnings valuation,
4. The relative valuation (comparable company & comparable transactions),
5. The discounted cash flow (*DCF*) valuation.

Let us review the modern **comparable transactions analysis** and **valuation methodologies** in the next few pages, considering the **Concept of Valuation**, which is increasingly important during the *M&A transaction process* origination and completion in *Schnoor (2006), Arzac (2005), Goedhart, Koller, Wessels (2005), Lie E, Lie H J (2002), Liu, Nissim, Thomas (2002)*:

1. *Public Equity Offerings*: How much is a company or division worth in the public markets?
2. *Debt Offerings*: What is the underlying value of the business or assets against which debt is being issued?
3. *Mergers and Acquisitions*: How much is the target company worth? What is the value of potential synergies?
4. *Divestitures*: How much can a company or division be sold for?
5. *Public Defense*: Is the company undervalued or overvalued in the event of a hostile bid?
6. *Fairness of opinions*: Is the price offered for a company or division fair from a financial point of view?
7. *Research*: Should the firm's clients buy, hold or sell positions in a particular security?

The **Valuation Methodologies** include in *Schnoor (2006)*:

1. *Publicly Traded Company Analysis Methodology*:
    a) Public market valuation,
    b) Value is based on market trading multiples of comparable companies,
    c) Can be regarded as a market's shortcut to a Discounted Cash Flow (DCF),
    d) This methodology does not usually reflect an M&A control premium.
2. *Comparable Transactions Analysis Methodology*:
    a) Focused on change of control situations,
    b) Value is based on multiples paid for the comparable companies in sale transactions,
    c) This methodology includes a control premium.
3. *Discounted Cash Flow Analysis Methodology*:

      a)       Represents the intrinsic value of a business,

      b)       Requires a detailed financial forecast (usually five to ten years),

      c)       Usually the most detailed and intensive methodology to prepare.

*4.*    *Leveraged Buyout Analysis Methodology*:

      a)       Variation on a Discounted Cash Flow (DCF) analysis with different assumptions,

      b)       Represents the value to a financial / LBO buyer,

      c)       Value is based on the Free Cash Flows, Debt Repayment and Return on Investment (*ROI*).

*5.*    *Other Methodologies*:

      a)       Liquidation Analysis,

      b)       Break up Analysis,

      c)       Greenfield / "Cost to Build" Analysis.

The two most common **Measures of Company Value** are in *Schnoor (2006)*

*1.*    **Equity Value**

      a)       Represents the value attributed to the common shareholders of the business

      b)       It is the value remaining after all debt and preferred stock obligations have been satisfied,

      c)       Often called **Market Capitalization,**

      d)       **Market Capitalization** = Diluted Shares O/S x Current Share Price.

*2.*    **Enterprise Value**

      a)       Sometimes referred to as "**Firm Value,**"

      b)       Represents the market value of all capital invested in a business,

      c)       Theoretically, debt and preferred securities should be valued at market, but this is often impractical,

      d)       **Enterprise Value** = Market Capitalization + Net Debt + Preferred Equity + Minority Interest – Long Term Investments,

      e)       Net Debt = Short Term Debt + Long Term Debt + Capitalized Leases – (Cash + Cash Equivalents).

The following diagram illustrates the **Minority Interest** and **Investments** situations in Company A and its three subsidiaries in *Schnoor (2006)*:

|  | **Subsidiary X** | **Subsidiary Y** | **Subsidiary Z** |
|---|---|---|---|
| Ownership Percentage | 100% | 30% | 75% |
| Accounting method | Consolidation | Equity Method | Consolidation |
| Does Subsidiary appear as a Long Term Investment on the Balance Sheet? | No | Yes | No |
| Does ownership create Minority Interest on Balance Sheet? | No | No | Yes |
| Amount of Minority Interest | 0% | 0% | 25% |
| Percent of EBITDA in Company A's EBITDA | 100% | 0% | 100% |
| Value of Subsidiary in Company A's Market Value | 100% | 30% | 75% |
| Are the previous two rows consistent? | Yes | No | No |
| Possible adjustments to reconcile | Not Required | Add 30% of Sub. Y's EBITDA to Co. A's EBITDA **or** reduce Co. A's Market Value by the value of Sub Y | Subtract 25% of Sub. Z's EBITDA from Co. A's EBITDA or increase Co. A's market Value by the value of sub. Z that is not owned |
| Decision Making criteria | Does Co. A have control over the subsidiary? | Does Co. A have control over the respective subsidiary? | Does Co. A have control over the respective subsidiary? |
| Typical Convention to resolve | None Required | No, so reduce Co. A's Market Value by the value of Sub. Y | Yes, increase Co. A's Market Value by the value of Sub. Z that is not owned |

***Tab. 1.*** *Minority Interest and Investments Situations in Company A and its Three Subsidiaries (after Schnoor (2006)).*

The *Equity Value* and *Enterprise Value* can be represented as in *Schnoor (2006)*:
1. Market Basis: ***Enterprise Value = Net Debt + Equity Value***,
2. Book Basis: ***Assets = Liabilities + Shareholders Equity***.

The *Equity Value Multiples* in *Schnoor (2006)*:
1. Certain ratios or values apply to equity holders only – ***Net Income*, *Cash Flow*, and *Book Value***,
2. Since these values are after debt, multiples applied to these values are based only on the value of the equity,
3. Some relevant equity value multiples are:
    a) ***Price /Earnings*** (P/E),
    b) ***Price / Book Value***,
    c) ***Price / Cash Flow***.

The *Enterprise Value Multiples* in *Schnoor (2006)*:
1. Certain ratios or values apply to all capital providers (including debt and equity) – ***Revenues*, *EBITDA*, *EBIT***,
2. These values are before the cost of debt, so the relevant multiples should be based on Enterprise Value,
3. Some relevant Enterprise value multiples are:
    a) Enterprise Value / Sales,
    b) Enterprise Value / EBITDA.

It is imperative to understand the differences between *Equity Value* and *Enterprise Value* and their respective multiples. The main difference between their use in various multiples resolves around the treatment of debt. A multiple with debt in the numerator must contain a value that is before interest in the denominator in *Schnoor (2006)*.

*Equity Value or Enterprise Value*

| Ratio Denominator | Has Interest been subtracted? | Numerator |
| --- | --- | --- |
| Book Value | Yes | Equity Value |
| Net Income | Yes | Equity Value |
| Cash Flow | Yes | Equity Value |
| Revenue | No | Enterprise Value |
| EBITDA | No | Enterprise Value |
| EBIT | No | Enterprise Value |

**Tab. 2.** *Equity value or enterprise value (after Schnoor (2006)).*

The *Publicly Traded Comparable Company Analysis* overview in *Schnoor (2006)*:

1. The analysis of publicly traded comparable companies usually consists of a comparison of several companies operating and trading statistics,
2. The exact ratios and values will vary from industry to industry,
3. There is no perfect, consistent template that can be used for any company,
4. Depending on the particular situation, multiples can be calculated in different ways:
    a) Valuation,
    b) Credit Analysis / Liquidity,
    c) Restructuring,
5. Each company has to be analyzed carefully to find adjustments and subtleties,
6. In addition, every senior banker may have preferences for specific ratios or calculations.

The *Pros and Cons of Comparables* are summarized in the table in *Schnoor (2006)*:

| PROS | CONS |
|---|---|
| Provides a benchmark to value a company by referring to other similar public companies | Does not account for "control premiums" nor potential synergies realized in an acquisition |
| Calculates valuation multiples based on current market conditions | May not reflect fundamental value in thinly traded, small capitalization or poorly followed stocks |
| Takes into account industry trends and growth prospects | For many companies, there are few, if any, good comparable companies |
| Provides insight into key valuation multiples for an industry | Does not explain market inefficiencies |
| Serves as a reliable value indicator for a minority investment | |

*Tab. 3. Pros and cons of comparables (after Schnoor (2006)).*

The **key to choosing appropriate comparables** or **compiling useful trading comparables** is to first identify *companies* that are considered comparable, based on the following criteria's in *Schnoor (2006)*:

1. *Operational Criteria's*:
    a) Industry,

- b) Products,
- c) Distribution Channels,
- d) Customers,
- e) Seasonality,
- f) Cyclicality,
- g) Geographic Location.

**2. *Financial Criteria's*:**
- a) Size,
- b) Leverage,
- c) Profit Margins,
- d) Growth Prospects,
- e) Shareholder Base,
- f) Risk Profile.

The ***Sources for Identifying Comparable Companies*** include in *Schnoor (2006)*:

1. SIC Code Screens,
2. Equity Research,
3. 1—K/Annual Report,
4. Proxy Statement,
5. Other Bankers,
6. Client.

The ***Time Period*** in a comparable analysis depends on the company, industry, and practices of different groups. Forecast multiples are generally more important than historical multiples, because investors pay for future earnings. However, historical data is usually more complete, and is often relied upon as well in *Schnoor (2006)*:

1. **LFY-1**: Latest fiscal year minus one
2. **LFY**: Latest fiscal year
3. **LTM**: Last twelve months
4. **LFY+1**: Forecast for the next fiscal year
5. **LFY+2**: Forecast for the year after next

The ***Time Period Adjustments***: *LTM Income Statement and Cash Flow Statement* values are calculated as follows in *Schnoor (2006)*:

***LTM** = Fiscal Year Ended 12/31/2005 – Q1 Ended 3/31/2005 + Q1 Ended 3/31/2006*

In future periods, if the companies don't all have the same year end, forecasts should be calendarized to make them comparable:

1. If Company A has a September 30 year end, and all its peers have a December 31 year end, Company A's forecast should be calendarized
2. Earnings per Share estimates for Company A are as follows:
    a) Fiscal 2006: $1.59
    b) Fiscal 2007: $1.86
3. Calendar 2006 EPS = (3/4 *1.59 + ¼ * 1.86) = $1.66

The *Income Statement Items* include (see the below Table) in *Schnoor (2006)*:

*Income Statement and Cash Flow Items*

| Item | Comments |
|---|---|
| **Revenues** | 1. Should only include revenue from the sale of the company's goods and services<br>2. Exclude interest and other income<br>3. If a company reports **Gross** and **Net Revenue**, use **Net Revenue** |
| **EBITDA** | 1. The most common performance measure among investment bankers<br>2. Serves as a pre-tax proxy for cash flow generated from operations<br>3. **EBITDA** and **EBIT** multiples attempt to normalize for differences in companies capital structures<br>4. The comparable analysis model calculates **EBITDA** by adding **EBIT** and **D&A** |
| **Depreciation and Amortization** | 1. Includes depreciation for **PP&E**, goodwill amortization, and items such as depletion for mining companies<br>2. Do not include amortization of debt issuance costs as these figures are typically included in interest expense |
| **EBIT** | 1. Must exclude special charges, non-recurring items and discontinued operations |
| **Earnings per Share** | 1. Usually looked at after preferred dividends and extraordinary items<br>2. Needs to be adjusted after-tax for any non-recurring items<br>3. Used for the calculation of **Price/Earnings** multiples |
| **Cash Flow** | 1. **Net Income + Deferred Taxes + D&A + Other non-cash items**<br>2. Measure of cash generated by a company after leverage and taxes, but generally before working capital items<br>3. A closer approximation of cash generated by a company's operations |

**Tab. 4.** *Income statement and cash flow items (after Schnoor (2006)).*

The ***Balance Sheet Items*** include (see the below Table) in *Schnoor (2006)*:

*Balance Sheet Items*

| Balance Sheet Items | Comments |
|---|---|
| **Cash and Marketable Securities** | 1. Includes cash & cash equivalents<br>2. Check for any long-term investments in marketable securities |
| **Short-Term Debt** | 1. Includes notes payable, commercial paper, lines of credit, bank overdrafts, current portion of long-term debt and capital leases |
| **Long-Term Debt** | 1. Includes long-term debt and capitalized lease obligations |
| **Minority Interest** | 1. Includes minority interest as it appears on balance sheet without adjustments<br>2. Represents the portion of earnings that are attributable to shareholders owning less than 50% of a subsidiary |
| **Preferred Shares** | 1. Includes Preferred Stock on the company's balance sheet that has debt-like characteristics |
| **Convertible Securities** | 1. If the security is in the money, treat as equity;<br>2. It the security is out of money, treat as debt |
| **Common Equity** | 1. Includes common stock, paid-in capital and retained earnings<br>2. Do not confuse with shareholders' equity, which generally includes preferred stock – confirm your group's definition of these categories |
| **Shareholders' Equity** | 1. Includes the Common Equity described above, plus the book value of preferred stock that is considered equity |

*Tab. 5. Balance sheet items (after Schnoor (2006)).*

The ***Shares Outstanding Values*** used in ***Comparable Company Analysis***. The following are the ***Shares Outstanding Values*** that are required in the comparable company analysis model in *Schnoor (2006)*:

*Shares Outstanding Definitions*

| **Shares Outstanding Category** | **Comments** |
|---|---|
| **Basic Shares Outstanding as of Comp date** | 1. Represents the total number of shares issued and outstanding as of the comp date<br>2. Start with the **Basic Shares Outstanding** as of the latest balance sheet date and check press releases to see if any shares have been issued or redeemed since the last balance sheet date<br>3. Used to calculate the company's market capitalization |
| **Fully Diluted Shares Outstanding as of the Comp Date** | 1. This number is calculated in the comp model<br>2. The basic shares outstanding is added together with the total number of in-the-money options to arrive at this value |
| **LTM Weighted Average Fully Diluted Shares Outstanding** | 1. This number is typically found in the notes to the financial statements<br>2. Used to calculate **Earnings per Share** and **Cash Flow per Share** |

*Tab. 6. Shares outstanding definitions (after Schnoor (2006)).*

The ***Performance Ratios*** include the following profitability ratios, which are often used when analyzing a company in *Schnoor (2006)*:

*Profitability Ratios*

| **Profitability Ratio** | **Definition** |
|---|---|
| **Return on Equity** | ROE = Net Income / Common Equity |
| **Gross Margin** | Gross Margin = Gross Profit / Net Sales |
| **EBITDA Margin** | EBITDA Margin = EBITDA / NET Sales |
| **EBIT Margin** | EBIT Margin = EBIT / Net Sales |
| **Net Income Margin** | NI Margin = Net Income / Net Sales |

*Tab. 7. Profitability ratios (after Schnoor (2006)).*

The ***Valuation Ratios*** include the following ratios, which are often used when analyzing a company in *Schnoor (2006)*:

*Valuation Ratios*

| Valuation Ratio | Definition |
|---|---|
| **Price / Earnings (P/E)** | P/E = Current Share Price / Fully Diluted EPS |
| **Price / Cash Flow** | P/CF = Current Share Price / (F/D) Operating CFPS |
| **Price / Book Value** | P/BV = Current Share Price / Book Value per Share |
| **Enterprise Value / Revenue** | EV/Rev = Enterprise Value / Revenue |
| **Enterprise Value / EBITDA** | EV / EBITDA = Enterprise Value / EBITDA |
| **Enterprise Value / EBIT** | EV/EBIT = Enterprise Value / EBIT |

***Tab. 8.*** *Valuation ratios (after Schnoor (2006))*.

The ***Credit Ratios*** include the ratios, which are often used to assess a company's debt capacity, and may also provide insight into a company's trading performance in *Schnoor (2006)*:

*Credit Ratios*

| Credit Ratios | Definition |
|---|---|
| **Net Debt / Total Cap (Book)** | Debt / Cap (Book) = Net Debt / Capitalization (Book) |
| **Net Debt / Total Cap (Market)** | Debt / Cap (Market) = Net Debt / Capitalization (Market) |
| **EBITDA / Interest** | EBITDA / Interest = EBITDA / Interest Expense |
| **(EBITDA – CAPEX) / Interest** | (EBITDA – CAPEX) / Interest = (EBITDA – CAPEX) / Interest |
| **Total Debt / EBITDA** | Debt / EBITDA = Total Debt / EBITDA |

***Tab. 9.*** *Credit ratios (after Schnoor (2006))*.

*Adjusting for the Operating Leases*: When calculating ***Credit Ratios***, you may need to capitalize a company's operating leases to make the company comparable with its peers in *Schnoor (2006)*:

1. An ***Operating Lease*** is a lease for which the lessee acquires the property for only small portion of its useful life.
2. A ***Capital Lease*** is a lease that meets one or more of the following criteria, and as such is classified as a purchase:

a) The lease term is greater than 75% of the property's estimated economic life
   b) The lease contains an option to purchase the property for less than fair market value
   c) Ownership of the property is transferred to the lessee at the end of the lease term
   d) The present value of the lease payments exceeds 90% of the fair market value of the property
3. An Operating Lease would be capitalized as follows:
   a) The operating lease expense (which can be found in the notes to the financial statements) would usually be multiplied by a ratio of 6.0x – 8.0x
   b) This capitalized value would get added to the company's debt
   c) The operating lease expense would then be subtracted from the company's costs to arrive at a higher EBITDA
   d) This adjustment is not normally done, when calculating trading multiples

*Adjusting for the Securitizations*: It may also be necessary to adjust for any securitized assets in *Schnoor (2006)*:
1. If a company has securitized some assets, you may need to add these assets back to the balance sheet
   a) This value can be found in the notes to the financial statements
   b) Add the assets back to the appropriate working capital item
   c) Subtract the corresponding amount from the cash line (or add the corresponding amount to the company's debt)
2. These adjustments are made for comparison purposes so that you are comparing companies on an apples-to-apples basis.

The *Important Reminders*, when preparing *Comparable Company Analysis* in *Schnoor (2006)*:
1. Eliminate non-recurring items that are recorded before NI:
   a) Restructuring Charges / one-time write offs
   b) Gains or losses on the sale of assets
   c) Read all footnotes and MD&A carefully to find these items
2. Tax-effect all adjustments:
   a) Check MD&A and footnotes for actual tax impact, if available

      b) If not available, use the company's marginal tax rate
3. D&A may be buried in COGS or SG&A, so look at the Cash Flow Statement
4. Make sure to understand the company's pension liability
5. Double check all calculations = perform reality checks
6. Use most current financials to check historical data
7. Include Marketable Securities in Cash
8. Subtract Long-Term Investments when calculating Enterprise Value
9. Calendarize earnings estimates

The *Analytical Trouble Shooting*: Read all disclosures carefully and be prepared to adjust for the following types of occurrences in *Schnoor (2006)*:
1. Stock splits, Dividends and Repurchases
2. Non-calendar year ends (EPS estimates)
3. Cash (Long-term investments)
4. Recent acquisitions and divestitures – pro forma numbers
5. Changes in earnings estimates
6. Differences in accounting treatment
7. Non-recurring items
8. Recent debt or equity offerings
9. Temporary sector reactions
10. Take-over activity among competitors
11. Conversion of convertible securities since the last reporting period
12. Differences in international accounting treatment.

Let us review the comparable transactions analysis by considering the *Comparable Transactions Analysis* in *Schnoor (2006):*
1. The *Comparable Transactions Analysis* provides some information on the transactions, which are completed in the same industry as the company being valued
2. *The Comparable Transaction Multiples* provide insight into:
    a) *Premiums*, which are paid by the acquirers to gain a control over the target companies, and
    b) *Potential for synergies*.
3. The *Comparable Transaction Multiples* preparation, may require the consideration of the following data:

1. The *industry of the company* being valued, in order to properly screen transactions,
2. The *time frame of specific transactions* – typically you want transactions that have been consummated in the past few years,
3. The *status of past transactions*:
    a) Successfully completed transaction,
    b) Pending transaction,
    c) Terminated transaction,
    d) Consideration: Cash vs. Stock exchange,
    e) Hostile vs. Friendly bids.

The attention has to be paid to the following **Transaction Issues** in *Schnoor (2006)*:
1. Two types of transaction multiples are usually calculated:
    a) **Financial Multiples** (Enterprise Value / LTM EBITDA)
    b) **Industry Specific** Multiples (Enterprise value / annual Production)
2. Comparable transaction analysis provides insight into the M&A activity in a specific industry
    a) Activity relative to the overall market
    b) Who is buying, and what are they buying?
    c) What kinds of premiums are buyers paying?
3. Provides an understanding of the events surrounding specific transactions:
    a) Hotly contested transactions
    b) Privately- negotiated friendly deals
    c) Major transactions that impact an entire industry
    d) Timing of transactions relative to specific business cycles

The **Sourcing Precedent Transactions**: The following sources can be used to identify appropriate comparable transactions in *Schnoor (2006)*:
1. M&A Databases – SDC, Internal databases
2. Previous analysis prepared by specific industry groups
3. Industry periodicals and news articles
4. Acquisitions footnotes in the annual reports of public companies
5. Tender offer documents
6. Equity research analysts

7. Senior investment bankers
8. Client

*The Use of Comparable Transaction Multiples (Comparables) to Derive Value* in *Schnoor (2006)*:

1. The first stage is to calculate the relevant multiples for each comparable company
2. Then, to determine which multiples or range of multiples justify a reasonable benchmark for valuing the specific target
3. Analyze the results to decide which companies are most comparable:
    a) Exclude outlying multiples
    b) Test for reasonableness and use common sense
4. Look at mean and median multiples, mean excluding high and low multiples, and multiples of specific companies that may be most relevant
5. Use judgement to determine, which ratios are most relevant given the company and the nature of its industry

The following table contains *Comparable Trading Multiples* for sample companies, which will be used to value a target company in *Schnoor (2006)*:

*Comparable Trading Multiples*

|  | Price 07/10/06 | Mkt. Cap. (US$MM) | EV (US$MM) | Debt/ Total Cap. | EV/ EBITDA 2005 | EV/ EBITDA LTM | EV/ EBITDA 2006E | P/BV LTM |
|---|---|---|---|---|---|---|---|---|
| Company 1 | $14.05 | $2,679 | $4,467 | 40% | 11.2x | 11.1x | 10.9x | 0.9x |
| Company 2 | $11.05 | $2,024 | $3,218 | 37% | 16.7x | 12.0x | 9.5x | 1.3x |
| Company 3 | $9.40 | $706 | $1,301 | 46% | 8.0x | 6.0x | 4.8x | 0.7x |
| Company 4 | $33.50 | $1,887 | $3,840 | 51% | 11.8x | 9.0x | 6.7x | 1.6x |
| Company 5 | $42.50 | $4,062 | $7,394 | 45% | 11.5x | 11.5x | 8.0x | 1.3x |

| | | | | 43% | 9.7x | 9.7x | 5.5x | 2.5x |
|---|---|---|---|---|---|---|---|---|
| Company 6 | 57.63 | $11,467 | $20,018 | | | | | |
| **Average** | | | | **44%** | **11.5** | **9.9x** | **7.6x** | **1.4x** |
| **Average Excluding Hi & Low** | | | | **43%** | **11.0x** | **10.3x** | **7.4x** | **1.3x** |

*Tab. 10. Comparable trading multiples (after Schnoor (2006)).*

The following table contains **Comparable Transaction Multiples** for sample transactions which will be used to value a target company in *Schnoor (2006)*:

*Comparable Transaction Multiples*

| Date | Acquirer | Target | EV Paid (US$MM) | Target's Capacity (000units) | LTM EBITDA (US$MM) | EV per Unit (US$/unit) | EV per EBITDA |
|---|---|---|---|---|---|---|---|
| 2005 | Company A | Company B | $995 | 885 | $80 | $1,124 | 12.4x |
| 2004 | Company C | Company D | $594 | 440 | $52 | $1,350 | 11.4x |
| 2003 | Company E | Company F | $882 | 875 | $91 | $1,008 | 9.7x |
| 2002 | Company G | Company H | $3,875 | 2,463 | $291 | $1,573 | 13.3x |
| 2002 | Company I | Company J | $450 | 250 | $31 | $1,800 | 14.5x |
| | **Median** | | | | | **$1,350** | **12.4x** |
| | **Mean** | | | | | **$1,371** | **12.3x** |
| | **Mean Excluding Hi & Low** | | | | | **$1,349** | **12.4x** |

*Tab. 11. Comparable transaction multiples (after Schnoor (2006)).*

The following table uses the *Comparable Trading Multiples* and **Comparable Transaction Multiples** from the previous pages to value a sample target company in *Schnoor (2006)*:

*Summary Value Table*

| Methodology | Target Company's Values | Multiple Range | Value (US$ Millions) |
|---|---|---|---|
| **Trading Multiples** LTM EBITDA 2006E EBITDA | $88 (US$/MM) $102 (US$/MM) | 9.5x – 10.5x 7.0x – 8.0xx | 9.5x88=836 – 924 7.0x102=714 – 816 |
| **Summary Enterprise Values** | | | (836+714)/2=**$775** -**$870** |
| **Transaction Multiples** LTM EBITDA Annual Capacity | $88 (US$/MM) 600 (000 units) | 12.0x – 12.5x $1,300-$1,400 | 12x88=1,056 – 1,100 1,300x600=780 – 840 |
| **Summary Enterprise Values** | | | **$918 - $970** |
| **Summary Enterprise Value Range** Less: Net Debt Equity Value SharesOutstanding(MM) | | | **$847 - $920** $250 847-250=$597 - $670 55.7 |
| **Equity Value perShare** | | | 597/55.7=**$10.70 -$12.00** |

*Tab. 12. Summary value table (after Schnoor (2006)).*

The *Important Considerations* have to be taken to the account during the above calculations see the *Summary Value Table* in *Schnoor (2006)*:

    *1.*    Multiply the *Operating Results* of the company to be valued by relevant comparable company *Multiples*:

    a) Use a few time periods – i.e. LTM, forecast years

    b) Select relevant multiple ranges for each period

    *2.*    Convert derived *Equity Values* to *Enterprise Values* by adding the target's ne debt

a) Net Income / EPS multiples

b) Book Value Multiples

***3.*** Convert derived ***Enterprise Values*** to ***Equity Values*** by subtracting the target's net debt

a) Sales Multiples

b) EBITDA Multiples

***4.*** A range of values, ***Value Range***, is usually calculated based on high, mean and low summary multiples.

Finally, let us summarize some important definitions of used terms:

***1. Consolidated income statement***: revenues (include product sales, contract research, royalty and interest income); Earnings Before Interest, Income Taxes, Depreciation and Amortization (EBITDA), depreciation and amortization, EBIT, earnings per share, cash flow from operating, investing, financing activities;

***2. Consolidated balance sheet***: total assets, total liabilities, total equity, net assets, cash and marketable securities, short term debt, long term debt, minority interest, preferred shares, convertible securities, common equity, shareholder equity;

***3. Consolidated statement of recognized income and expense***: currency translation on foreign currency net investments, amounts charged to hedging reserve, actuarial gains/losses on pension schemes, current tax on items taken directly to equity, deferred tax on items taken directly to equity, net income recognized directly in equity, profit, total recognized income and expense;

***4. Reconciliation of underlying earnings per share***: profit, preference dividends, net financing credit, market value movements on derivatives, amortization of assets;

***5. Performance ratios***: return on equity, profit margin (it represents after-tax income as percentage of revenues), gross margin, EBITDA margin, EBIT margin, net income margin;

***6. Valuation ratios***: price / earnings, price cash / flow, price / book value, enterprise value / revenue, enterprise value / EBITDA, enterprise value / EBIT;

***7. Credit ratios***: net debt / total capitalization (book), net debt / total capitalization (market), EBITDA / interest, EBITDA – CAPEX / Interest, total debt / EBITDA;

***8. Other ratios***: consumer financial obligations ratio measures all consumer credits, including credit cards, auto loans, durable goods payment plans (published by Federal Reserve); mortgage financial obligations ratio measures mortgage debt as a share of personal disposable income.

Let us summarize all the information on the *business valuation methodologies*, which can be used during the *M&A transaction strategy* implementation.

First of all, let us point out that to the fact that we derived the following formula to describe the *successful merger*

$$V_C^0 \geq \left(V_A^0 + V_B^0\right) - P_B^0,$$

where $V_C^0$ is the value of the combined firm in the period 0, $V_A^0$ is the value of the firm *A* in the period 0, $V_B^0$ is the value of the firm *B* in the period 0, $P_B^0$ is the acquisition price for the firm *B*.

The formula for the calculation of the *combined firm value with the multiple business processes* in the given time moment is *in Lenz (2010)*

$$V_C^0 = \sum_{n=1}^{N} \sum_{t=1}^{T} \frac{NCF_t^n}{(1+r)^t},$$

where $V_C^0$ is the value of the combined firm in the period 0, $NCF_t^n$ is the net cash flow of the business process *n* in the period *t*, *r* is the discount rate (*Weighted Average Cost of Capital*), *n* is the number of business processes, *t* is the period of time.

Fig. 1 shows the *corporate system C scheme* in *Lenz (2010)*

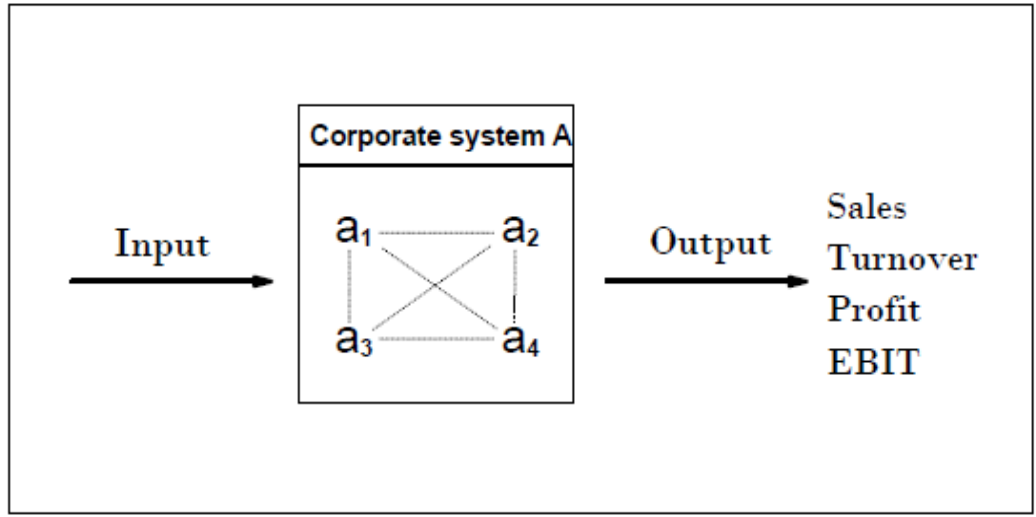

***Fig. 1.*** *Corporate system scheme (after Lenz (2010)).*

Fig. 2 depicts the *business evaluation methodology* of the corporate system *C* in *Lenz (2010)*.

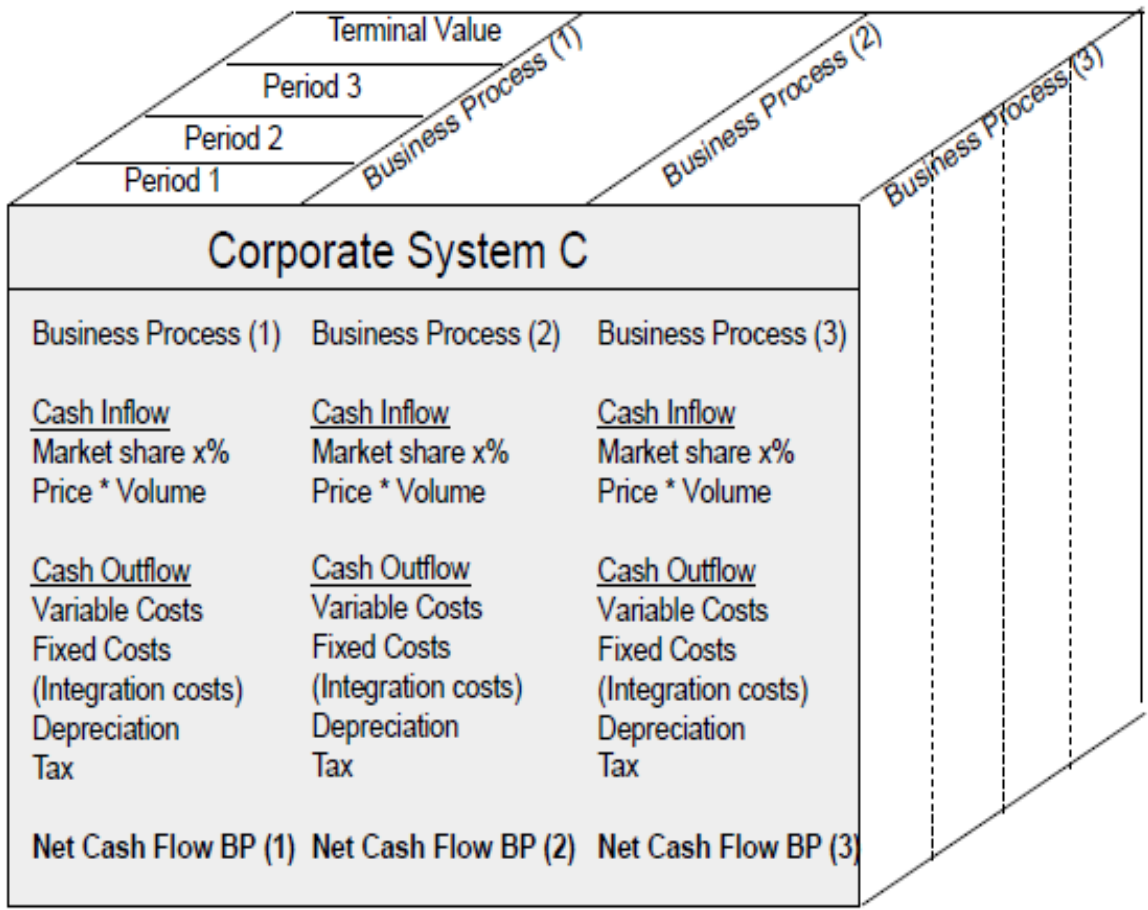

***Fig. 2.*** *Evaluation of corporate system C (after Lenz (2010)).*

# Mergers and acquisitions transactions strategies in diffusion - type financial systems in highly volatile global capital markets with nonlinearities

Let us discuss the current state of research on the *M&A transactions*. The extended knowledge base on the *M&A transactions* has been created as a result of completion of research programs by many world renowned scientists.

*Damiani, Pompei (2008)* write: "**A first group of studies**, focusing on international comparisons, has explored the role of *corporate governance systems*, investor protection laws and other countries' regulatory institutions as the main determinants of *takeovers* around the world (see, for instance, *Rossi and Volpin (2004)*). The underlying claim of these studies is that, **in better-regulated systems, it is easier and less expensive to raise capital and to finance corporate acquisitions**."

*Damiani, Pompei (2008)* continue to explain: "**A second group of contributions** (*Andrade et. (2001), Mitchell and Mulherin (1996); Jovanovic and Rousseau (2001, 2002)*) has attributed a central role to variations in industry composition, documenting that**, in each country, mergers occur in waves and within each wave clustering by industry is observed**. In this field of research, industry level shocks (due to technological and regulatory changes) play a central role in explaining takeovers and their evolution in time."

Figs. 3 - 5 illustrate the *M&A transactions* activities in *Damiani, Pompei (2008)*.

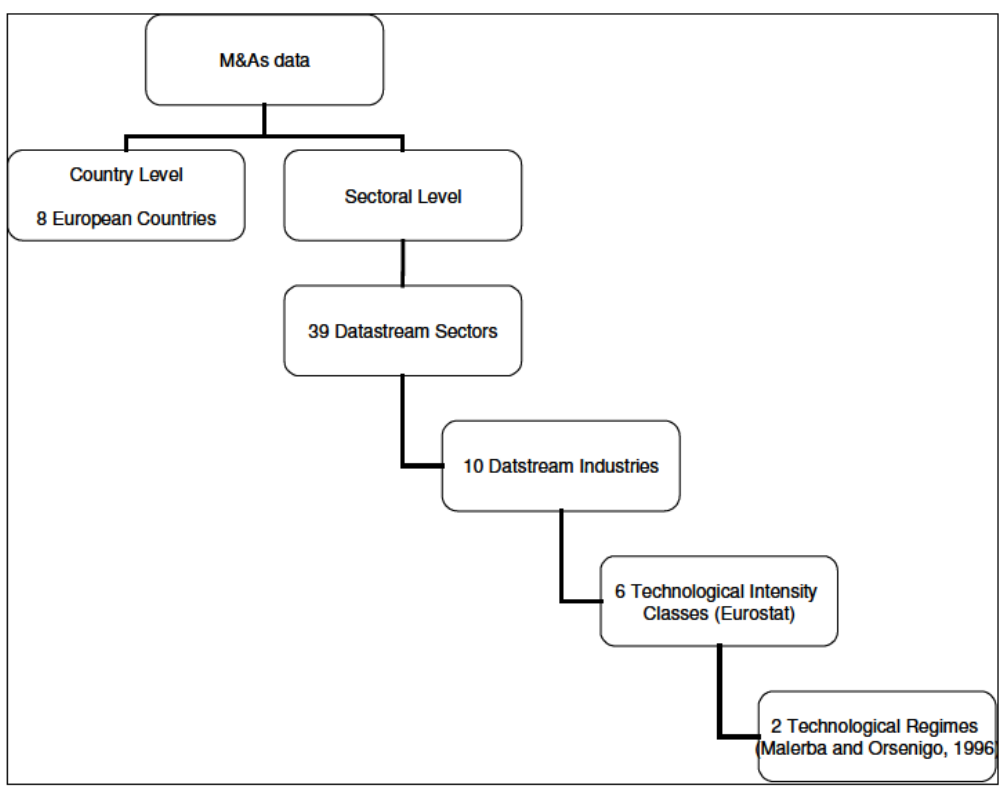

***Fig. 3.*** *Classifications of M&A data (after Damiani, Pompei (2008)).*

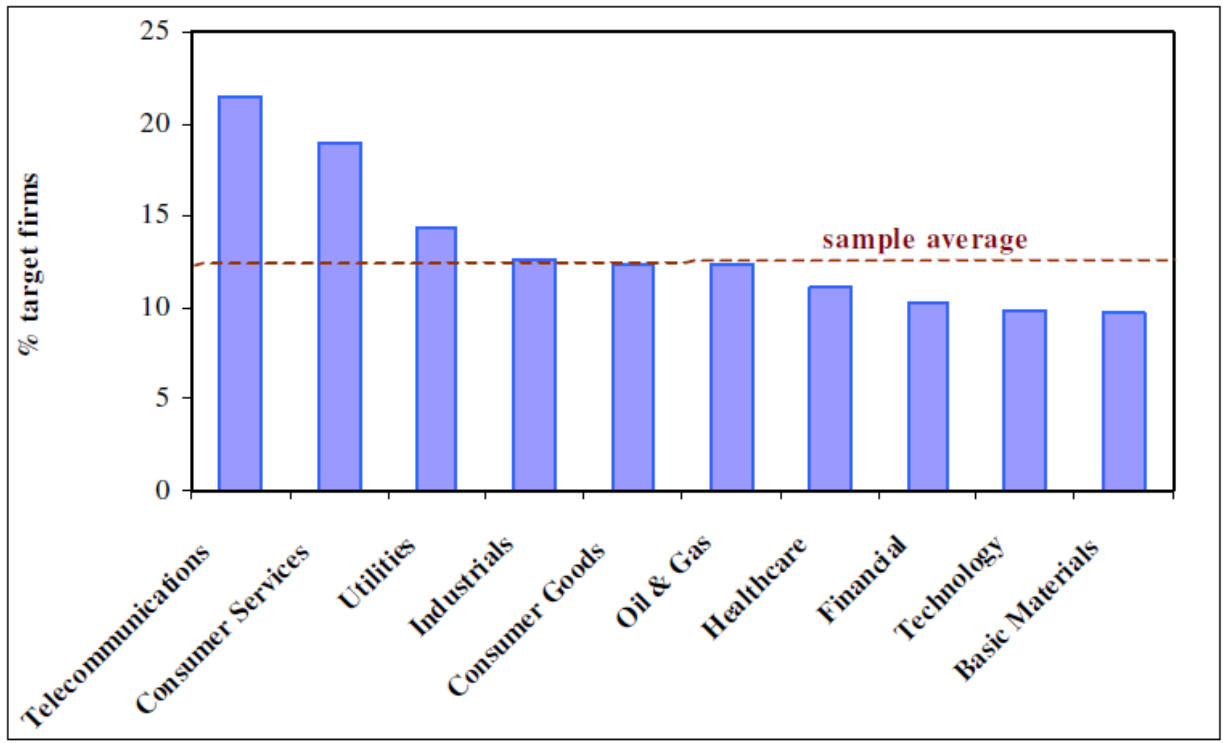

*Fig. 4.* M&A activity in eight European countries: Incidence of takeovers by 2 digit sectors (% incidence of deals on the total number of firms for each sector) (after Damiani, Pompei (2008).

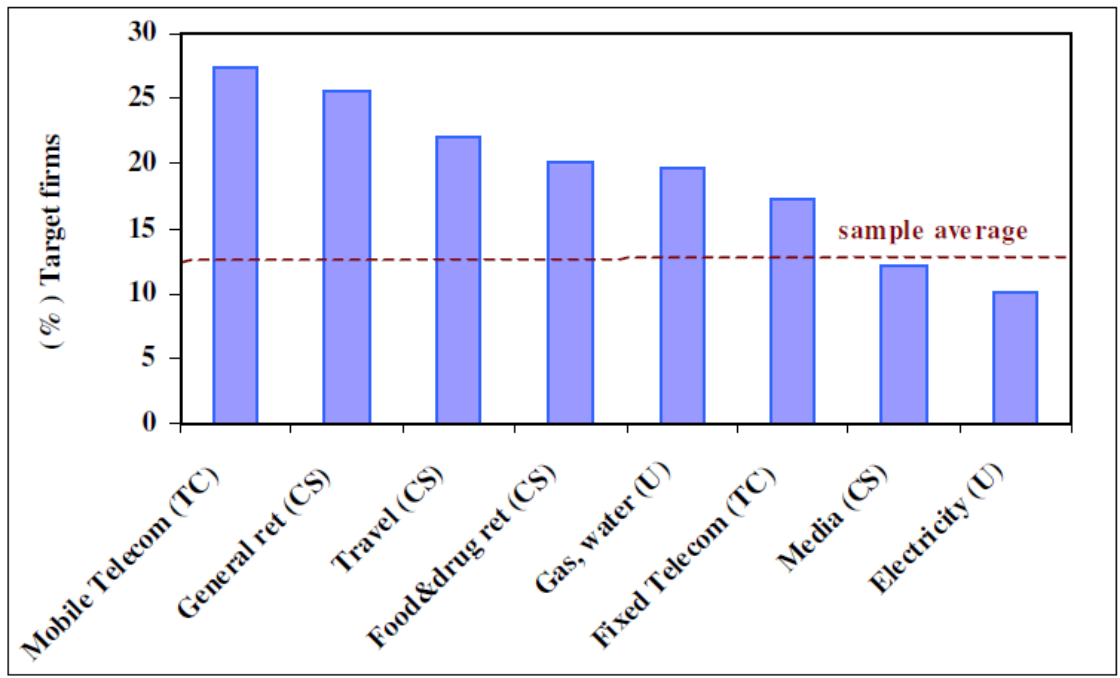

*Fig. 5.* Top markets for corporate control in eight European countries: Frequency of M&A by four digit sectors in 2002-2005 (after Damiani, Pompei (2008).

*Damiani, Pompei (2008)* derive the formula to evaluate both: *1) the volume of M&A activity* and *2) the determinants of various patterns*, observed by the countries and by the sectors

$$M\&A_{ij} = \beta_0 + \sum_{m=1}^{5}\beta_{1m}I_{i.m} + \sum_{n=1}^{2}\beta_{2n}S_{ijn} + \sum_{l=1}^{2}\beta_{3l}TEC_{ijl} + \sum_{z=1}^{2}\beta_{4z}\left(TEC_{ijz}TR_z\right) + \varepsilon_{ij},$$

where $i = 1,...8$ Countries; $j = 1, ...39$ Four digit sectors; $m = 1,...5$ Institutional variables (I); $n = 1,2$ Sectoral variables (S); $l = 1,2$ Technological variables (TEC); $z = 1,2$ Technological Regimes dummy variables (TR). Tab. 13 shows the *estimates of takeover frequencies* in *Damiani, Pompei (2008)*.

| Dependent Variable: M&A frequency | Obs. 175 Column a | Obs. 162 Column b | Obs. 175 Column c | Obs 175 Column d | Obs. 175 Column e | Obs. 175 Column f | Obs. 175 Column g | Obs. 175 Column h |
|---|---|---|---|---|---|---|---|---|
| Log GDP per capita | 0.717*** | 0.780*** | 0.200 | 0.679*** | 0.700*** | 0.663*** | 0.650*** | 0.608*** |
|  | (0.185) | (0.251) | (0.151) | (0.172) | (0.161) | (0.151) | (0.133) | (0.146) |
| Concentrated Ownership | 0.547** |  | 0.394 | 0.570** | 0.526** | 0.439** | 0.500*** | 0.507** |
|  | (0.236) |  | (0.455) | (0.228) | (0.215) | (0.214) | (0.197) | (0.210) |
| Widely held firms |  | -0.005** |  |  |  |  |  |  |
|  |  | (0.002) |  |  |  |  |  |  |
| Takeover regulation | 0.121*** | 0.162*** |  | 0.114*** | 0.111*** | 0.104*** | 0.101*** | 0.100*** |
|  | (0.030) | (0.049) |  | (0.027) | (0.025) | (0.024) | (0.021) | (0.023) |
| Antidirector Rights (index revised) |  |  | -0.009 |  |  |  |  |  |
|  |  |  | (0.061) |  |  |  |  |  |
| Market to Book value of Equity (PBR) | -0.028*** | -0.033*** | -0.035*** | -0.029** | -0.025** | -0.022** | -0.017* | -0.023** |
|  | (0.012) | (0.012) | (0.013) | (0.012) | (0.012) | (0.013) | (0.010) | (0.011) |
| Shock in the sectoral growth rate |  |  |  | 0.011 | 0.012* | 0.013** | 0.008 | 0.009 |
|  |  |  |  | (0.007) | (0.007) | (0.006) | (0.006) | (0.006) |
| R&D |  |  |  |  | -0.441* |  |  |  |
|  |  |  |  |  | (0.260) |  |  |  |
| R&D *SMI |  |  |  |  |  | -0.340 |  |  |
|  |  |  |  |  |  | (0.309) |  |  |
| R&D *SMII |  |  |  |  |  | -10.472*** |  |  |
|  |  |  |  |  |  | (2.940) |  |  |
| Innovation |  |  |  |  |  |  | -0.017*** |  |
|  |  |  |  |  |  |  | (0.006) |  |
| Innovation * SMI |  |  |  |  |  |  |  | -0.010* |
|  |  |  |  |  |  |  |  | (0.006) |
| Innovation* SMII |  |  |  |  |  |  |  | -0.102*** |
|  |  |  |  |  |  |  |  | (0.029) |
| Sectoral Dummies | Yes | Yes | Yes | Yes | Yes | Yes | Yes | Yes |
| Constant | -2.79*** | -2.724*** | -0.757 | -2.648*** | -2.673*** | -2.501*** | -2.385*** | -2.342 |
|  | (0.736) | (0.973) | (0.655) | (0.670) | (0.627) | (0.587) | (0.522) | (0.570) |
| Chi2 Test (Prob>Chi2) | 0.0112 | 0.0114 | 0.130 | 0.009 | 0.002 | 0.001 | 0.000 | 0.001 |

**Tab. 13.** *Estimates of takeover frequencies: the role institutional, sectoral and technological factors (after Damiani, Pompei (2008).*

Let us provide the definitions of various types of the *M&A* transactions Tab. 14 in *Lehto, Böckerman (2006)*.

| Variables | Definition/measurement |
|---|---|
| Types of M&As: | |
| Cross-border M&A | Cross-border M&As are defined as cases where the acquiring firm is foreign. "Foreign" here means that the firm, which is reported by the magazine *Talouselämä* to be the acquirer, is not located in Finland at the time of the M&A. (Source: the magazine *Talouselämä*) |
| Domestic M&A (Domestic owner, located in Finland) | We separate two types of domestic M&As based on ownership. (These types are added together in Figs. 1-2). First, there are domestic M&As, where the acquirer is domestically-owned and located in Finland.) (Source: the magazine *Talouselämä*) |
| Domestic M&A (Foreign owner, located in Finland) | Second, there are domestic M&As, where the acquirer is foreign-owned, but located in Finland. (Hence, cross-border M&As are defined as cases where transaction truly occurs across national borders.) (Source: the magazine *Talouselämä*) |
| Internal restructuring | Internal restructurings involve cases of transformation of a firm's organizational form without the involvement of another company. For instance, management buy-outs that have been popular through the 1990s and a smaller number of cases where an individual Finnish investor is buying the firm belong to this category of M&As. (Source: the magazine *Talouselämä*) |

***Tab. 14.*** *Definitions of various types of M&A transactions (after Lehto, Böckerman (2006)).*

Let us illustrate the *change dynamics of M&A transactions* in *Finland* as shown in Fig. 6 in *Lehto, Böckerman (2006)*.

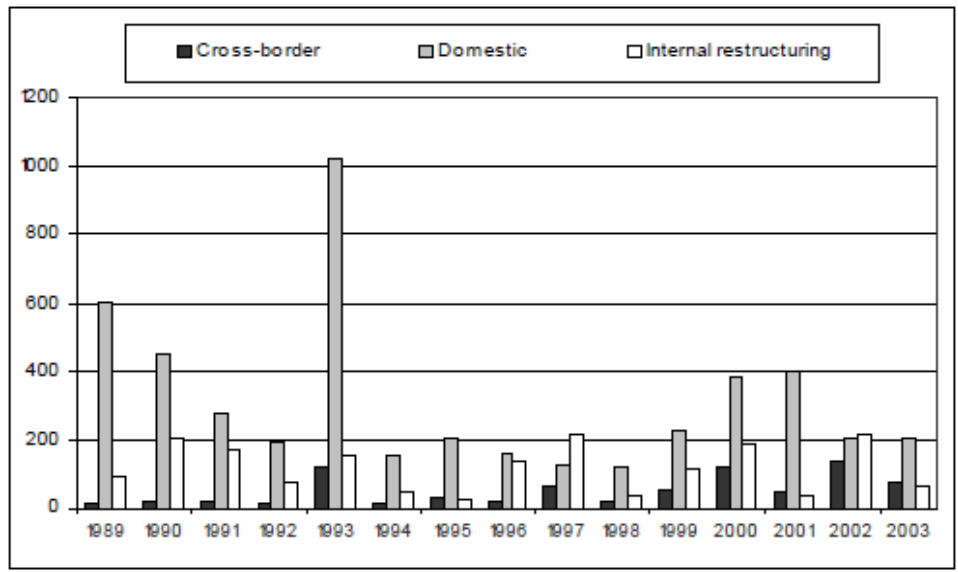

***Fig. 6.*** *The number of different types of M&As in Finland over the period 1989-2003. Two types of domestic M&As are added together (after Lehto, Böckerman (2006)).*

Let us show a sectoral division of the different types of *M&A transactions* in *Finland* in Fig. 7 in *Lehto, Böckerman (2006)*.

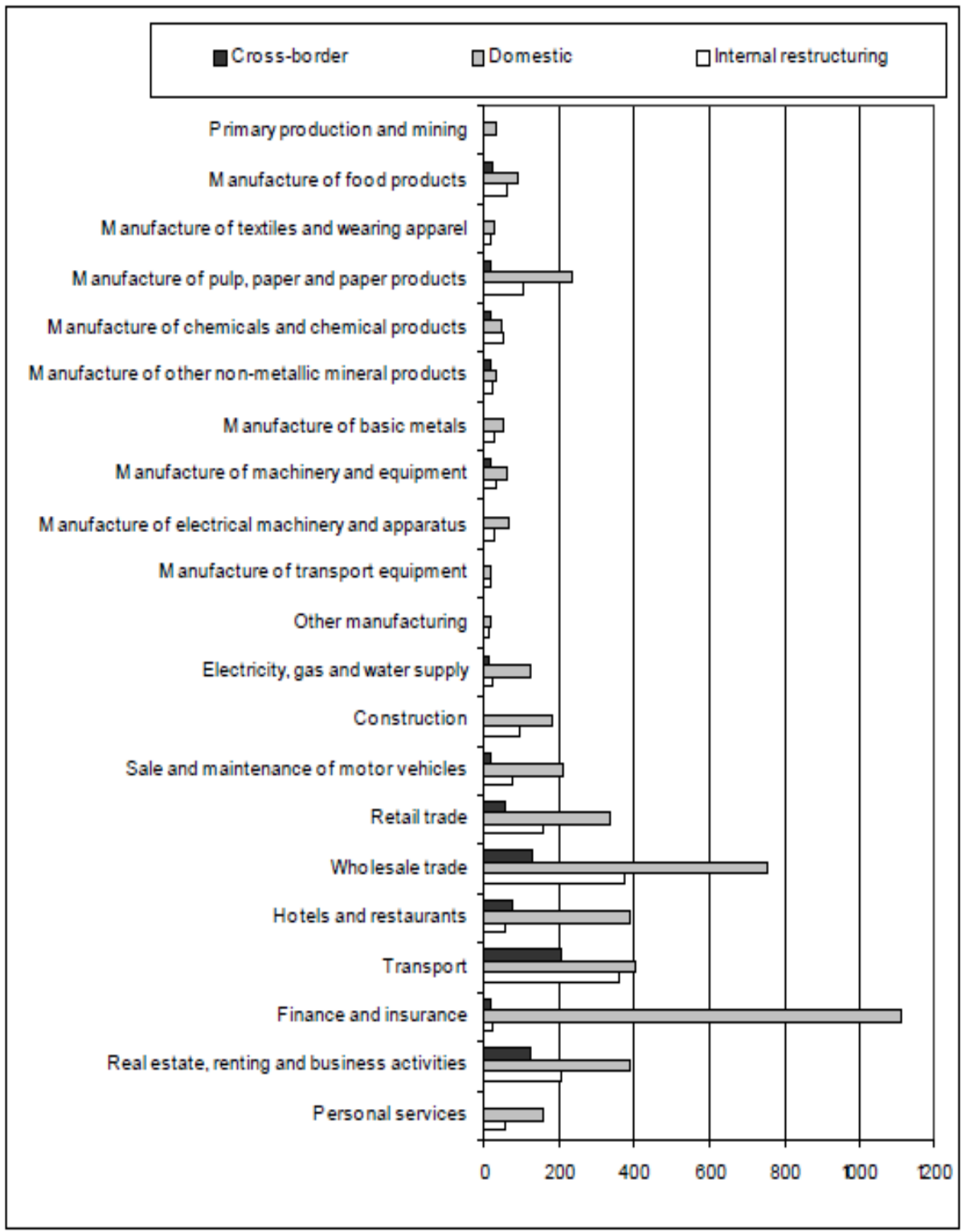

*Fig. 7.* *The sectoral division of different types of M&As in Finland. The figures are reported as sums over the period 1989-2003 and two types of domestic M&As are added together (after Lehto, Böckerman (2006)).*

Tab. 15 demonstrates the main types of the *European M&As* in the banking industry in *Beckmann, Eppendorfer, Neimke (2002)*, and Tab. 16 shows the motives and risks with the four types of the *Mergers & Acquisitions (M&As)* in *Beckmann, Eppendorfer, Neimke (2002)*.

|  | Within banking industry | Cross industry |
|---|---|---|
| **Within border** | **Domestic bank M&As**<br>M&As involving credit institutions located in the same country.<br><br>Examples:<br>• Banco Santander/Banco Central Hispanoamericano (1999)<br>• BNP/Paribas (1999)<br><br>The most dominant type of M&As in the European banking industry: 60.7% of all European banking M&A transactions between 1990 and 1999 took place within the domestic banking industry (1990: 44%, 1995: 45.8%, 1999: 61.4%) | **Domestic conglomeration**<br>M&As involving credit institutions and insurances and/or other financial institutions all located in the same country.<br><br>Examples:<br>• Credit Suisse/Winterthur (1997)<br>• Citicorp/Travelers Group (1998)<br><br>Less dominant than domestic M&As within the banking industry: 16.1% of all banking M&A transactions between 1990 and 1999 were domestic cross industry transactions; after an increase in the mid-nineties (1990: 16%, 1995: 22.9%, 1999: 12.5) there are declining shares in the last years. |
| **Cross-border** | **International bank M&As**<br>M&As involving credit institutions located in different countries, one of which is an EU country.<br><br>Examples:<br>• Deutsche Bank/Bankers Trust (1998)<br>• HypoVereinsbank/Bank Austria (2000)<br>• Nordea Group (2000)<br><br>Less dominant than domestic bank M&As but the most increasing type of M&As involving the banking industry: 16.1% of all banking M&A transactions between 1990 and 1999 were cross-border bank M&As (1990: 30%, 1991: 10%, 1993: 10.8%, 1995:20.1%, 1999: 19.5) | **International conglomeration**<br>M&As involving credit institutions located in an EU country and insurances and/or other financial institutions located in another EU or third country.<br><br>Examples:<br>• Deutsche Bank/Morgan Grenfell (1997)<br>• Dresdner Bank/Kleinwort Benson (1995)<br><br>Less dominant type of European M&A transactions in recent years and at present: 7.1% of all transactions involving the European banking industry between 1990 and 1999 were international cross industry transactions; after increasing shares in the mid-nineties (1990: 8%, 1995: 11.2%) there is a decline during the last years (1997: 9.6%, 1998: 8.5%, 1999: 6.5%). |

*Tab. 15.* *European M&As types (after Beckmann, Eppendorfer, Neimke (2002)).*

|  | Within banking industry | Cross industry |
|---|---|---|
| **Within border** | **Domestic bank M&As**<br>Main motives:<br>• Cost benefits from economies of scale, e.g. by reduction of surplus staff and overlapping branches or mutual use of administrative functions;<br>• Increasing market share;<br><br>Main risks:<br>• Pricing of the strategic risks;<br>• Operational risks after the transaction, mainly related to the integration of personnel, information and risk management, customer and account systems etc. | **Domestic conglomeration**<br>Main motives:<br>• Cost benefits from economies of scope through cross-selling;<br>• Revenue enhancement due to product diversification;<br>• Risk diversification;<br><br>Main risks:<br>• Increased ex ante risks because of different business area;<br>• Reputation risks (failure of one company may lead to declining reputation of the whole conglomerate);<br>• Increased integration difficulties due to different fiscal and accounting treatment etc. |
| **Cross-border** | **International bank M&As**<br>Main motives:<br>• Achieving access and presence in an international market with a larger customer base (market share);<br>• Possibility of reaching the critical mass to offer specific services;<br><br>Main risks:<br>• Increased ex ante risks because of cultural barriers or differences (unknown market, regulations and practices);<br>• Different fiscal and accounting treatment and reporting requirements; | **International conglomeration**<br>Main motives:<br>• Cost benefits from economies of scope through cross-selling;<br>• Access and presence in international financial markets;<br>• International product diversification;<br><br>Main risks:<br>• Increased ex ante risks due to a different business area;<br>• Different fiscal and accounting treatment and reporting requirements;<br>• Reputation risks; |

*Tab. 16.* *M&As motives and risks (after Beckmann, Eppendorfer, Neimke (2002)).*

Tab. 17 shows the *BNP Paribas* strategies in *Beckmann, Eppendorfer, Neimke (2002)*.

| General Company Data | <ul><li>One of the European top-ten-banks (market capitalisation at 31 Dec 2000: € 41.9 bn;</li><li>Operations in 87 countries in Europe, Asia and United States;</li><li>80,000 employees, including 61,000 in Europe, including 49,000 in France;</li><li>Network of 2,200 branches in France, 13.1 million retail customers in the EU;</li></ul> | | | |
|---|---|---|---|---|
| Pan-European strategy | <ul><li>Set-up of Multichannel Bank;</li><li>Cross-selling between the various business units and with other divisions of the Group;</li></ul> | | | |
| Recent expansion activities | | | | |
| ▪ Target Countries | France | Germany | Spain | Pan-European |
| ▪ Date of first entry | 2000 | 2000 | 1980's | 1990's |
| ▪ Entry method | <ul><li>Merger between BNP and Paribas to create BNP Paribas;</li><li>1998 merger between Compagnie Financière de Paribas, Banque Paribas and Compagnie Bancaire to form Paribas;</li><li>1993 Privatisation of BNP</li></ul> | <ul><li>Joint venture between BNP Paribas and Dresdner Bank</li></ul> | <ul><li>Establishing of a branch net (sold in 2000)</li></ul> | Different market access strategies (e.g. "Cardif-model", "Cetelem-model") |
| ▪ Products offered | Full banking service | Consumer finance | Private Banking (only wealthy customers) | Full banking service |
| ▪ Distribution channel | Multi-channel | Multi-channel | Traditional | Multi-channel |

***Tab. 17.*** *BNP Paribas business strategies (after Beckmann, Eppendorfer, Neimke (2002)).*

Fig. 8 demonstrates the *M&A* activity in European banking sector *(1990-2004) in Asimakopoulos, Athanasoglou (2011))*

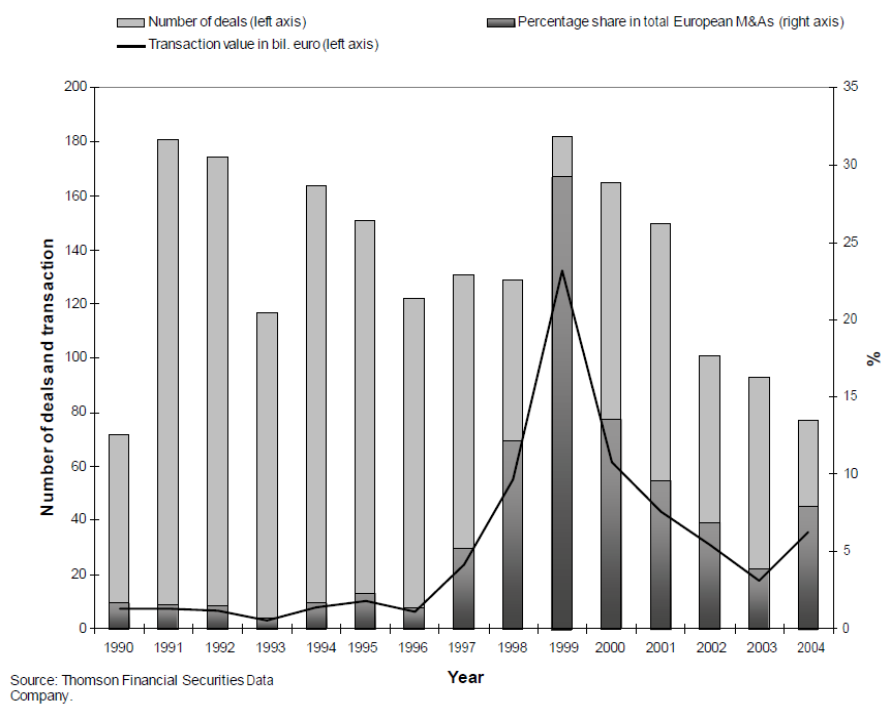

***Fig. 8.*** *M&A activity in European banking sector (1990-2004) (after Asimakopoulos, Athanasoglou (2011)).*

Let us discuss the *M&A* performance shortly. *Beltratti, Paladino (2011)* write: "*Acquisitions* in the banking sector may be driven by a multiplicity of factors among which are: (i) exploiting economies of scale associated with centralizing functions like *IT*, cash management personnel, (ii) exerting market power and imposing better pricing conditions on customers, (iii) pursuing geographical diversification that brings benefits in terms of risk reduction, (iv) taking advantage from implicit subsidies connected with a too-big-to-fail (*TBTF*)[1] status, (v) managers maximizing their own utility function rather than the shareholders utility function."

*Focarelli, Panetta, Salleo (2003)* state: "In summary, acquisitions appear to be aimed at increasing the value of the passive bank by improving the quality of its loan portfolio, while mergers apparently reflect a strategy of increasing the reach of the active bank's services."

Fig. 9 shows the *M&A transaction success factors* in *Jones (2009)*.

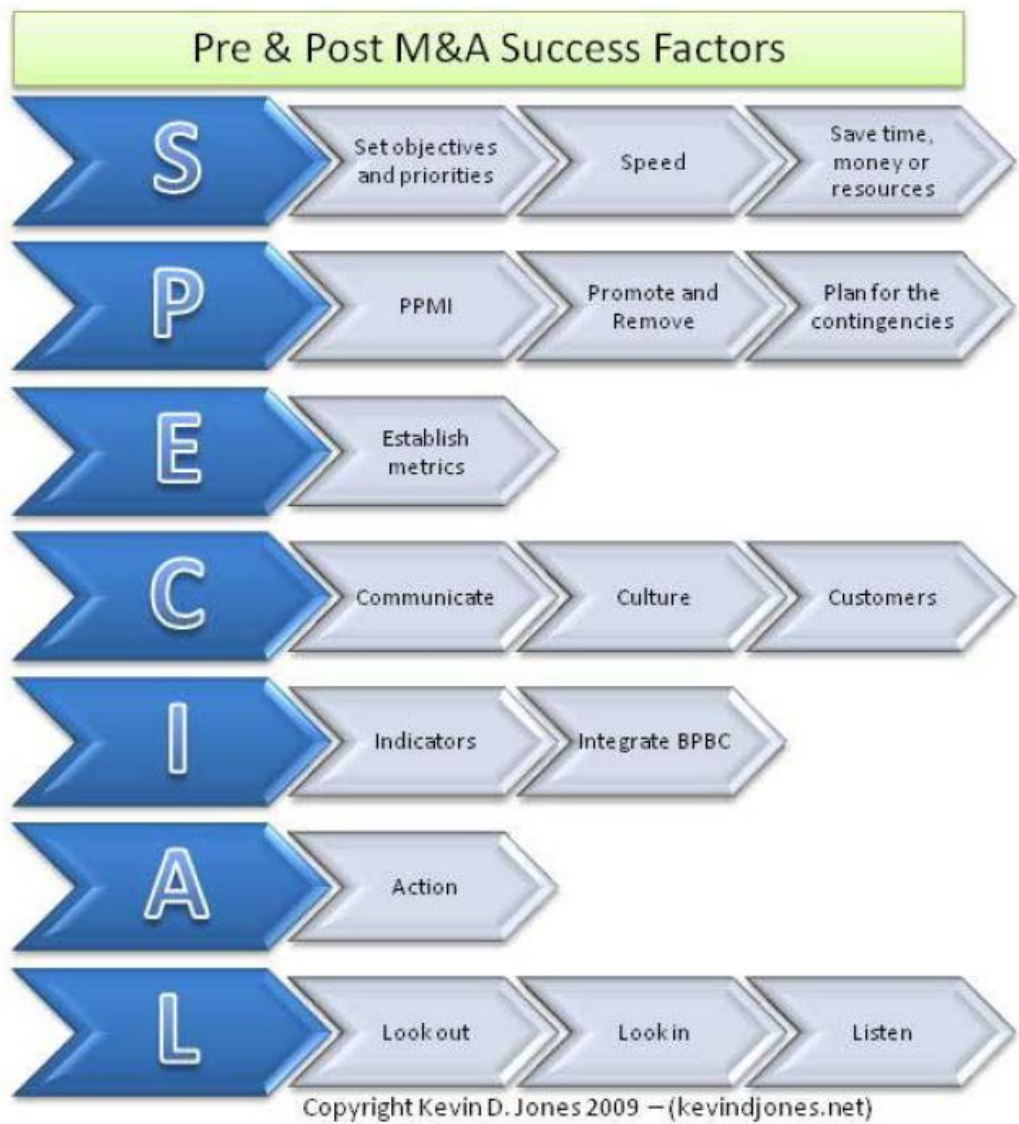

***Fig. 9.*** *M&A transaction success factors (after Jones (2009)).*

Tab. 18 displays the categorization of studies of the *acquisition performance* by the *performance metrics* in *Zollo, Meier (2008)*.

| Reference | Integration Process Performance | Overall Acquisition Performance | Employee Retention | Customer Retention | Accounting Performance | Long-Term Financial Performance | Short-Term Financial Performance | Acquisition Survival | Innovation Performance | Knowledge Transfer | Systems Conversion | Variation in Market Share |
|---|---|---|---|---|---|---|---|---|---|---|---|---|
| Agrawal et al., 1992 | | | | | | X | | | | | | |
| Ahuja and Katila, 2001 | | | | | | | | | X | | | |
| Amit and Livnat, 1988 | | | | | X | | | | | | | |
| Anand and Singh, 1997 | | | | | X | | | | | | | |
| Barber and Lyon, 1997 | | | | | | X | X | | | | | |
| Beckman and Haunschild, 2002 | | | | | | | X | | | | | |
| Berger and Ofek, 1995 | | | | | | | X | | | | | |
| Bergh, 2001 | | | | | | | | X | | | | |
| Bresman et al., 1999 | X | | | | | | | | | X | | |
| Brush, 1996 | | | | | X | | | | | | | X |
| Bruton et al., 1994 | | X | | | | | | | | | | |
| Buono et al., 1985 | X | X | | | | | | | | | | |
| Cannella and Hambrick, 1993 | | X | | | X | | | | | | | |
| Capon et al., 1988 | | | | | X | | | | | | | |
| Capron, 1999 | X | X | | | | | | | | | | |
| Capron and Pistre, 2002 | | | | | | | X | | | | | |
| Carow et al., 2004 | | | | | | X | X | | | | | |
| Chang, 1996 | | | | | X | | | | | | | |
| Chatterjee, 1986 | | | | | | | X | | | | | |
| Chatterjee, 1991 | | | | | | | X | | | | | |
| Chatterjee, 1992 | | | | | | X | | | | | | |
| Chatterjee et al., 1992 | | | | | | | X | | | | | |
| Clark and Ofek, 1994 | | | | | X | X | | | | | | |
| Covin et al., 1997 | | | X | | | | | | | | | |
| Datta, 1991 | X | X | | | | | | | | | | |
| Datta and Grant, 1990 | | X | | | | | | | | | | |
| DeLong and DeYoung, 2007 | | | | | X | | X | | | | | |
| Eckbo, 1983 | | | | | | | X | | | | | |
| Feea and Thomas, 2004 | | | | | X | | X | | | | | |
| Fowler and Schmidt, 1989 | | | | | X | X | | | | | | |
| Franks et al., 1991 | | | | | | | X | | | | | |
| Haleblian and Finkelstein, 1999 | | | | | | | X | | | | | |
| Hambrick and Cannella, 1993 | | | X | | | | | | | | | |
| Harris and Ravenscraft, 1991 | | | | | | | X | | | | | |
| Harrison et al., 1991 | | | | | X | | | | | | | |
| Harrison et al., 2005 | | | | | | X | X | | | | | |
| Hayward, 2002 | | X | | | | | X | | | | | |
| Hayward and Hambrick, 1997 | | | | | | | X | | | | | |
| Heron and Lie, 2002 | | | | | X | | | | | | | |
| Hitt et al., 1991 | | | | | | | | | X | | | |
| Hitt et al., 1996 | | | | | X | | | | X | | | |
| Hitt et al., 1998 | | | | | X | | | | X | | | |
| Holl and Kyriazis, 1997 | | | | | | | X | | | | | |
| Homburg and Bucerius, 2006 | | X | | | | | | | | | | |
| Hoskisson et al., 1993 | | | | | X | X | | | | | | |
| Hunt, 1990 | X | X | | | | | | | | | | |
| Jensen and Ruback, 1983 | | | | | | | X | | | | | |
| Kapoor and Lim, 2005 | | | | | | | | | X | | | |
| Krishnan et al., 1997 | | | | | X | | | | | | | |
| Kroll et al., 1997 | | | | | | | X | | | | | |
| Krug and Hegarty, 2001 | | | X | | | | | | | | | |
| Kusewitt, 1985 | | | | | X | X | | | | | | |
| Lahey and Conn, 1990 | | | | | | | X | | | | | |
| Larsson and Finkelstein, 1999 | X | X | | | | | | | | | | |
| Loughran and Vijh, 1997 | | | | | | X | | | | | | |
| Lubatkin, 1987 | | | | | | X | X | | | | | |
| Lubatkin et al., 1997 | | | | | | X | X | | | | | |

| Reference | Integration Process Performance | Overall Acquisition Performance | Employee Retention | Customer Retention | Accounting Performance | Long-Term Financial Performance | Short-Term Financial Performance | Acquisition Survival | Innovation Performance | Knowledge Transfer | Systems Conversion | Variation in Market Share |
|---|---|---|---|---|---|---|---|---|---|---|---|---|
| Markides and Ittner, 1994 | | | | | | | X | | | | | |
| Moeller et al., 2004 | | | | | | | X | | | | | |
| Montgomery and Wilson, 1986 | | | | | | | | X | | | | |
| Morck et al., 1988 | | | | | X | | | | | | | |
| Morosini et al., 1998 | | | | | X | | | | | | | |
| Palich et al., 2000 | | | | | X | X | X | | | | | |
| Pangarkar, 2004 | | | | | | | X | | | | | |
| Pennings et al., 1994 | | | | | | | | X | | | | |
| Puranam et al., 2006 | | X | | | | | | | | | | |
| Ramaswamy, 1997 | | | | | X | | | | | | | |
| Ravenscraft and Scherer, 1987 | | | | | X | | | | | | | |
| Schweiger and Denisi, 1991 | | | X | | | | | | | | | |
| Seth, 1990 | | | | | | | X | | | | | |
| Seth et al., 2002 | | | | | | | X | | | | | |
| Shanley and Correa, 1992 | X | X | | | | | | | | | | |
| Shelton, 1988 | | | | | | | X | | | | | |
| Shahrur, 2005 | | | | | | | X | | | | | |
| Singh and Montgomery, 1987 | | | | | | X | | | | | | |
| Slusky and Caves, 1991 | | | | | | | X | | | | | |
| Thakor, 1999 | | | | | | | | | | | X | |
| Travlos, 1987 | | | | | | | X | | | | | |
| Travlos and Waegelein, 1992 | | | | | | | X | | | | | |
| Vermeulen and Barkema, 1996 | | | | | | | | X | | | | |
| Walker, 2000 | | | | | | | X | | | | | |
| Wansley et al., 1983 | | | | | | | X | | | | | |
| Walsh, 1988 | | | X | | | | | | | | | |
| Walsh, 1989 | | | X | | | | | | | | | |
| Weber, 1996 | X | | | | X | | | | | | | |
| Zollo, in press | | | | | | X | | | | | | |
| Zollo and Reuer, in press | | | | | X | X | | | | | | |
| Zollo and Singh, 2004 | | | | | X | | | | | | | |
| Total | 8 (9%) | 12 (14%) | 6 (7%) | 0 (0%) | 25 (28%) | 17 (19%) | 35 (40%) | 4 (5%) | 5 (6%) | 1 (1%) | 1 (1%) | 1 (1%) |

**Tab. 18.** *Categorization of studies of acquisition performance by performance metric (after Zollo, Meier (2008)).*

*Zollo, Meier (2008)* explain: "Results of factor and structural equations analysis reveal that: *(a) M&A performance* is a multifaceted construct; there is no one overarching factor capturing all the different ways used to proxy it,

*(b)* there is a path linking integration process performance to long-term firm performance (both accounting and financial returns) via customer retention and overall synergy realization, and

*(c)* short-term window event studies are not linked to any of the other performance metrics."

*Zollo, Meier (2008)* propose to differentiate the *performance measurements* at the three possible levels:

"*1. The task level*. The integration process, with its different components related to multiple tasks necessary to reach the desired level of integration between the two organizations (e.g., alignment of control systems, conversion of the IT systems, transferring sales practices, etc.). Each of these tasks generates its own performance, which can be aggregated in a more general notion of integration process performance, that is, *the degree to which the targeted level of integration between the two organizations has been achieved across all of its task dimensions in a satisfactory manner*.

*2. The transaction level*. The performance of the entire acquisition encompasses all the phases of the acquisition process and focuses on the actual value creation eventually generated by the acquisition. The transaction performance construct can thus be defined as *the amount of value, in cost efficiencies and revenue growth, generated by the complete transaction process*, from the completion of the negotiation to the execution of the business plan. Note that this notion of performance centers around the realization of the value creation objectives, as they have been envisioned at the time of the transaction.

*3. The firm level*. The performance of the combined entity, over and above the value generated by the transaction itself, can be defined as *the variation in firm performance that occurred during the period of relevance for the execution of the business plan connected to the acquisition*. Needless to say, this construct is the broadest of the three, and includes the effects of the acquisition on the performance of other business processes simultaneously ongoing within the firm during the period in consideration."

*King, Slotegraaf, Kesner (2008)* explain: "Strategic management literature in general (*Daily (1994), Daily et al. (2002)*) and acquisition research specifically (*Javidan et al. (2004)*) do not provide a consensus for measuring firm performance. Generally, *M&A* research focuses on financial performance using either accounting or stock market measures. We elected to avoid using accounting measures of performance because they tend to have a historical focus (*Chakravarthy (1986)*) and, in the case of return on assets, can be biased by the method of accounting for an acquisition (*Ravenscraft and Scherer (1987), Sirower (1997)*). This left a choice between *short or long-term stock market measures of performance*." *King, Slotegraaf, Kesner (2008)* continue the discussion: "Therefore, we selected a long-term stock measure. *Jensen's alpha (Jensen (1968))*, a variation of the two-parameter market model previously used *(Farjoun (1998); Hoskisson et al. (1993, 1994))*, was our primary performance measure. An advantage of *Jensen's alpha* is that it compares the return of an acquiring firm with a benchmark from a common starting point."

$$R_{it} = \alpha_i + \beta_i (R_{mt}) + \varepsilon_{it},$$

where $R_{it}$ is the monthly rate of return of firm *i* during month *t*,

$\alpha_i$ is the *Jensen's alpha* for firm *i*,

$\beta_i$ is the firm *i*'s stock price variance relative to the variance of market benchmark (*m*),

$R_{mt}$ is the monthly rate of return of the market benchmark (*m*) during month *t*,

$\varepsilon_{it}$ is the random error term.

*King, Slotegraaf, Kesner (2008)* conclude: 'Positive values of *Jensen's alpha* indicate that an acquiring firm outperformed the market benchmark or *S&P 500*. Comparing a firm's stock performance with a benchmark portfolio offers several benefits, including:

*(1)* comparing an acquirer with a benchmark of multiple firms that eliminates matching firms to calculate abnormal returns, and

*(2)* calculating the average abnormal return of investing in a firm against a benchmark over the same time period. This eliminates a preannouncement estimation period needed for the capital asset pricing model to estimate a firm's normal return.

However, long-term stock market measures of firm performance are also subject to criticism because of the potential for confounding events (*Williams and Siegel (1997)*)."

The abnormal return for firm *j* on day *t* is defined as in *King, Slotegraaf, Kesner (2008)*:

$$AR_{jt} = R_{jt} - (\alpha_j + \beta_j R_{mt}),$$

where $R_{jt}$ is the rate of return on the common stock of the *j*$^{th}$ firm on the time window *t*,

$R_{mt}$ is the rate of return of a market index on the time period *t*,

$\alpha_i$ is the *Jensen's alpha* for firm *j*,

$\beta_i$ is the parameter that measures the sensitivity of $R_{jt}$ to the market index.

*Van Beers, Dekker (2009)* make interesting comments to think about: "The main findings of this study are as follows.

First, innovating firms are significantly more involved in *acquisition activities* than non-innovating firms, which suggests that acquisitions are a strategy to gain access to new technologies or knowledge.

Second, lack of knowledge as a barrier to innovate increases the chance of *acquiring assets* of other firms although not significantly. Lack of finance as a barrier to innovate increases significantly the chance of divesting assets.

Third, *acquisitions* motivated by knowledge barriers in the innovation process affect the probability of positive innovative sales positively while acquisitions motivated by other reasons than innovation barriers affect this probability negatively. No effect of knowledge barriers induced *acquisitions* on the level of the innovative sales could be found."

Tab. 19 shows the plant and firm level studies of the effects of the *mergers and acquisitions* on the employment and wages in *Siegel, Simons (2008)*.

| Authors | Unit of Analysis | Country | Type of Transaction | Empirical Results |
|---|---|---|---|---|
| Lichtenberg and Siegel (1987) | Plant | U.S. | All M&A | Lower Labor Input Growth Rates Before the Transaction; Slightly Higher After the Transaction |
| Mitchell and Mulherin (1989) | Firm | U.S. | Corporate Takeovers | Only a Few Takeovers Resulted in a Termination of a Pension Fund |
| Rosett (1990) | Firm | U.S. | Corporate Takeovers | Gains to Shareholders Arising From Corporate Takeovers Are Not the Result of Losses to Employees |
| Bhagat, Shleifer, and Vishny (1990) | Firm | U.S. | Hostile Takeovers | 45% of the Companies Involved in a Hostile Takeover Laid Off Workers |
| Pontiff, Shleifer, and Weisbach (1990) | Firm | U.S. | Tender Offers (Corporate Takeovers) | 15% of Hostile Takeover Bids and 8% of Friendly Takeover Bids Led to a Pension Fund Termination |
| Lichtenberg and Siegel (1990a) | Plant | U.S. | LBOs and MBOs of Divisions and Firms | After an LBO or MBO, Non-Production Employment and Wages Declined (Not for Production Workers) |
| Lichtenberg and Siegel (1990b) | Plant | U.S. | All M&A Involving Manufacturing and Auxiliary Establishments | Employment and Wage Growth is Lower in Auxiliary ("Central Office") Establishments Changing Owners; Much Smaller Effects at Production Establishments |
| McGuckin, and Nguyen (2001) | Plant | U.S. | All M&A | Wages and Employment Increase After M&A; Effects Worse For Workers in Large Plants |
| Conyon, Girma, Thompson, Wright (2002a) | Firm | U.K. | Related and Unrelated Mergers | 19% Decline in Employment for Related Mergers; 8% Decline in Employment for Unrelated Mergers |
| Conyon, Girma, Thompson, Wright (2004) | Firm | U.K. | Related and Unrelated Mergers | Increases in Wages For All Mergers, But Especially for Related Mergers |
| Gugler and Yurtoglu (2004) | Firm | U.S. & Europe | Mergers | Mergers Reduced Labor Demand in Europe, But Not in the U.S. |
| Harris, Siegel, and Wright (2005) | Plant | U.K. | MBOs | MBOs Resulted in A Substantial Decline in Plant Employment |
| Siegel, Simons, and Lindstrom (2007) | Plant | Sweden | Partial and Full Acquisitions and Divestitures, Related and Unrelated Acquisitions | Plants Involved in Full Acquisitions and Divestitures and Unrelated Acquisitions Experience Increases in Average Employee Age, Experience, and the Percentage of Employees With a College Education |

***Tab. 19.*** *Plant and firm level studies of the effects of mergers and acquisitions on employment and wages (after Siegel, Simons (2008)).*

The *M&A transactions strategies* and related financial issues in the *diffusion – type financial systems* in the *global capital markets* have been extensively researched in *Nelson (1959), Nelson, Winter (1978), Penrose (1959), Marris (1964), Manne (1965), Thompson (1967), Levy, Sarnat (1970), Lewellen (1971), Weston, Mansinghka (1971), Fama, Miller (1972), Fama (1980), Kirzner (1973), Higgins, Schall (1975), Jensen, Meckling (1976), Jensen, Ruback (1983), Jensen (1986, 1987, 1991, 1993), Jensen, Murphy (1990), Steiner (1976), Kim, McConnell (1977), Lucas (1978), Salter, Weinhold (1979), Bradley (1980), Grossman, Hart (1980), Mueller (1980, 1984, 1987, 1989), Scherer (1980), Amihud, Baruch (1981), Asquith, Kim (1982), Asquith (1983), Asquith, Bruner, Mullins (1983), Hennart (1982, 1988), Hennart, Park (1993), Hennart, Reddy (1997), Stapleton (1982), Eckbo (1983, 2009), Eger (1983), Lubatkin (1983, 1987), Lubatkin, Shrieves (1986), Lubatkin, Srinivasan, Merchant (1997), Lubatkin, Schweiger, Weber (1998), Lubatkin, Schulze, Mainkar, Cotterill (2001), Malatesta (1983), Salant, Switzer, Reynolds (1983), Wansley, Lane, Yang (1983), Blake, Monton (1984), Dess, Robinson (1984), Geroski (1984), Lobue (1984), Sales, Mirvis (1984), Stewart, Harris, Carleton (1984), Buono, Bowditch, Lewis (1985), Duhaime (1985), Hasbrouch (1985), Kusewitt (1985), Malatesta, Thompson (1985), Perry, Porter (1985), Walkling, Edmister (1985), Chatterjee (1986, 1991, 1992), Chatterjee, Lubatkin (1990), Chatterjee, Lubatkin, Schweiger, Weber (1992), Dennis, McConnell (1986), Jemison, Sitkin (1986a, b), Montgomery, Wilson (1986), Palepu (1986), Pastena, Ruland (1986), Roll (1986), Shleifer, Vishny (1986, 1990, 1991, 2001, 2003), Shleifer, Summers (1988), Shrivastava (1986), De, Duplichan (1987), Hansen (1987), Haspeslagh, Farquhar (1987), Huang, Walkling (1987), James, Wier (1987), Neely (1987), Ravenscraft, Scherer (1987a, b), Sicherman, Pettway (1987), Singh, Montgomery (1987), Travlos (1987), Travlos, Waegelein (1992), Trifts, Scanlon (1987), Varaiya, Ferris (1987), Auerbach (1988, 1989), Barney (1988, 1991), Bradley, Desai, Kim (1988), Brown, Medoff (1988), Hall (1988, 1990, 1999), Morck, Shleifer, Vishny (1988, 1990), Nahavandi, Malekzadeh (1988), Schipper, Thompson (1983), Shelton (1988), Walsh (1988, 1989), Bertin, Ghazanfari, Torabzadeh (1989), Bulow, Rogoff (1989), Cantwell (1989), Dubofsky, Fraser (1989), Fishman (1989), Fowler, Schmidt (1989), Golbe, White (1989), Hannan, Wolken (1989), Hayn (1989), Jacquemin, Slade (1989), Kaen, Tehranian (1989), Ravenscraft, Scherer (1989), Wall, Gup (1989), Baradwaj, Fraser, Furtado (1990), Baradwaj, Dubofsky, Fraser (1992), Bhagat, Shleifer, Vishny (1990), Borenstein (1990), Datta, Grant (1990), Datta (1991), Datta, Iskander-Datta, Raman(1992), Datta, Iskandar-Datta (1995), Datta, Iskandar-Datta, Raman (2001), Farrell, Shapiro (1990), Hawawini, Swary (1990), Healy, Palepu, Ruback (1990), Hunt (1990), Farrell, Shapiro (1990), Hitt, Hoskisson, Ireland (1990), Hitt, Hoskisson, Ireland, Harrison*

*(1991), Hitt, Hoskisson, Johnson, Moesel (1996), Hitt, Harrison, Ireland, Best (1998), Hitt, Harrison, Ireland (2001), Holmes, Schmitz (1990), Kamien, Zang (1990 1991, 1993), Lahey, Conn (1990), Loderer, Martin (1990), Rosett (1990), Seth (1990), Seth, Song, Pettit (2002), Shastri (1990), Smith (1990), Cornett, De (1991), Franks, Harris, Titman (1991), Haspeslagh, Jemison (1991), Harris, Ravenscraft (1991), Harrison, Hitt, Hoskisson, Ireland (1991), Mann, Sicherman (1991), Schweiger, Denisi (1991), Servaes (1991), Slusky, Caves (1991), Willig (1991), Agrawal, Jaffe, Mandelker (1992), Agrawal, Jaffe (2000, 2002), Alford (1992), Berger, Humphrey (1992), Berger, Mester (1997), Berger, Saunders, Scalise, Udell (1997, 1998), Berger (1998), Berger, DeYoung, Hesna, Udell (1999), Berger, Demsetz, Strahan (1999), Brush, Vanderwerf (1992), Brush (1996), Gaudet, Salant (1992), Grandstrand, Bohlin, Oskarsson, Sjoberg (1992), Healey, Palepu, Ruback (1992), Ingham, Kran, Lovestam (1992), Loderer, Martin (1992), Shanley, Correa (1992), Srnivasan (1992), United States Department of Justice (1992), Cannella, Hambrick (1993), Hambrick, Cannella (1993), Cotterill (1993, 2002), Gibbs (1993), Kim, Singal (1993), Linder, Crane (1993), Neven, Nuttall, Seabright (1993), Rhoades (1993), Rhoades (1994), Schrantz (1993), Song, Walking (1993), Bruton, Oviatt, White (1994), Chakrabarti, Hauschildt, Suverkup (1994), Clark, Ofek (1994), Houston, Ryngaert (1994), Houston, James, Ryngaert (2001), Hudson (1994), Kesner, Shapiro, Sharma (1994), Madura, Wiant (1994), Markides, Ittner (1994), Pablo (1994), Rhoades (1994), Smith, Kim (1994), Werden, Froeb (1994), Christensen, Rosenbloom (1995), Collins, Kemsley, Shackelford (1995), Collins, Kemsley, Shackelford (1995), Derhy (1995), Gerpott (1995), Gilbert, Sunshine (1995), Gilbert, Tom (2001), Gilbert (2006), Kaplan, Ruback (1995), Long, Vousden (1995), Sudarsanam (1995), Zhang (1995), Brush (1996), Esty, Narasimhan, Tufano (1996), Franks, Mayer (1996), Freixas, Rochet (1996), Gerstein (1996), Mitchell, Mulherin (1996), Pilloff (1996), Schwert (1996, 2000), Servaes (1996), Shapiro (1996), Vennet (1996), Vermeulen, Barkema (1996), Weber (1996), Akhavein, Berger, Humphrey (1997), Anand, Singh (1997), Bergh (1997), Capron, Mitchell (1997, 1998), Capron, Dussauge, Mitchell (1998), Capron (1999a, b), Capron, Hulland (1999), Capron, Pistre (2002), Collins, Maydew, Weiss (1997), Covin, Kolenko, Sightler, Tudor (1997), Fubini, Price, Zollo (1997), Hayward, Hambrick (1997), Hayward (2002), Head, Ries (1997), Holl, Kyriazis (1997), Klemperer (1997), Kroll, Wright, Toombs, Leavell (1997), Krug, Hegarty (1997, 2001), Loughran, Vijh (1997), Megginson, Morgan, Nail (1997), Powell (1997), Ramaswamy (1997), Rosenberg (1997), Siegel, Waldman, Youngdahl (1997), Siegel (1999), Siegel, Simons, Lindstrom (2005), Sirower (1997), Stavros (1997), Walraven (1997), Barkema, Vermeulen (1998), Boyd, Graham (1998), Bulow, Huang, Klemperer (1998, 1999), Chang (1998), Cole, Walraven (1998), Falvey (1998),*

*Schwartz (2008), Maksimovic, Phillips, Prabhala (2008), Neto, Brandão, Cerqueira (2008a, b), Norbäck, Persson (2008a, b), Norbäck, Persson, Vlachos (2009), Toxvaerd (2008), Varma (2008), Venkiteswaran (2008), Balcaen, Buyze, Ooghe (2009), DeYoung, Evanoff, Molyneux (2009), Elnathan, Gavious, Hauser (2009), Jones K (2009), Ornaghi (2009), Panetta, Schivardi, Shum (2009), Patel, Bourgeois (2009), Van Beers, Dekker (2009), Van Lelyveld, Knot (2009), Wong (2009), Aharon, Gavious, Yosef (2010), Byrne (2010), Lenz (2010), Spyrou, Siougle (2010), Asimakopoulos, Athanasoglou (2011), Beltratti, Paladino (2011), Brealey, Myers, Allen (2011), Chaudhuri (2011), Chen, Chou, Lee (2011), Jie, Song, Walkling (2011), Netter, Stegemoller, Wintoki (2011), Vallascas, Hegendorff (2011), Binder (2012), Fornalczyk (2012), Mehwish, Kayani, Javid (2012), Rus (2012), Weitzel, Kling (2012), Friedman (2013), Mukherjee, Chowdhury (2013), Pfister, Kerler, Valk, Prien, Peyer, Kuhn, Hintermann, Ljaskowsky, Arnet (2013), Shumska, Stepanenko-Lypovyk (2013), UNCTAD (2013), Wikipedia (2013).*

## Analysis of mergers and acquisitions transactions strategies in diffusion - type financial system in Switzerland

We analyzed the *M&A transactions* in *Switzerland* in such industries segments as the *chemicals; commodities; consumer markets; financial services; industrial markets; pharmaceuticals & life sciences; power & utilities; private equity; technology, media & telecommunications* over the last few decades, using the data in Tabs. 13 – 30 in *Pfister, Kerler, Valk, Prien, Peyer, Kuhn, Hintermann, Ljaskowsky, Arnet (2013)* and some other data sources.

We would like to make the following research comments:

*1. We think that the globalization has a strong influence on the M&A deals completion in Switzerland.* This research observation is in a good agreement with the data, which has been early documented in *Lenz (2010)*: "*Globalization* and economies of scale effects fuel *corporate merger* activities in nearly every sector. Nevertheless empirical studies have shown that most *mergers and acquisitions* fail to be successful and destroy shareholders wealth. If the firm's management has a clear picture of the relationship between the take-over premium and the requested return of synergy in the future, some irrational decisions can be avoided." A total number of forthcoming *M&A* transactions and their cumulative value in *Switzerland*, namely the M&A transactions, which would be initiated by the growing businesses from the emerging economies, could increase exponentially in coming years as a direct result of globalization.

*2. We believe that the fluctuating dependence of the M&A transactions number over the certain time period is quasi-periodic and cyclic.* This research observation is in full agreement

with the research results in *Chaudhuri (2011)*: "Over the decades, due to *globalization*, *merger waves* have taken on an increasingly international dimension. The volume of *cross-border mergers* has grown over time such that, in *2000*, the ratio of the value of *global cross-border mergers* to the value of *global foreign direct investment (FDI)* was about *80%* (*UNCTAD (2000)*). There is considerable evidence that *cross-border mergers* tend to occur in *waves* (see, for example, *Gaughan (2002), Gugler et al (2003) and UNCTAD.s World Investment Report (2004)*)." The oscillating nature of the *M&A transactions number* was also reported in *Derhy (1995)*: "The re-structurizations which occurred in several western countries analyzed over a long period of time, reveal the existence of *mergers and acquisitions (M&A) waves*." Moreover, the time fluctuation of *M&A transactions number* was reported in *Brakman, Garretsen, Van Marrewijk (2005)* write: "Analyzing *M&As* in a *General Oligopolistic Equilibrium (GOLE)* model incorporating strategic interaction between firms in a general equilibrium setting, we argue that: *M&As* follow revealed comparative advantage as measured by the *Balassa index*, and *M&As* come in waves." The evidences on the oscillating wave nature of the M&A transactions were presented in *Gorton, Khal, Rosen (2005), Martynova, Renneboog (2005), Toxvaerd (2008)*.

**3. We think that there are many factors, which can generate the quasi periodic oscillations of the M&A transactions number in the time domain, for example: the stock market bubble effects.** *Aharon, Gavious, Yosef (2010)* analyzed the *stock market bubble effects* on the *mergers and acquisitions (M&A) transactions*, considering the changes in the pricing of *M&A* transactions throughout the four sub-periods. *Aharon, Gavious, Yosef (2010)* used the *univariate analysis* as well as the *multivariate analysis*, taking to the account the fact that the differences in the valuation multiples can derive not only from the *industrial affiliation*, *time period* or *market trend*, but also from the *target firm's profitability*, *risk* and *growth*. *Aharon, Gavious, Yosef (2010)*: "We document a considerable increase in the prevalence of *M&A transactions* during the bubble for all sectors, followed by a reduction to pre-bubble levels at the bursting of the bubble, and further reduction in sub-sequent post-burst in years, despite recovery in the capital markets during that time." Applying *the integrative creative collateral design thinking principles* to analyze *the M&A transactions behaviour trends*, we propose that the *econophysics* techniques can be used to accurately characterize the *M&A transactions number quasi-periodic oscillations in various industrial sectors over the selected time period*. In our research approach, the *M&A transactions number quasi-periodic oscillations* depend on the following factors *1) the changes of companies valuations over the time; 2) the changes of companies stock market shares prices over the time*; *3) the changes of companies knowledge absorption capacities in the process of integration over the time*; and *4) the changes of the*

*particular industry sector performance index over the time.* **We performed the research of the nonlinearities in the M&A transactions number quasi-periodic oscillations in Matlab, including the following dependences: 1) the ideal dependence of the M&A transactions number quasi-periodic oscillations over the time, 2) the linear dependence of the M&A transactions number quasi-periodic oscillations over the time, 3) the quadratic dependence of the M&A transactions number quasi-periodic oscillations over the time, 2) the exponential dependence of the M&A transactions number quasi-periodic oscillations over the time** in *Ledenyov D O, Ledenyov V O (2013j).*

*4.* Researching the *M&A transactions* in *Europe, North America* and *Asia – Australia regions* over the long time period of observation, **we discovered the Ledenyov effect: the average of a sum of random numbers in the M&A transactions time series represents a time series with the quasi periodic systematic oscillations, which can be finely approximated by the polynomial numbers.** We think that the **Ledenyov effect** in the case of the *M&A transactions quasi periodic cycles* can be considered as similar to the **Slutsky-Yule effect** in *Slutsky (1927a), Yule (1927)* in the cases of the *business cycles* and the *solar activity cycles*.

*5.* The *absorption theory* has been created and applied to understand the *nature of absorption processes* by the *different chemical compounds* in the various *physical – chemical systems* in the *physics* and *chemistry* at the research institutions and top league universities in the *XX – XXI* centuries, for example, in the *nuclear physics* in *Ledenyov O P, Neklyudov (2013), Neklyudov, Dovbnya, Dikiy, Ledenyov O P, Lyashko (2013), Neklyudov, Ledenyov O P, Fedorova, Poltinin (2013a, b), Neklyudov, Fedorova, Poltinin, Ledenyov O P (2013), Ledenyov O P, Neklyudov, Poltinin, Fedorova (2012a, b), Neklyudov, Ledenyov O P, Fedorova, Poltinin (2012).* In the *econophysics*, the *absorption process* plays an important role in *Cohen, Levinthal (1990).* Therefore, it is a subject of intensive research in *Cohen, Levinthal (1990).* For instance, the *absorptive capacity* in the *M&A deals* in *Germany* have been investigated in *Hussinger (2010, 2012).* **We think that, in the course of the M&A transaction implementation, the ability by the companies to absorb the newly acquired knowledge and to create the new innovative knowledge base, is a key pre-determinant of the M&A deal completion success. In our opinion, the integrative collateral creative design thinking has a direct impact on the company's knowledge absorption properties toward the new innovative knowledge base formation in the highly competitive global markets.**

*6.* The strategy selection problems have been extensively researched in *Porter (1987, 1996, 2008), Besanko, Shanley, Dranove (2007), Gavetti, Rivkin (2007), Teece, Winter (2007).* **We would like to state that the winning virtuous mergers and acquisitions transactions strategies** *in diffusion - type financial systems in highly volatile global capital markets with nonlinearities can only be selected through the decision making process on the M&A choices, applying the econophysical econometrical analysis in Amemiya (1985), Greene (2008) with the use of the inductive, deductive and abductive logics in Martin (1998-1999, 2005-2006).*

*List of M&A transactions in Switzerland in 2012*

**Chemicals**

| Announced date | Target | Stake | Target country | Bidder | Bidder country | Seller | Seller country | Value (USDm) |
|---|---|---|---|---|---|---|---|---|
| Feb 2012 | Aluflexpack d.o.o. | 100 | Croatia | Montana Tech Components AG (in consortium with others) | Switzerland | Hypo Alpe-Adria-Bank AG | Austria | 65.0 |
| Mar 2012 | BASF SE (IMEX offset printing inks business) | 100 | Germany | Quantum Kapital AG | Switzerland | - | - | n/a |
| Apr 2012 | APAG Holding AG | 90 | Switzerland | Kanoria Chemicals & Industries Ltd | India | - | - | 8.5 |
| May 2012 | Foamalite Ltd | 100 | Ireland | 3A Composites Holding AG | Switzerland | - | - | n/a |
| Jun 2012 | Gascogne Laminates Switzerland | 100 | Switzerland | UPM Raflatac Oy | Finland | - | - | n/a |
| Aug 2012 | DuPont Professional Products insecticide business | 100 | United States | Syngenta AG | Switzerland | DuPont | United States | 125.0 |
| Sep 2012 | Pasteuria Bioscience Inc | 100 | United States | Syngenta AG | Switzerland | Various | Various | 113.0 |
| Sep 2012 | Devgen NV | 100 | Belgium | Syngenta AG | Switzerland | Various | Various | 463.0 |
| Oct 2012 | RUAG Coatings AG | 100 | Switzerland | Impreglon SE | Germany | - | - | n/a |
| Nov 2012 | Staerkle & Nagler AG | 100 | Switzerland | DKSH Holding Ltd. | Switzerland | - | - | n/a |
| Nov 2012 | Dobroplast Sp zoo SKA | 100 | Poland | AFG Arbonia-Forster-Holding AG | Switzerland | - | - | n/a |
| Dec 2012 | Emulsion Business, Paper Specialities, Textile Chemicals | 100 | Switzerland | SK Capital Partners LP | United States | Clariant AG | Switzerland | 550.0 |

*Tab. 20. List of M&A transactions in chemicals in Switzerland in 2012*

*(after Pfister, Kerler, Valk, Prien, Peyer, Kuhn, Hintermann, Ljaskowsky, Arnet (2013)).*

**Commodities**

| Announced date | Target | Stake | Target country | Bidder | Bidder country | Seller | Seller country | Value (USDm) |
|---|---|---|---|---|---|---|---|---|
| Jan 2012 | Terminal de Carvao da Matola | 35 | Mozambique | Vitol Anker International BV | Switzerland | Grindrod Limited | South Africa | 68.0 |
| Feb 2012 | Xstrata PLC | 66 | Switzerland | Glencore International PLC | Switzerland | Various | Various | 40,212.6 |
| Feb 2012 | Chemoil Energy Limited | 38 | Hong Kong | Singfuel Investment Pte Ltd (Glencore) | Switzerland | - | - | 174.0 |
| Feb 2012 | Montanwerke Brixlegg AG | 100 | Austria | Umcor AG | Switzerland | - | - | 365.6 |
| Feb 2012 | CHAO Kolos | 100 | Ukraine | Glencore International PLC | Switzerland | Cisp Ltd | Ukraine | 80.0 |
| Mar 2012 | Trevali Mining Corp | 9 | Canada | Glencore International PLC | Switzerland | - | - | 18.2 |
| Mar 2012 | Coking coal deposit from Talisman Energy | 100 | Canada | Xstrata | Switzerland | - | - | 500.0 |
| Mar 2012 | GKE Metal Logistics Pte Ltd | 51 | Singapore | Louis Dreyfus Commodities Asia Pte Ltd. | Switzerland | Van Der Horst Energy Limited | Singapore | n/a |
| Mar 2012 | Viterra Inc | 100 | Canada | Glencore International PLC | Switzerland | Various | Various | 6,120.8 |
| Mar 2012 | Ello-Puma Distribuidora de Combustiveis S.A. | 100 | Brazil | AleSat Combustiveis S.A. | Brazil | Trafigura Beheer B.V., Grupo Tavares de Melo, Grupo JB, Portus | Switzerland | n/a |
| Mar 2012 | Cocamar Cooperativa Agroindustrial (Paranavai orange juice plant) | 100 | Brazil | Louis Dreyfus Commoditites B.V. | Switzerland | Cocamar Cooperativa Agroindustrial | Brazil | n/a |
| Mar 2012 | Optimum Coal Holdings Limited | 37 | South Africa | Glencore International plc | Switzerland | - | - | 414.0 |
| Mar 2012 | Cockett Marine Oil Limited | 50 | UK | Vitol Holding B.V. | Switzerland | Grindrod Limited | South Africa | n/a |
| May 2012 | Imperial Sugar Company | - | United States | Louis Dreyfus Commodities LLC | Switzerland | - | - | 206.0 |
| May 2012 | Petroplus Refining Cressier SA | 100 | Switzerland | Varo Holding SA | Switzerland | - | - | n/a |
| May 2012 | KenolKobil Ltd | 50 | Kenya | Puma Energy LLC | Switzerland | - | - | n/a |
| May 2012 | Mineracao Caraiba SA | 29 | Brazil | Glencore International PLC | Switzerland | Branford RJ Participacoes SA | Brazil | 118.5 |
| May 2012 | Samref Overseas / Samref Congo | 24 | Panama | Glencore International PLC | Switzerland | Gorupo Bazano S.p.r.l/ High Grade Minerals S.A. | Congo | 480.0 |
| May 2012 | Ingolstadt Refinery | 100 | Germany | Gunvor Group | Switzerland | - | - | n/a |
| Jul 2012 | Ecoval Dairy Trade | - | Netherlands | Louis Dreyfus Commoditites B.V. | Switzerland | C.V. Datrex Beheer, Prominter N.V./S.A | Belgium | n/a |
| Jul 2012 | Vale Manganese Norway AS | 100 | Norway | Glencore International PLC | Switzerland | - | - | 168.4 |
| Jul 2012 | Vale Manganese France SASU | 100 | France | Glencore International PLC | Switzerland | - | - | 160.5 |
| Jul 2012 | Vale Manganese France/ Norway SAS | - | France | Glencore International PLC | Switzerland | Vale S.A. | Brazil | 160.0 |
| Jul 2012 | Northern Iron Limited | 100 | Australia | Prominvest AG | Switzerland | - | - | 646.0 |
| Aug 2012 | Canadian Fertilizers Limited | 34 | Canada | CF Industries Holdings, Inc. | United States | Glencore International plc | Switzerland | 914.0 |
| Aug 2012 | Kolmar Management Company LLC | 60 | Russia | Gunvor Group | Switzerland | - | - | n/a |

*Tab. 21. List of M&A transactions in commodities in Switzerland in 2012 (after Pfister, Kerler, Valk, Prien, Peyer, Kuhn, Hintermann, Ljaskowsky, Arnet (2013)).*

| Announced date | Target | Stake | Target country | Bidder | Bidder country | Seller | Seller country | Value (USDm) |
|---|---|---|---|---|---|---|---|---|
| Sep 2012 | 38 Shallow Water Drilling Rigs (Transocean Ltd.) | 100 | Switzerland | Shelf Drilling International Holdings, Ltd. | United Arab Emirates | Transocean Ltd. | Switzerland | 1,050.0 |
| Sep 2012 | ACCL Pty Ltd | - | Australia | Glencore Grain Pty Ltd | Switzerland | AACL Holdings Limited | Australia | 0.5 |
| Sep 2012 | Griffiths Energy-Mangara and B | 25 | Chad | Glencore International PLC | Switzerland | - | - | 300.0 |
| Sep 2012 | Griffiths Energy-Exploration A | 33 | Chad | Glencore International PLC | Switzerland | - | - | 31.0 |
| Sep 2012 | LN Metals International Ltd | 100 | United Kingdom | MRI Trading AG | Switzerland | - | - | 12.3 |
| Sep 2012 | Melior Resources Inc | 66 | Canada | Pala Investments Holdings Ltd | Switzerland | - | - | 12.6 |
| Sep 2012 | Kazzinc Ltd | 19 | Kazakhstan | Glencore International plc | Switzerland | Verny Capital JSC | Kazakhstan | 1,395.0 |
| Oct 2012 | Oil Terminal in Port of Antwerp | 100 | Belgium | Mercuria Energy Asset Management B.V. | Switzerland | Nafta (B) N.V. | Belgium | n/a |
| Oct 2012 | Mercator Minerals Ltd | 18 | Canada | Nevada Copper Corp | Canada | Pala Investments Holdings Ltd | Switzerland | 25.9 |
| Oct 2012 | Asian Mineral Resources Ltd | 32 | Canada | Investor Group | Switzerland | - | - | 10.2 |
| Oct 2012 | Medco Sarana Kalibaru PT | 64 | Indonesia | Puma Energy International BV | Switzerland | - | - | n/a |
| Oct 2012 | CEC North Star Energy Ltd | 22 | Canada | Octagon 88 Resources Inc | Switzerland | - | - | 63.0 |
| Oct 2012 | Vesta Terminals B.V. | 50 | Switzerland | Sinomart KTS Development Limited | Bermuda | Mercuria Energy Asset Management B.V. | Switzerland | 163.8 |
| Oct 2012 | Hana Mining Ltd | 82 | Canada | Cupric Canyon Capital LLC | United States | Pala Investments Holdings Ltd | Switzerland | 67.1 |
| Oct 2012 | Belgian Refinery | 100 | Belgium | Gunvor Group | Switzerland | | | n/a |
| Dec 2012 | Usina Sao Carlos storage facility and rights (Biosev S.A.) | 100 | Brazil | Sao Martinho SA | Switzerland | Louis Dreyfus (Biosev S.A.) | Switzerland | 96.0 |

*Tab. 22. List of M&A transactions in commodities in Switzerland in 2012 (after Pfister, Kerler, Valk, Prien, Peyer, Kuhn, Hintermann, Ljaskowsky, Arnet (2013)).*

**Consumer Markets**

| Announced date | Target | Stake | Target country | Bidder | Bidder country | Seller | Seller country | Value (USDm) |
|---|---|---|---|---|---|---|---|---|
| Jan 2012 | La Morella Nuts SA | 100 | Spain | Barry Callebaut AG | Switzerland | - | - | n/a |
| Jan 2012 | Smart Grids AG | 100 | Germany | Sarchem AG | Switzerland | - | - | 24.0 |
| Jan 2012 | Convenience Concept GmbH | 100 | Germany | Valora Holding AG | Switzerland | - | - | n/a |
| Feb 2012 | Kolos AO | 100 | Ukraine | Glencore International plc | Switzerland | Cisp Ltd | Ukraine | 80.0 |
| Feb 2012 | Mösli Fleischwaren AG | 100 | Switzerland | Orior AG | Switzerland | - | - | n/a |
| Mar 2012 | Prothor Holdings SA | 100 | Switzerland | Citizen Holdings Co Ltd | Japan | - | - | 70.7 |
| Mar 2012 | Fly (Schweiz) AG | 100 | Switzerland | Mobilier Europeen SA | France | - | - | n/a |
| Mar 2012 | De Grisogono SA | 45 | Switzerland | Angolan crude oil investors | Angola | - | - | n/a |
| Apr 2012 | Simon Et Membrez SA | 100 | Switzerland | Swatch Group AG | Switzerland | - | - | n/a |
| Apr 2012 | SKINS Global Holding AG | 36 | Switzerland | Jaimie Fuller | United Kingdom | Equity Partners Pty Ltd | Australia | 30.0 |
| Apr 2012 | United Coffee | 100 | Switzerland | UCC Holdings Co Ltd | Japan | CapVest Limited; Harkjaer 1 Limited | United Kingdom | 615.0 |
| Apr 2012 | Pfizer Nutrition | 100 | United States | Nestle SA | Switzerland | Pfizer Inc | United States | 11,850.0 |
| May 2012 | Gastro Star AG | 100 | Switzerland | Hilcona AG | Liechtenstein | - | - | n/a |
| May 2012 | Cash+Carry Angehrn | 50 | Switzerland | Migros-Genossenschafts-Bund | Switzerland | - | - | n/a |
| May 2012 | Contashop AG | 100 | Switzerland | Spar Handels AG | Switzerland | Oettinger Imex AG | Germany | n/a |
| Jun 2012 | Hess Natur-Textilien GmbH | 100 | Germany | CapVis Equity Partners AG | Switzerland | - | - | n/a |
| Jun 2012 | Alliance Boots GmbH | 45 | Switzerland | Walgreen Company | United States | AB Acquisitions Holdings Limited | Gibraltar | 6,665.5 |
| Jun 2012 | Charles Fueglister AG | 100 | Switzerland | Tobi Seeobst AG | Switzerland | - | - | n/a |
| Jul 2012 | Acrotec SA | - | Switzerland | Quilvest SA | Luxembourg | - | - | 32.0 |
| Jul 2012 | Kaiku Corporacion Alimentaria | 23 | Spain | Emmi AG | Switzerland | - | - | n/a |
| Jul 2012 | MMH Holding AG | 100 | Switzerland | Weita Holding AG | Switzerland | - | - | n/a |
| Jul 2012 | Distrimondo AG | 100 | Switzerland | Bunzl Plc | United Kingdom | Markus Meier (Private Investor); Reto Hofmann (Private Investor); Daniel Meier (Private Investor) | Switzerland | n/a |
| Jul 2012 | Telcos Srl | 15 | Italy | Interfines Ag | Switzerland | Almaviva Technologies Srl | Italy | n/a |
| Jul 2012 | Korolevskaya Voda | 100 | Russia | Nestle SA | Switzerland | - | - | 50.0 |
| Aug 2012 | Limoni SpA | 50 | Italy | Orlando Italy Management SA | Switzerland | - | - | 49.6 |
| Aug 2012 | Vilebrequin International SA | 100 | Switzerland | VBQ Acquisition BV | Netherlands | - | - | 133.9 |
| Aug 2012 | Venchiaredo SpA | 26 | Italy | Emmi AG | Switzerland | - | - | n/a |
| Aug 2012 | Hogrefe AG-Bookstores, Magazine | 100 | Switzerland | Lehmanns Media GmbH | Germany | - | - | 15.0 |
| Aug 2012 | s.Oliver Bernd Freier GmbH | 100 | Germany | Schild AG | Switzerland | - | - | n/a |
| Aug 2012 | Domino's Pizza Switzerland AG | 100 | Switzerland | Domino's Pizza UK & IRL PLC | United Kingdom | - | - | 8.0 |
| Sep 2012 | Folli Follie SA | 51 | Greece | Dufry AG | Switzerland | - | - | n/a |
| Sep 2012 | Peter Millar | 100 | United States | Compagnie Financiere Richemont SA | Switzerland | - | - | n/a |

***Tab. 23**. List of M&A transactions in commodities in Switzerland in 2012 (after Pfister, Kerler, Valk, Prien, Peyer, Kuhn, Hintermann, Ljaskowsky, Arnet (2013)).*

## Consumer Markets

| Announced date | Target | Stake | Target country | Bidder | Bidder country | Seller | Seller country | Value (USDm) |
|---|---|---|---|---|---|---|---|---|
| Jan 2012 | La Morella Nuts SA | 100 | Spain | Barry Callebaut AG | Switzerland | - | - | n/a |
| Jan 2012 | Smart Grids AG | 100 | Germany | Sarchem AG | Switzerland | - | - | 24.0 |
| Jan 2012 | Convenience Concept GmbH | 100 | Germany | Valora Holding AG | Switzerland | - | - | n/a |
| Feb 2012 | Kolos AO | 100 | Ukraine | Glencore International plc | Switzerland | Cisp Ltd | Ukraine | 80.0 |
| Feb 2012 | Mösli Fleischwaren AG | 100 | Switzerland | Orior AG | Switzerland | - | - | n/a |
| Mar 2012 | Prothor Holdings SA | 100 | Switzerland | Citizen Holdings Co Ltd | Japan | - | - | 70.7 |
| Mar 2012 | Fly (Schweiz) AG | 100 | Switzerland | Mobilier Europeen SA | France | - | - | n/a |
| Mar 2012 | De Grisogono SA | 45 | Switzerland | Angolan crude oil investors | Angola | - | - | n/a |
| Apr 2012 | Simon Et Membrez SA | 100 | Switzerland | Swatch Group AG | Switzerland | - | - | n/a |
| Apr 2012 | SKINS Global Holding AG | 36 | Switzerland | Jaimie Fuller | United Kingdom | Equity Partners Pty Ltd | Australia | 30.0 |
| Apr 2012 | United Coffee | 100 | Switzerland | UCC Holdings Co Ltd | Japan | CapVest Limited; Harkjaer 1 Limited | United Kingdom | 615.0 |
| Apr 2012 | Pfizer Nutrition | 100 | United States | Nestle SA | Switzerland | Pfizer Inc | United States | 11,850.0 |
| May 2012 | Gastro Star AG | 100 | Switzerland | Hilcona AG | Liechtenstein | - | - | n/a |
| May 2012 | Cash+Carry Angehrn | 50 | Switzerland | Migros-Genossenschafts-Bund | Switzerland | - | - | n/a |
| May 2012 | Contashop AG | 100 | Switzerland | Spar Handels AG | Switzerland | Oettinger Imex AG | Germany | n/a |
| Jun 2012 | Hess Natur-Textilien GmbH | 100 | Germany | CapVis Equity Partners AG | Switzerland | - | - | n/a |
| Jun 2012 | Alliance Boots GmbH | 45 | Switzerland | Walgreen Company | United States | AB Acquisitions Holdings Limited | Gibraltar | 6,665.5 |
| Jun 2012 | Charles Fueglister AG | 100 | Switzerland | Tobi Seeobst AG | Switzerland | - | - | n/a |
| Jul 2012 | Acrotec SA | - | Switzerland | Quilvest SA | Luxembourg | - | - | 32.0 |
| Jul 2012 | Kaiku Corporacion Alimentaria | 23 | Spain | Emmi AG | Switzerland | - | - | n/a |
| Jul 2012 | MMH Holding AG | 100 | Switzerland | Weita Holding AG | Switzerland | - | - | n/a |
| Jul 2012 | Distrimondo AG | 100 | Switzerland | Bunzl Plc | United Kingdom | Markus Meier (Private Investor); Reto Hofmann (Private Investor); Daniel Meier (Private Investor) | Switzerland | n/a |
| Jul 2012 | Telcos Srl | 15 | Italy | Interfines Ag | Switzerland | Almaviva Technologies Srl | Italy | n/a |
| Jul 2012 | Korolevskaya Voda | 100 | Russia | Nestle SA | Switzerland | - | - | 50.0 |
| Aug 2012 | Limoni SpA | 50 | Italy | Orlando Italy Management SA | Switzerland | - | - | 49.6 |
| Aug 2012 | Vilebrequin International SA | 100 | Switzerland | VBQ Acquisition BV | Netherlands | - | - | 133.9 |
| Aug 2012 | Venchiaredo SpA | 26 | Italy | Emmi AG | Switzerland | - | - | n/a |
| Aug 2012 | Hogrefe AG-Bookstores, Magazine | 100 | Switzerland | Lehmanns Media GmbH | Germany | - | - | 15.0 |
| Aug 2012 | s.Oliver Bernd Freier GmbH | 100 | Germany | Schild AG | Switzerland | - | - | n/a |
| Aug 2012 | Domino's Pizza Switzerland AG | 100 | Switzerland | Domino's Pizza UK & IRL PLC | United Kingdom | - | - | 8.0 |
| Sep 2012 | Folli Follie SA | 51 | Greece | Dufry AG | Switzerland | - | - | n/a |
| Sep 2012 | Peter Millar | 100 | United States | Compagnie Financiere Richemont SA | Switzerland | - | - | n/a |

**Tab. 24**. *List of M&A transactions in commodities in Switzerland in 2012*

*(after Pfister, Kerler, Valk, Prien, Peyer, Kuhn, Hintermann, Ljaskowsky, Arnet (2013)).*

**Consumer Markets**

| Announced date | Target | Stake | Target country | Bidder | Bidder country | Seller | Seller country | Value (USDm) |
|---|---|---|---|---|---|---|---|---|
| Jan 2012 | La Morella Nuts SA | 100 | Spain | Barry Callebaut AG | Switzerland | - | - | n/a |
| Jan 2012 | Smart Grids AG | 100 | Germany | Sarchem AG | Switzerland | - | - | 24.0 |
| Jan 2012 | Convenience Concept GmbH | 100 | Germany | Valora Holding AG | Switzerland | - | - | n/a |
| Feb 2012 | Kolos AO | 100 | Ukraine | Glencore International plc | Switzerland | Cisp Ltd | Ukraine | 80.0 |
| Feb 2012 | Mösli Fleischwaren AG | 100 | Switzerland | Orior AG | Switzerland | - | - | n/a |
| Mar 2012 | Prothor Holdings SA | 100 | Switzerland | Citizen Holdings Co Ltd | Japan | - | - | 70.7 |
| Mar 2012 | Fly (Schweiz) AG | 100 | Switzerland | Mobilier Europeen SA | France | - | - | n/a |
| Mar 2012 | De Grisogono SA | 45 | Switzerland | Angolan crude oil investors | Angola | - | - | n/a |
| Apr 2012 | Simon Et Membrez SA | 100 | Switzerland | Swatch Group AG | Switzerland | - | - | n/a |
| Apr 2012 | SKINS Global Holding AG | 36 | Switzerland | Jaimie Fuller | United Kingdom | Equity Partners Pty Ltd | Australia | 30.0 |
| Apr 2012 | United Coffee | 100 | Switzerland | UCC Holdings Co Ltd | Japan | CapVest Limited; Harkjaer 1 Limited | United Kingdom | 615.0 |
| Apr 2012 | Pfizer Nutrition | 100 | United States | Nestle SA | Switzerland | Pfizer Inc | United States | 11,850.0 |
| May 2012 | Gastro Star AG | 100 | Switzerland | Hilcona AG | Liechtenstein | - | - | n/a |
| May 2012 | Cash+Carry Angehrn | 50 | Switzerland | Migros-Genossenschafts-Bund | Switzerland | - | - | n/a |
| May 2012 | Contashop AG | 100 | Switzerland | Spar Handels AG | Switzerland | Oettinger Imex AG | Germany | n/a |
| Jun 2012 | Hess Natur-Textilien GmbH | 100 | Germany | CapVis Equity Partners AG | Switzerland | - | - | n/a |
| Jun 2012 | Alliance Boots GmbH | 45 | Switzerland | Walgreen Company | United States | AB Acquisitions Holdings Limited | Gibraltar | 6,665.5 |
| Jun 2012 | Charles Fueglister AG | 100 | Switzerland | Tobi Seeobst AG | Switzerland | - | - | n/a |
| Jul 2012 | Acrotec SA | - | Switzerland | Quilvest SA | Luxembourg | - | - | 32.0 |
| Jul 2012 | Kaiku Corporacion Alimentaria | 23 | Spain | Emmi AG | Switzerland | - | - | n/a |
| Jul 2012 | MMH Holding AG | 100 | Switzerland | Weita Holding AG | Switzerland | - | - | n/a |
| Jul 2012 | Distrimondo AG | 100 | Switzerland | Bunzl Plc | United Kingdom | Markus Meier (Private Investor); Reto Hofmann (Private Investor); Daniel Meier (Private Investor) | Switzerland | n/a |
| Jul 2012 | Telcos Srl | 15 | Italy | Interfines Ag | Switzerland | Almaviva Technologies Srl | Italy | n/a |
| Jul 2012 | Korolevskaya Voda | 100 | Russia | Nestle SA | Switzerland | - | - | 50.0 |
| Aug 2012 | Limoni SpA | 50 | Italy | Orlando Italy Management SA | Switzerland | - | - | 49.6 |
| Aug 2012 | Vilebrequin International SA | 100 | Switzerland | VBQ Acquisition BV | Netherlands | - | - | 133.9 |
| Aug 2012 | Venchiaredo SpA | 26 | Italy | Emmi AG | Switzerland | - | - | n/a |
| Aug 2012 | Hogrefe AG-Bookstores, Magazine | 100 | Switzerland | Lehmanns Media GmbH | Germany | - | - | 15.0 |
| Aug 2012 | s.Oliver Bernd Freier GmbH | 100 | Germany | Schild AG | Switzerland | - | - | n/a |
| Aug 2012 | Domino's Pizza Switzerland AG | 100 | Switzerland | Domino's Pizza UK & IRL PLC | United Kingdom | - | - | 8.0 |
| Sep 2012 | Folli Follie SA | 51 | Greece | Dufry AG | Switzerland | - | - | n/a |
| Sep 2012 | Peter Millar | 100 | United States | Compagnie Financiere Richemont SA | Switzerland | - | - | n/a |

*Tab. 25.* *List of M&A transactions in consumer markets in Switzerland in 2012 (after Pfister, Kerler, Valk, Prien, Peyer, Kuhn, Hintermann, Ljaskowsky, Arnet (2013)).*

| Announced date | Target | Stake | Target country | Bidder | Bidder country | Seller | Seller country | Value (USDm) |
|---|---|---|---|---|---|---|---|---|
| Sep 2012 | Brezelbaeckerei Ditsch GmbH | 100 | Germany | Valora Holding AG | Switzerland | - | - | 107.0 |
| Oct 2012 | Hellenic Duty Free Shops S.A. (Travel Retail Business) | 51 | Switzerland | Dufry Group | Switzerland | Hellenic Duty Free Shops S.A. | Greece | 258.0 |
| Oct 2012 | Tegut-Retail Business | 100 | Germany | Genossenschaft Migros Zuerich | Switzerland | - | - | n/a |
| Oct 2012 | Douglas Holding AG | 100 | Germany | Beauty Holding Three AG | Germany | Bank Sarasin & Cie AG; Dr August Oetker KG | Switzerland | 1,940.0 |
| Oct 2012 | DocMorris NV | 100 | Netherlands | Zur Rose AG | Switzerland | - | - | 32.4 |
| Oct 2012 | Lacoste S.A. | 30 | France | Maus Freres SA | Switzerland | - | - | 388.3 |
| Nov 2012 | Givaudan SA (Vegetables, wine and vinegar extracts business) | 100 | Switzerland | DIANA Group SA | France | Givaudan SA. | Switzerland | n/a |
| Nov 2012 | Lacoste S.A. | 28 | France | Maus Freres SA | Switzerland | - | - | 357.6 |
| Nov 2012 | Faberge Ltd | 100 | Switzerland | Gemfields PLC | United Kingdom | Pallinghurst Resources LLP | United Kingdom | 133.0 |
| Nov 2012 | Fnac Italia SpA | 100 | Italy | Orlando Italy Management SA | Switzerland | FNAC SA | France | n/a |
| Nov 2012 | Petra Foods-Cocoa Ingredients | 100 | Singapore | Barry Callebaut AG | Switzerland | Petra Foods Limited | Singapore | 950.0 |

*Tab. 26. List of M&A transactions in consumer markets in Switzerland in 2012 (after Pfister, Kerler, Valk, Prien, Peyer, Kuhn, Hintermann, Ljaskowsky, Arnet (2013)).*

**Financial Services**

| Announced date | Target | Stake | Target country | Bidder | Bidder country | Seller | Seller country | Value (USDm) |
|---|---|---|---|---|---|---|---|---|
| Jan 2012 | Mirabaud Finanzas Sociedad | 75 | Spain | Mirabaud & Cie Banquiers Prives | Switzerland | - | - | n/a |
| Jan 2012 | EFG Bank Dänemark AB | 100 | Denmark | SEB AG | Germany | EFG Banken Gruppe | Switzerland | n/a |
| Jan 2012 | Wegelin & Co. | 100 | Switzerland | Raiffeisen Schweiz | Switzerland | Wegelin & Co. | Switzerland | n/a |
| Jan 2012 | SIF Swiss Investment Funds SA | 100 | Switzerland | CACEIS (Switzerland) SA | Switzerland | - | | n/a |
| Feb 2012 | Swisspartners Investment Network AG | 67 | Switzerland | unknown | Unknown | Liechtensteinische Landesbank (LLB) | Liechtenstein | n/a |
| Feb 2012 | Arkos Capital SA | 100 | Switzerland | GAM Group AG | Switzerland | - | - | n/a |
| Feb 2012 | CMB Banque Privee (Suisse) SA | 100 | Switzerland | PKB Privatbank AG | Switzerland | - | - | n/a |
| Feb 2012 | Nexar Capital SAS | 100 | France | Union Bancaire Privee(UBP) | Switzerland | - | - | n/a |
| Mar 2012 | Ariel-Credit and Surety Ops | 100 | Switzerland | Arch Capital Group Ltd | Bermuda | - | - | n/a |
| Mar 2012 | Walkers Management Services | 100 | Cayman Islands | Intertrust Group Holding SA | Switzerland | - | - | n/a |
| Mar 2012 | Marble Bar Asset Management | - | United Kingdom | Investor Group | United Kingdom | EFG International AG | Switzerland | 31.4 |
| Mar 2012 | Clariden-ILS Business | 100 | Switzerland | LGT Capital Management AG | Liechtenstein | - | - | n/a |
| Mar 2012 | Arbuthnot AG | 100 | Switzerland | Ducartis Holding AG | Switzerland | - | - | 2.2 |
| Apr 2012 | Securis Investment Partners LLP (Majority Stake) | 100 | United Kingdom | Northill Capital LLP | Jersey | Swiss Re AG | Switzerland | n/a |
| May 2012 | Catam Asset Management / Asserta Asset Management / F.I.T. G | 100 | Switzerland | Management schweizer Unternehmung | Switzerland | Cat Group AG | Switzerland | n/a |
| May 2012 | Reassure America Life Insurance Company | 100 | United States | Jackson National Life Insurance Company | United Kingdom | Swiss Re AG | Switzerland | 600.0 |
| Jun 2012 | Banco Santander (Suisse) SA | 100 | Spain | Union Bancaire Privee, UBP SA | Switzerland | Banco Santander, S.A. | Spain | n/a |
| Jun 2012 | PT Asuransi Jaya Proteksi | 100 | Indonesia | ACE Limited | Switzerland | - | - | 130.0 |
| Jul 2012 | GAN Eurocourtage | 100 | France | Helvetia Holding AG | Switzerland | Groupama SA | France | 49.5 |
| Jul 2012 | Swiss Re Private Equity | 100 | Switzerland | BlackRock Inc | United States | - | - | n/a |
| Jul 2012 | Bank Sarasin & Cie AG | 40 | Switzerland | Grupo Safra SA | Switzerland | Various | Various | 699.3 |
| Aug 2012 | BOA Merrill Lynch-Wealth Mgmt | 100 | Switzerland | Julius Baer Group Ltd | Switzerland | Bank of America | United States | 883.3 |
| Sep 2012 | Fianzas Monterrey SA | 100 | Mexico | ACE Ltd | Switzerland | - | - | 285.0 |
| Oct 2012 | Clariden Leu(Europe)Ltd | 100 | United Kingdom | Falcon Private Bank Ltd | Switzerland | - | - | n/a |
| Oct 2012 | ABA Seguros | 100 | Mexico | ACE Limited | Switzerland | Ally Financial Inc. | United States | 865.0 |
| Nov 2012 | Accion Investments in Microfinance, SPC (Controlling stake) | 100 | United States | Bamboo Finance S.a.r.l. | Switzerland | ACCION International | United States | 105.0 |
| Nov 2012 | Chiara Vita SpA | 30 | Italy | Helvetia Holding AG | Switzerland | Banco Di Desio e Della Brianza | Italy | 28.8 |
| Nov 2012 | Chiara Assicurazioni SpA | 51 | Italy | Helvetia Holding AG | Switzerland | - | - | 21.8 |
| Dec 2012 | Intertrust Group Holding SA | 100 | Switzerland | Blackstone Group LP | United States | Waterland Private Equity Investments BV | Netherlands | 867.5 |
| Dec 2012 | Oslo Clearing ASA | 100 | Norway | SIX Group AG | Switzerland | VPS Holding | Norway | 32.1 |

*Tab. 27. List of M&A transactions in financial services in Switzerland in 2012 (after Pfister, Kerler, Valk, Prien, Peyer, Kuhn, Hintermann, Ljaskowsky, Arnet (2013)).*

**Industrial Markets**

| Announced date | Target | Stake | Target country | Bidder | Bidder country | Seller | Seller country | Value (USDm) |
|---|---|---|---|---|---|---|---|---|
| Jan 2012 | Aerowatt SA | 70 | France | Kleinkraftwerk Birseck AG | Switzerland | Viveris Management SAS; Credit Agricole Private Equity; Demeter Partners SA | France | 49.9 |
| Jan 2012 | Rioglass Solar SA | 49 | Spain | Partners Group Holding; Ventizz Capital Fund IV L.P. | Switzerland | - | - | 129.7 |
| Jan 2012 | Gohlke Elektronik GmbH | 100 | Germany | CCS Customer Care & Solutions Holding AG | Switzerland | - | - | n/a |
| Jan 2012 | Thomas & Betts Corp | 100 | United States | ABB Ltd | Switzerland | Various | Various | 3,901.3 |
| Jan 2012 | payment solution AG | 59 | Germany | Bluehill ID AG | Switzerland | - | - | 9.0 |
| Feb 2012 | IMA Automation Berlin GmbH | 100 | Germany | Mikron Holding AG | Switzerland | - | - | n/a |
| Feb 2012 | TLM-TVP doo | - | Croatia | Montana Tech Components AG | Switzerland | - | - | 4.7 |
| Feb 2012 | Maag Pump Systems AG | 100 | Switzerland | Dover Corp | United States | - | - | 303.9 |
| Feb 2012 | Uster Technologies AG | 50 | Switzerland | Toyota Industries Corporation | Japan | Various | Various | 318.7 |
| Feb 2012 | CIEFFE Holding SpA | 100 | Italy | 1C Industries Zug AG | Switzerland | - | - | n/a |
| Mar 2012 | Pyromex AG | 70 | Switzerland | PowerHouse Energy Group Plc | United Kingdom | Peter Jeney (Private Investor) | - | 52.0 |
| Mar 2012 | Oerlikon Solar Holding AG | 100 | Switzerland | Tokyo Electron Ltd | Japan | The Oerlikon Group | Switzerland | 275.6 |
| Mar 2012 | AKAtech | 55 | Austria | Zurmont Madison Private Equity LP | Switzerland | - | - | n/a |
| Mar 2012 | Stadler Rail AG | 20 | Switzerland | Peter Spuhler (Private Investor) | Switzerland | Capvis Equity Partners AG | Switzerland | n/a |
| Mar 2012 | Rauscher & Stoecklin AG | 100 | Switzerland | CGS Management AG (Funding) | Jersey | - | - | n/a |
| Mar 2012 | HERZING + SCHROTH GmbH | 100 | Germany | Feintool International Holding AG | Switzerland | - | - | n/a |
| Mar 2012 | Aseptomag AG | 100 | Switzerland | GEA Group AG | Germany | - | - | n/a |
| Mar 2012 | Pentair Inc | 53 | United States | Tyco Flow Control | Switzerland | Various | Various | 5,230.0 |
| Apr 2012 | CT-Concept Technologie AG | 100 | Switzerland | Power Integrations Inc | United States | - | - | 116.2 |
| Apr 2012 | Leybold Optics GmbH | 100 | Germany | Buehler AG | Switzerland | - | - | n/a |
| Apr 2012 | Pilatus Flugzeugwerke AG | 14 | Switzerland | Investor Group | Switzerland | OC Oerlikon Corp AG | Switzerland | n/a |
| Apr 2012 | Port-A-Cool LLC | 100 | United States | Walter Meier AG | Switzerland | - | - | n/a |
| Apr 2012 | Connectors Verbindungstechnik | 100 | Switzerland | NORMA Group AG | Germany | - | - | n/a |
| May 2012 | Trimco International Holdings Limited (Majority Stake) | 100 | China | Partners Group Holding | Switzerland | Navis Asia Fund IV L.P. | Malaysia | 11.2 |
| May 2012 | Risi AG-Civil Engineering | 100 | Switzerland | Johann Mueller AG | Switzerland | - | - | n/a |
| May 2012 | Indo Schottle Auto Parts Pvt | 45 | India | SFS Intec Holding AG | Switzerland | - | - | 30.0 |
| May 2012 | WinGroup AG | 36 | Switzerland | Nordstjernan Industriutveckling AB | Sweden | - | - | n/a |
| May 2012 | Almatec AG | 100 | Switzerland | Knill Gruppe | Austria | VTC Partners GmbH; Jurg J. Spieler (Private individual); Alfred Hertli (Private investor) | Germany | n/a |
| May 2012 | e2v Microsensors SA | 100 | Switzerland | SGX Sensortech Ltd | United Kingdom | - | - | 24.0 |

*Tab. 28.* List of M&A transactions in industrial markets in Switzerland in 2012 (after Pfister, Kerler, Valk, Prien, Peyer, Kuhn, Hintermann, Ljaskowsky, Arnet (2013)).

| Announced date | Target | Stake | Target country | Bidder | Bidder country | Seller | Seller country | Value (USDm) |
|---|---|---|---|---|---|---|---|---|
| May 2012 | Independent Pipe Products Inc | 100 | United States | Georg Fischer Piping Systems Ltd | Switzerland | - | - | n/a |
| May 2012 | Bumotec SA | 100 | Switzerland | Starrag Group Holding AG | Switzerland | - | - | n/a |
| May 2012 | Pago Holding AG | 100 | Switzerland | Fuji Seal International Inc | Japan | - | - | 128.5 |
| May 2012 | Petrowell Ltd | 100 | United Kingdom | Weatherford International Ltd | Switzerland | - | - | n/a |
| Jun 2012 | Graph-Tech AG | 81 | Switzerland | Domino Printing Sciences PLC | United Kingdom | - | - | 30.0 |
| Jun 2012 | Revue Thommen AG | 100 | Switzerland | ZAO "Tranzas" | Russia | - | - | n/a |
| Jun 2012 | Soutec AG | 100 | Switzerland | Andritz AG | Austria | VTC Partners GmbH | Germany | n/a |
| Jul 2012 | OC Oerlikon Corporation AG | 100 | Switzerland | Mizar Holding Company, Inc. | United States | OC Oerlikon Corporation AG | Switzerland | n/a |
| Jul 2012 | CGM AB | - | Sweden | ABB Ltd | Switzerland | - | - | n/a |
| Jul 2012 | WMF Württembergische Metallwarenfabrik AG | 100 | Germany | KKR Kohlberg, Kravis, Roberts & Co. | United Kingdom | CapVis Equity Partners AG | Switzerland | 305.9 |
| Jul 2012 | Nedis BV | 100 | Netherlands | Daetwyler Holding AG | Switzerland | Konig Corporate Holding BV; Gilad BVBA | Netherlands; Belgium | n/a |
| Jul 2012 | Bartec GmbH | 100 | Germany | Charterhouse Capital Partners LLP | United Kingdom | Partners Group AG | Switzerland | n/a |
| Jul 2012 | RGM S.p.A. (Rail Vehicle Power Business) | 100 | Italy | ABB Ltd | Switzerland | RGM S.p.A. | Italy | 14.0 |
| Aug 2012 | Allpack Group AG | 100 | Switzerland | Rhenochem AG | Switzerland | - | - | n/a |
| Aug 2012 | TSK Pruefsysteme GmbH | 100 | Germany | Komax Holdings AG | Switzerland | - | - | n/a |
| Aug 2012 | Temple Tag, Inc. | 100 | United States | Datamars SA | Switzerland | | | 6.4 |
| Aug 2012 | Tornos SA | 5 | Switzerland | Walter Fust | Switzerland | - | - | n/a |
| Aug 2012 | Unisteel Technology Ltd | 100 | Singapore | SFS Intec Holding AG | Switzerland | - | - | n/a |
| Aug 2012 | Microoled SAS | - | France | STMicroelectronics NV | Switzerland | - | - | 7.5 |
| Sep 2012 | Anhui Zhongding Taike Auto | 50 | China | Daetwyler Holding AG | Switzerland | - | - | 63.7 |
| Sep 2012 | Xiril AG | 100 | Switzerland | Sias AG | Switzerland | - | - | n/a |
| Sep 2012 | Swisslog Holding AG | 10 | Switzerland | Grenzebach Maschinenbau GmbH | Germany | - | - | n/a |
| Sep 2012 | Nexus Marine AB | 100 | Sweden | Garmin Ltd | Switzerland | - | - | n/a |
| Sep 2012 | Wetzel Holding AG | 100 | Switzerland | Matthews International Corp | United States | - | - | 55.5 |
| Sep 2012 | Ionbond AG | 100 | Switzerland | IHI Corp | Japan | - | - | n/a |
| Oct 2012 | FHS Frech-Hoch AG | 100 | Switzerland | ESTECH Industries Holding AG | Switzerland | - | - | 12.0 |
| Oct 2012 | Saia-Burgess Controls AG | 100 | Switzerland | Honeywell International Inc | United States | - | - | 129.7 |
| Oct 2012 | KVT-Fastening Branch | 100 | Switzerland | Bossard Holding AG | Switzerland | - | - | 214.5 |
| Oct 2012 | Pergo AG | 100 | Switzerland | Mohawk Industries Inc | United States | - | - | 150.0 |
| Nov 2012 | Trinecke Zelezarny as | 11 | Czech Republic | Moravia Steel as | Czech Republic | Commercial Metals GmbH | Switzerland | 29.0 |
| Nov 2012 | Swissmetal Luedenscheid GmbH | 100 | Germany | LBIS SA | Switzerland | - | - | n/a |
| Nov 2012 | MWH Barcol-Air AG | 80 | Switzerland | Walter Meier AG | Switzerland | - | - | n/a |
| Nov 2012 | TE Connectivity Ltd-Magnetics | 100 | Switzerland | Bel Fuse Inc | United States | - | - | 22.4 |

**Tab. 29**: *List of M&A transactions in industrial markets in Switzerland in 2012 (after Pfister, Kerler, Valk, Prien, Peyer, Kuhn, Hintermann, Ljaskowsky, Arnet (2013)).*

| Announced date | Target | Stake | Target country | Bidder | Bidder country | Seller | Seller country | Value (USDm) |
|---|---|---|---|---|---|---|---|---|
| Nov 2012 | BT Magnet Technologie GmbH | 100 | Germany | Quantum Kapital AG | Switzerland | Robert Bosch GmbH; TDK Electronics Europe GmbH | Germany | n/a |
| Dec 2012 | OC Oerlikon-Natural Textiles | 100 | Switzerland | Jiangsu Jinsheng Industry Co Ltd | China | - | - | 702.5 |
| Dec 2012 | Volvo Baumaschinen Bayern GmbH | 100 | Germany | Robert Aebi AG | Switzerland | Volvo Construction Equipment | Belgium | n/a |
| Dec 2012 | Filtrox Engineering AG | 100 | Switzerland | Bucher Unipektin AG | Switzerland | - | - | n/a |
| Dec 2012 | Ismeca Semiconductor Holding | 100 | Switzerland | Cohu Inc | United States | - | - | 54.5 |
| Dec 2012 | Aceros y Techos S.A.; Galvanizadora Peruana, S.A.C. | 75 | Peru | Duferco S.A. | Switzerland | - | - | 8.0 |
| Dec 2012 | Apollo Construction Equipment | 70 | India | Ammann Group | Switzerland | - | - | 51.3 |
| Dec 2012 | OELHYDRAULIK ALTENERDING | 100 | Germany | Bucher Industries AG | Switzerland | - | - | n/a |
| Dec 2012 | Astrolab Inc. | 100 | United States | HUBER+SUHNER AG | Switzerland | - | - | n/a |

*Tab. 30.* List of M&A transactions in industrial markets in Switzerland in 2012 (after Pfister, Kerler, Valk, Prien, Peyer, Kuhn, Hintermann, Ljaskowsky, Arnet (2013)).

**Pharmaceuticals & Life Sciences**

| Announced date | Target | Stake | Target country | Bidder | Bidder country | Seller | Seller country | Value (USDm) |
|---|---|---|---|---|---|---|---|---|
| Jan 2012 | Sonnenhof AG-Hospitals(2) | - | Switzerland | Stiftung Lindenhof Bern | Switzerland | - | - | 51.0 |
| Apr 2012 | Actavis Group | 100 | Switzerland | Watson Pharmaceuticals Inc | United States | Novator Partners LLP | United Kingdom | 5,806.0 |
| May 2012 | Fougera Pharmaceuticals Inc | 100 | United States | Sandoz AG | Switzerland | Nordic Capital; DLJ Merchant Banking Partners; Avista Capital Partners, L.P. | United States | 1,525.0 |
| May 2012 | Neodent | 49 | Brazil | Straumann Holding AG | Switzerland | Private Investors | Various | 275.3 |
| May 2012 | Alliance Medical Products, Inc. | 100 | United States | Siegfried Holding AG | Switzerland | Various | Various | 58.0 |
| Jul 2012 | Ondal Medical Systems GmbH | 100 | Germany | Capvis Equity Partners AG | Switzerland | Findos Investor GmbH | Germany | n/a |
| Jul 2012 | Micro-Macinazione SA | - | Switzerland | Cross Equity Partners AG | Switzerland | - | - | 13.9 |
| Jul 2012 | Klinik Lindberg AG | 100 | Switzerland | Privatklinik Bethanien AG | Switzerland | - | - | n/a |
| Aug 2012 | OLIC Ltd. | | Switzerland | Fuji Pharma | Japan | DKSH Holding AG | Switzerland | 54.1 |
| Sep 2012 | F2G Ltd | 100 | United Kingdom | Novartis Bioventures Ltd.; Advent Life Sciences | Switzerland | - | - | 30.0 |
| Sep 2012 | NeuroSearch A/S-Huntexil | 100 | Denmark | Ivax International GmbH | Switzerland | - | - | 35.4 |
| Nov 2012 | SENIOcare AG | - | Switzerland | Waterland Private Equity Investments BV | Netherlands | Akina Partners Limited (Euro Choice II fund) | Switzerland | n/a |
| Nov 2012 | Straumann Holding AG | 10 | Switzerland | Government of Singapore Investment Corp Pte Ltd{GIC} | Singapore | - | - | n/a |
| Nov 2012 | Peptidream Inc | - | Japan | Novartis AG | Switzerland | - | - | 7.5 |
| Dec 2012 | Spirig Pharma AG | 100 | Switzerland | Galderma Pharma SA | Switzerland | - | - | n/a |

*Tab. 31.* List of M&A transactions in pharmaceuticals and life sciences in Switzerland in 2012 (after Pfister, Kerler, Valk, Prien, Peyer, Kuhn, Hintermann, Ljaskowsky, Arnet (2013)).

**Power & Utilities**

| Announced date | Target | Stake | Target country | Bidder | Bidder country | Seller | Seller country | Value (USDm) |
|---|---|---|---|---|---|---|---|---|
| Mar 2012 | Borkum Riffgrund 1 | 32 | Germany | Kirkbi AG | Switzerland | Dong A/S | Denmark | 7.0 |
| Apr 2012 | Gemeinde Oberhofen-Electric | 100 | Switzerland | Energie Thun AG | Switzerland | - | - | 8.1 |
| Apr 2012 | Alpiq Holding AG (Energieversorgungs-technik Division) | 100 | Switzerland | VINCI Energies S.A. | United Kingdom | Alpiq Holding AG | Switzerland | 392.0 |
| May 2012 | Repartner Produktions AG | - | Switzerland | Energie Wasser Luzern(ewl) | Switzerland | Various | Various | 53.8 |
| May 2012 | Wind farm in Aveyron | 100 | France | Terravent AG | Switzerland | Direct Energy | France | n/a |
| Jul 2012 | Windreich AG-Wind Turbines(7) | 100 | Germany | Swisspower AG | Switzerland | - | - | n/a |
| Jul 2012 | etwag | 100 | Switzerland | Erdgas Zuerich AG | Switzerland | - | - | 15.1 |
| Jul 2012 | Eight wind farms in France | 100 | Switzerland | Groupe E SA; EOS Holding SA; SI-REN SA | Switzerland | Eolfi Asset Management | France | 147.0 |
| Sep 2012 | Nant de Drance SA | 15 | Switzerland | IWB Industriele Werke Basel | Switzerland | Alpiq AG | Switzerland | 323.7 |
| Dec 2012 | Repower AG | 25 | Switzerland | Investor Group | Switzerland | Alpiq Holding AG | Switzerland | n/a |
| Dec 2012 | Romande Energie Holding SA | 6 | Switzerland | Romande Energie Holding SA | Switzerland | Alpiq Holding AG | Switzerland | 85.4 |
| Dec 2012 | Photovoltaic plant in Totana | 100 | Spain | Lufin Partners AG | Switzerland | - | - | 39.6 |

*Tab. 32*: *List of M&A transactions in power utilities in Switzerland in 2012 (after Pfister, Kerler, Valk, Prien, Peyer, Kuhn, Hintermann, Ljaskowsky, Arnet (2013)).*

## Technology, Media & Telecommunications

| Announced date | Target | Stake | Target country | Bidder | Bidder country | Seller | Seller country | Value (USDm) |
|---|---|---|---|---|---|---|---|---|
| Jan 2012 | Synchronica PLC | 100 | United Kingdom | Myriad Group AG | Switzerland | - | - | 37.6 |
| Jan 2012 | Medium GmbH | 100 | Germany | The ALSO-Actebis Group | Switzerland | - | - | n/a |
| Jan 2012 | Plancal AG | 100 | Switzerland | Trimble Navigation Limited | United States | - | - | n/a |
| Jan 2012 | Petroplus Holdings AG | - | Switzerland | Creditors | Switzerland | - | - | n/a |
| Jan 2012 | Reize Optik AG | - | Switzerland | Essilor International SA | France | - | - | n/a |
| Jan 2012 | Trivon AG | - | Switzerland | Virgin Media Inc | United Kingdom | Private Investors | Various | 106.0 |
| Feb 2012 | Cofina SGPS SA | 17 | Portugal | Credit Suisse | Switzerland | - | - | 14.7 |
| Feb 2012 | Seiler Hotels Zermatt AG | 91 | Switzerland | Christian Seiler | Switzerland | | - | n/a |
| Feb 2012 | Zustellgeschäft der AWZ AG | 100 | Switzerland | Die schweizerische Post | Switzerland | - | - | n/a |
| Mar 2012 | Ascom Holding AG-Defence Unit | 100 | Switzerland | Ruag Holding AG | Switzerland | - | - | 18.8 |
| Mar 2012 | Suhrkamp Verlag GmbH & CO KG | 61 | Germany | Medienholding Winthertur AG | Switzerland | - | - | n/a |
| Mar 2012 | BCI Cherganovo EOOD | 100 | Bulgaria | H1 Venture Swiss Holding AG | Switzerland | - | - | 6.6 |
| Apr 2012 | Andermatt Gotthard Sportbahnen AG | - | Switzerland | Andermatt Swiss Alps AG | Switzerland | - | - | 8.5 |
| Apr 2012 | Birkhauser Verlag AG | 100 | Switzerland | Walter de Gruyter GmbH & Co KG | Germany | - | - | 12.4 |
| Apr 2012 | 4M Wireless Ltd | 100 | United Kingdom | U-Blox AG | Switzerland | - | - | 9.0 |
| May 2012 | Acetrax AG | 100 | Switzerland | British Sky Broadcasting Group Plc | United Kingdom | - | - | 24.2 |
| May 2012 | AZ Medien AG-Radio 32 | 50 | Switzerland | Investor Group | Switzerland | - | - | n/a |
| May 2012 | GPC Global Project,RunningBall | 100 | Switzerland | Perform Media Services Ltd | United Kingdom | Private Investors | Various | 155.3 |
| May 2012 | Swisslog Holding AG | 11 | Switzerland | Grenzebach Maschinenbau GmbH | Germany | - | - | n/a |
| Jun 2012 | Grupa Onet.pl SA | 75 | Poland | Ringier Axel Springer Media AG | Switzerland | Various | Various | 272.4 |
| Jun 2012 | NRS Printing Solutions AG | 100 | Switzerland | ALSO Schweiz AG | Switzerland | Peter Nyffenegger (Private Investor); John Zahm (Private Investor) | Switzerland | n/a |
| Jun 2012 | Namics AG | 91 | Switzerland | Investor Group | Switzerland | - | - | n/a |
| Jun 2012 | GE Healthcare Ltd. (Nurse Call business) | 100 | United Kingdom | Ascom Holding AG | Switzerland | GE Healthcare Ltd. | United Kingdom | 22.0 |
| Jun 2012 | Kreis AG | 100 | Switzerland | Groupe Ecotel Chomette Favor | France | - | - | n/a |
| Jun 2012 | Cognovo Ltd | 100 | United Kingdom | U-Blox AG | Switzerland | - | | 16.6 |
| Jun 2012 | SwissQual AG | 100 | Switzerland | Rohde & Schwarz GmbH & Co. KG | Germany | - | - | n/a |
| Jun 2012 | Le Plaza Basel AG | 86 | Switzerland | Credit Suisse Funds AG | Switzerland | - | - | n/a |
| Jul 2012 | Sportradar AG | - | Switzerland | EQT Expansion Capital II | Guernsey | - | - | 54.2 |
| Jul 2012 | NRS Printing Solutions AG | 100 | Switzerland | ALSO Schweiz AG | Switzerland | - | - | n/a |
| Jul 2012 | GMC Software AG | 100 | Switzerland | Neopost SA | France | Private Investors | Various | 152.0 |

*Tab. 33.* List of M&A transactions in technology, media and telecommunications in Switzerland in 2012 (after Pfister, Kerler, Valk, Prien, Peyer, Kuhn, Hintermann, Ljaskowsky, Arnet (2013)).

| Announced date | Target | Stake | Target country | Bidder | Bidder country | Seller | Seller country | Value (USDm) |
|---|---|---|---|---|---|---|---|---|
| Jul 2012 | Velocity Technology Solutions, Inc. | 100 | United States | Partners Group Holding; Silver Lake Sumeru; Northleaf Capital Partners Ltd. | Switzerland | Spire Capital Partners, LLC; Tudor Ventures; EisnerAmper LLP | United States | n/a |
| Aug 2012 | Starhome BV | 100 | Switzerland | Fortissimo Capital Ltd | Israel | Comverse Technology Inc; Gemini Israel Ventures; Azini Capital Partners LLP | United States | 80.3 |
| Aug 2012 | Travel24.com AG | 46 | Germany | Metzler Corporate Finance; LOET Trading AG | Switzerland | Unister Holding GmbH | Germany | 30.0 |
| Sep 2012 | Bleuel Electronic AG | 100 | Switzerland | Sennheiser Electronic GmbH & Co KG | Germany | - | - | n/a |
| Sep 2012 | TransMedia Communications SARL | - | Switzerland | Time Equity Partners SAS | France | - | - | 7.7 |
| Sep 2012 | Telecom Liechtenstein AG | 75 | Liechtenstein | Swisscom AG | Switzerland | | - | n/a |
| Sep 2012 | jobs.ch AG | 100 | Switzerland | Tamedia / Ringier | Switzerland | Global Tiger Management | United States | 416.3 |
| Sep 2012 | Vigil Software Ltd | - | United Kingdom | Infinigate Holding AG | Switzerland | - | - | n/a |
| Sep 2012 | Teleclub AG | 33 | Switzerland | CT Cinetrade AG | Switzerland | Ringier AG | Switzerland | 6.4 |
| Oct 2012 | C1 FinCon GmbH-AdviceManager | 100 | Germany | Crealogix Holding AG | Switzerland | - | - | n/a |
| Oct 2012 | Zephyr Associates, Inc. | 100 | United States | Informa Group Plc | Switzerland | Kemmons Wilson Inc | United States | 62.0 |
| Oct 2012 | Improve Digital BV | 85 | Netherlands | PubliGroupe SA | Switzerland | - | - | n/a |
| Oct 2012 | Daetwyler-Cabling Solutions | 100 | Switzerland | Pema Holding AG | Switzerland | - | - | 20.0 |
| Oct 2012 | Fastrax Oy | 100 | Finland | u-blox AG | Switzerland | - | - | 17.0 |
| Nov 2012 | Red Universal de Marketing y | 100 | Spain | Bravofly SA | Switzerland | Telefonica, S.A.; Orizonia Corporacion SL | Spain | 95.1 |
| Nov 2012 | Comfriends S.A. | 100 | Switzerland | Yvan Vuignier (Private Investor) | Switzerland | Tamedia AG | Switzerland | n/a |
| Nov 2012 | Radisson Blu Hotel and Casino | 100 | Switzerland | ACRON HELVETIA X Immobilien AG | Switzerland | - | - | 62.1 |
| Nov 2012 | MetroXpress Denmark AS | 100 | Denmark | 20 Minuten AG | Switzerland | Metro International S.A. | Norway | n/a |
| Dec 2012 | Calendaria AG | 100 | Switzerland | Media Print Group GmbH | Germany | - | - | n/a |
| Dec 2012 | Betty Bossi AG | 50 | Switzerland | Coop Genossenschaft | Switzerland | - | - | n/a |

***Tab. 34.*** *List of M&A transactions in technology, media and telecommunications in Switzerland in 2012 (after Pfister, Kerler, Valk, Prien, Peyer, Kuhn, Hintermann, Ljaskowsky, Arnet (2013)).*

## Other Industries

| Announced date | Target | Stake | Target country | Bidder | Bidder country | Seller | Seller country | Value (USDm) |
|---|---|---|---|---|---|---|---|---|
| Jan 2012 | Wabern – Restaurant Bären | - | Switzerland | Glanzmann&Dreifuss AG | Switzerland | PSP Swiss Property | Switzerland | n/a |
| Jan 2012 | CIMM Tecnologias y Servicios SA | 100 | Chile | SGS SA | Switzerland | - | - | 37.0 |
| Jan 2012 | Direct Mail Company AG | 50 | Switzerland | Schweizerische Post | Switzerland | - | - | n/a |
| Jan 2012 | DDT GmbH | 100 | Germany | Weiss + Appetito Holding AG | Switzerland | - | - | n/a |
| Jan 2012 | Stopinc AG | 50 | Switzerland | RHI AG | Austria | - | - | n/a |
| Feb 2012 | Bauprojekt Lindbergh-Allee Opfikon Glattbrugg | - | Switzerland | Credit Suisse 1a Immo PK | Switzerland | Steiner AG | Switzerland | 178.9 |
| Feb 2012 | Überbauung "Seven" Altstätten (SG) | - | Switzerland | UBS "Sima" | Switzerland | - | - | 42.6 |
| Feb 2012 | Grundstück | - | Switzerland | Politische Gemeinde Rüschlikon | Switzerland | SBB | Switzerland | 21.1 |
| Feb 2012 | Businesspark Grünau Wabern | - | Switzerland | Jürg Guggisberg | Switzerland | Werner Hofmann | Switzerland | n/a |
| Feb 2012 | Cementval Materiales de | 20 | Spain | Holcim Ltd | Switzerland | - | - | n/a |
| Feb 2012 | Helios MPPD B.V. | 100 | Netherlands | deSter Holding BVBA / gategroup Holding | Switzerland | - | - | 28.7 |
| Feb 2012 | Fashion Days Holding AG | - | Switzerland | MIH | Switzerland | - | - | 56.6 |
| Feb 2012 | Lindbergh-Allee | 100 | Switzerland | Credit Suisse | Switzerland | - | - | 178.9 |
| Feb 2012 | R Haesler AG | - | Switzerland | Constellation Capital AG | Switzerland | - | - | n/a |
| Mar 2012 | VSN, Inc. | 100 | Japan | Adecco SA | Switzerland | SBI Holdings, Inc.; Ant Capital Partners Co Ltd | Japan | 123.7 |
| Mar 2012 | WerkZwei Areal in Arbon | - | Switzerland | HRS Investment AG | Switzerland | Oerlikon Saurer Arbon AG | Switzerland | 37.3 |
| Mar 2012 | Hotel und Thermalbald Vals AG (Hoteba), ohne Felsentherme | - | Switzerland | Stoffelpart AG (Remo Stoffel) | Switzerland | Gemeinde Vals | Switzerland | n/a |
| Mar 2012 | SRG Hochhaus Giacomettistrasse, Bern | - | Switzerland | Mobiliar | Switzerland | SRG | Switzerland | n/a |
| Mar 2012 | Montena EMC SA-Testing | 100 | Switzerland | Electrosuisse, SEV Verband fuer Elektro-, Energie- und Informations | Switzerland | - | - | n/a |
| Mar 2012 | Selecta Italia SpA | 100 | Italy | IVS Group Holding SpA | Italy | Selecta Management AG | Switzerland | 9.4 |
| Mar 2012 | Estudios Tecnicos SA | 100 | Colombia | SGS SA | Switzerland | - | - | n/a |
| Mar 2012 | LEONI Studer Hard AG | 100 | Switzerland | Synergy Health PLC | United Kingdom | - | - | 63.0 |
| Mar 2012 | Baunova AG | 51 | Switzerland | Strabag SE | Austria | BH Holding AG | Switzerland | n/a |
| Mar 2012 | Zwahlen & Mayr SA | 70 | Switzerland | Cimolai SpA | Italy | - | - | 30.6 |
| Mar 2012 | Ejobs Group SRL | 70 | Romania | Ringier Holding AG | Switzerland | - | - | 12.9 |
| Apr 2012 | Novotel Nathan Road Kowloon Hotel | 100 | China | Partners Group Holding; CSI Properties Ltd.; Gaw Capital Partners | Switzerland | LaSalle Investment Management Inc | Germany | 305.0 |
| Apr 2012 | TE Connectivity-Professionals | 100 | Switzerland | BlueStream Professional Services LLC | United States | - | - | 23.5 |
| May 2012 | LPG Tecnicas en Extincion de | 100 | Spain | Tyco International Ltd | Switzerland | - | - | n/a |
| May 2012 | ASC International House SA | 100 | Switzerland | Argos Soditic SA | Switzerland | - | - | n/a |

*Tab. 35.* *List of M&A transactions in other industries in Switzerland in 2012*

*(after Pfister, Kerler, Valk, Prien, Peyer, Kuhn, Hintermann, Ljaskowsky, Arnet (2013)).*

| Announced date | Target | Stake | Target country | Bidder | Bidder country | Seller | Seller country | Value (USDm) |
|---|---|---|---|---|---|---|---|---|
| May 2012 | Flightcare S.L.; Flightcare Belgium S.A./N.V. | 100 | Spain | Swissport International AG | Switzerland | F.C.C Versia S.A. | Spain | 175.0 |
| May 2012 | Aqualux Products Holding Ltd | - | British Indian Ocean Territory | Fetim BV | Netherlands | AFG Arbonia-Forster-Holding AG | Switzerland | n/a |
| May 2012 | Oxygen Aviation Ltd | 100 | United Kingdom | Perfect Holding SA | Switzerland | - | - | n/a |
| May 2012 | Global Blue SA | 100 | Switzerland | Silver Lake Management LLC, Partners Group | United States | Equistone Partners Europe | - | 1,258.7 |
| Jun 2012 | 2 Kommerzielle Liegenschaften (Dietlikon/Egerkingen) | - | Switzerland | UBS AST-KIS | Switzerland | - | - | 58.6 |
| Jun 2012 | BrainNet Supply Management Group AG | 100 | Switzerland | KPMG | Switzerland | - | - | n/a |
| Jun 2012 | Locher Bauunternehmer AG | 100 | Switzerland | Implenia AG | Switzerland | - | - | n/a |
| Jul 2012 | Steinentorstrasse 8, Basel / Businessappartements & Kino | - | Switzerland | UBS "Sima" | Switzerland | - | - | 28.8 |
| Jul 2012 | Stadthotels Swissôtel Zürich | 100 | Switzerland | CS REF Hospitality | Switzerland | - | - | n/a |
| Jul 2012 | Merchandise Mart Ppty | 100 | Canada | Informa PLC | Switzerland | - | - | 52.2 |
| Jul 2012 | Catering Businesses of Qantas Airways Ltd | - | Australia | Gate Gourmet Holding AG | Switzerland | - | - | n/a |
| Jul 2012 | Höhenklinik Valbella Davos Genossenschaft | 100 | Switzerland | HRS Real Estate AG | Switzerland | - | - | 70.4 |
| Jul 2012 | Campus Böblingen | 100 | Germany | Deka Immobilien GmbH | Germany | HPI Helvetic Private Investments AG | Switzerland | 38.6 |
| Aug 2012 | Swissôtel Le Plaza Basel | - | Switzerland | CS REF Hospitality | Switzerland | - | - | 79.9 |
| Aug 2012 | Hotel Atlantis / Zürich | - | Switzerland | Neue Hotel Atlantis AG | Switzerland | Rosebud-Gruppe | Israel | n/a |
| Aug 2012 | Chemlube International, Inc, Sopetra AG | 50 | Switzerland | Chemoil Energy Limited | Singapore | - | - | 16.0 |
| Aug 2012 | Swissprinters AG | - | Switzerland | SWP Holding AG | Switzerland | - | - | n/a |
| Sep 2012 | Geschäftssitz in Muttenz/BL | - | Switzerland | Sitex Properties | Switzerland | Valora | Switzerland | n/a |
| Sep 2012 | Hauptsitz AFG in Arbon TG | - | Switzerland | Credit Suisse Fond | Switzerland | AFG Arbonia-Forster-Holding AG | Switzerland | n/a |
| Sep 2012 | Haus «Metropol» Börsenstrasse / Zürich | - | Switzerland | SNB | Switzerland | Credit Suisse | Switzerland | n/a |
| Sep 2012 | Wincasa | 100 | Switzerland | Swiss Prime Site AG | Switzerland | Credit Suisse | Switzerland | n/a |
| Sep 2012 | Hotels in Basel und Zürich | - | Switzerland | Credit Suisse Hospitality | Switzerland | Swissôtel Hotels & Resorts | Switzerland | n/a |
| Sep 2012 | Lodestone Management Consultants AG | 100 | Switzerland | Infosys Ltd | India | Private Investors | Various | 348.8 |
| Sep 2012 | Bahnhofstrasse 53 / Zürich | 100 | Switzerland | Axa Winterthur | Switzerland | Credit Suisse | Switzerland | n/a |
| Sep 2012 | "The Chedi" / Andermatt | 100 | Switzerland | Acuro Immobilien AG | Switzerland | Orascom | Egypt | 132.5 |
| Sep 2012 | Sersa Group AG | 100 | Switzerland | Rhomberg Rail Holding GmbH | Austria | - | - | n/a |
| Sep 2012 | NSI NV-Real Estate Portfolio | 100 | Switzerland | Undisclosed Acquiror | Unknown | - | - | 100.8 |

*Tab. 36.* List of M&A transactions in other industries in Switzerland in 2012
*(after Pfister, Kerler, Valk, Prien, Peyer, Kuhn, Hintermann, Ljaskowsky, Arnet (2013)).*

| Announced date | Target | Stake | Target country | Bidder | Bidder country | Seller | Seller country | Value (USDm) |
|---|---|---|---|---|---|---|---|---|
| Oct 2012 | Basler Zeitung (Aeschenplatz / Basel Hochbergerstrasse in Kleinhüningen) | - | Switzerland | Robestate (C. Blocher) | Switzerland | Basler Zeitung | Switzerland | 69.3 |
| Oct 2012 | Trafo / Baden | | Switzerland | UBS Swissreal | Switzerland | Lochreal BV | Netherlands | 51.1 |
| Oct 2012 | Kardan NV-Real Estate(7) | 100 | Switzerland | Undisclosed Acquiror | Unknown | - | - | 78.8 |
| Oct 2012 | Jerusalem Economy-RE Portfolio | 100 | Switzerland | Undisclosed Acquiror | Unknown | - | - | 77.0 |
| Oct 2012 | TMC Group N.V. | 100 | Netherlands | Gilde Buy Out Partners | Switzerland | - | - | 89.0 |
| Oct 2012 | OEBB-Wien Westbahnhof A3 Build | 100 | Austria | ACRON AG | Switzerland | - | - | 112.4 |
| Nov 2012 | Uetlihof Office Complex Zürich | 100 | Switzerland | Norges Bank Investment Management (NBIM) | Norway | Credit Suisse | Switzerland | 1,064.7 |
| Nov 2012 | Frutt-Lodge & Spa | - | Switzerland | Yunfeng Gao | China | Eberli Generalunternehmung | Switzerland | 54.4 |
| Nov 2012 | Kameha Grand Zürich (Hotel) | - | Switzerland | UBS "Sima" | Switzerland | Mettler2Invest AG | Switzerland | n/a |
| Nov 2012 | W.T. Burdens Ltd. (Property, Stock and Vehicle assets) | 100 | United Kingdom | Wolseley Plc | Switzerland | W.T. Burdens Ltd. | United Kingdom | 40.0 |
| Nov 2012 | Bravofly SA | - | Switzerland | Investor Group | France | - | - | 25.7 |
| Nov 2012 | Riverside Business Park | 100 | Switzerland | Swiss Prime Site AG | Switzerland | - | - | 95.5 |
| Nov 2012 | Ally Financial Inc. (Europe, Latin America and China Operations) | 100 | Switzerland | General Motors Financial Company Inc | United States | Ally Financial Inc. | United States | 4,200.0 |
| Nov 2012 | Hotel frutt LODGE & SPA | 100 | Switzerland | Yunfeng Gao (Private investor) | China | Eberli Holding AG | Switzerland | 53.0 |
| Dec 2012 | City Halle / Hauptbibliothek ZHAW Winterthur | - | Switzerland | CS REF Hospitality | Switzerland | Implenia | Switzerland | 53.3 |
| Dec 2012 | "Les Portes de Fribourg" | 100 | Switzerland | UBS "Swissreal" | Switzerland | - | - | 20.6 |
| Dec 2012 | Grand Hotel Bellevue in Gstaad | - | Switzerland | Daniel Koetser und Rudolf Maag | Switzerland | Thomas Straumann | Switzerland | n/a |
| Dec 2012 | Robert Aebi AG-Falsework Unit | 100 | Switzerland | Implenia AG | Switzerland | - | - | n/a |
| Dec 2012 | Gujarat Apollo Industries Limited (Road construction equipment business) | 70 | India | Ammann Group | Switzerland | Gujarat Apollo Industries Limited | India | 73.2 |
| Dec 2012 | Eastern Property Holdings Ltd | - | Switzerland | Aurora Value Fund | Liechtenstein | - | - | n/a |

**Tab. 37**: *List of M&A transactions in other industries in Switzerland in 2012 (after Pfister, Kerler, Valk, Prien, Peyer, Kuhn, Hintermann, Ljaskowsky, Arnet (2013)).*

# Conclusion

The *M&A transactions* represent a wide range of unique business optimization opportunities in the corporate transformation deals, which are usually characterized by the high level of total risk. The *M&A transactions* can be successfully implemented by taking to an account the *size of investments, purchase price, direction of transaction, type of transaction*, and using the *modern comparable transactions analysis* and the *business valuation techniques* in the *diffusion – type financial systems* in the *finances*.

We think that there are many factors, which can generate the quasi periodic oscillations of the *M&A transactions* number in the time domain, for example: the *stock market bubble effects*. We performed the research of the *nonlinearities* in the *M&A transactions number quasi-periodic oscillation*s in *Matlab*, including the *ideal, linear, quadratic, and exponential dependences*.

We discovered that the average of a sum of random numbers in the *M&A transactions* time series represents a time series with the *quasi periodic systematic oscillations*, which can be finely approximated by the *polynomial numbers*.

We think that, in the course of the *M&A transaction* implementation, the ability by the companies to *absorb the newly acquired knowledge* and to *create the new innovative knowledge bases*, is a key pre-determinant of the *M&A deal* completion success. In our opinion, the *integrative collateral creative design thinking* has a direct impact on the *new innovative knowledge bases formation* by companies in the highly competitive global markets.

We analyzed the *M&A transactions* in *Switzerland* in *2012* in various industrial segments. We think that the globalization has a strong influence on the successful *M&A deals* completion in *Switzerland*. We believe that the fluctuating dependence of *M&A transactions* number over the certain time period is quasi-periodic.

We would like to state that the *winning virtuous mergers and acquisitions transactions strategies* in the *diffusion - type financial systems* in the *highly volatile global capital markets* with the *nonlinearities* can only be selected through the decision making process on the various available *M&A choices*, applying the *econophysical econometrical analysis* with the use of the *inductive, deductive* and *abductive logics*.

We developed the *MicroM&A software program* with the *embedded optimized near-real-time artificial intelligence algorithm* to create the *winning virtuous M&A strategies,* using the financial performance characteristics of the involved firms, and to estimate the *probability of the M&A transaction completion success*.

**Acknowledgement**

This *research article* is a result of our scientific collaborations with a number of leading academicians with the specializations in the *economics, finances, mathematics, and physics* from many leading universities in the *Western Europe, North America, Asia and Australia* over the last *25* years. The initial research ideas on the *M&A transactions strategies* have been outlined in the form of short notes by the *first author* many years ago, then as it frequently happens, the original research ideas have been intelligently re-worked for a few times and finally forgotten somewhere on the *first author's* research desk at the *James Cook University* in *Townsville* in *Australia*. Several decisive attempts have been recently undertaken to make a certain kind of order from the chaotic mess of numerous research papers, reports and books on the *first author's* research desk at *JCU* in *Townsville* in *Australia*. As a result, the once forgotten research article on the *M&A transactions strategies* has been re-written and significantly updated with the new research materials added. Of course, the main line of thinking on the *M&A transactions strategies* in the *diffusion - type financial systems* in the *highly volatile global capital markets* with the *nonlinearities* is pretty much the same as it was originally chosen by the authors, however the text of research article has been extensively complemented by the new research findings with the insightful analysis on the *M&A transactions strategies* included. The final variant of research article has been edited at the *JCU* in *Townsville* in *Queensland* in *Australia*.

The *first author's knowledge* on the origins of the *nonlinearities* in the complex systems in the electrical, electronic, and computer engineering as well as the financial engineering has been strongly influenced by the intensive scientific collaboration with *Prof. Janina E. Mazierska, Personal Chair, Electrical and Computer Engineering Department, James Cook University, Townsville, Australia* and *former IEEE Director and IEEE Fellow*. *Prof. Janina E. Mazierska* helped the first author to develop both the logical mathematical analysis skills and the abstract scientific thinking ability to tackle the complex scientific problems on *the nonlinearities in the microwave superconductivity* as well as on *the nonlinearities in the finances*, applying the interdisciplinary scientific knowledge together with the advanced computer modeling skills during the innovative research projects at *James Cook University* in *Townsville, Australia* in *2000 – 2014* after the graduation from *V. N. Karazyn Kharkov National University* in *Kharkov, Ukraine* in *1994 – 1999*.

The *second author* would like to appreciate a long hours discussion on the *M&A transactions* by the multinationals in the conditions of globalization in the *19th* to the *21st* centuries with *Prof. Geoffrey Jones, Harvard University,* which had place at the *University of Toronto* in *Canada* in 2006.

In addition, the *second author* would like to thank the *Presidents of the Empire Club of Canada, Economic Club of Toronto, Canadian Club* and the *Prof. Roger L. Martin, former Dean of the Rotman School of Management, University of Toronto* for their valuable help in the organization of many hundreds of business meetings with the *senior executives* from the *Canadian* and *American multinational corporations, insurance companies, international banks, investment funds and universities*, which have been conducted with the aim to discuss the modern *M&A transactions strategies selection and implementation techniques* together with the *industry specific trends* in *North America* over the last decades. The obtained valuable information has been collected, researched, analyzed and partly used in preparation of this research article. The *second author* takes this opportunity to thank all the researchers and business partners, who were involved in the intensive research on the *M&A transactions*, which has been conducted for more than *20* years since the time of the graduation from *V. N. Karazyn Kharkov National University* in *Kharkov, Ukraine* in *1988 – 1993*.

We think that our initial research findings highlight both: *1)* the complexity of considered research problems, and *2)* the novelty of obtained research results. Therefore, our next research phase will certainly improve our evolving understanding on the *mergers and acquisitions transactions selection and implementation strategies* in the *diffusion - type financial systems* in the *highly volatile global capital markets* with the *nonlinearities*, aiming to continue to accumulate the necessary critical mass of knowledge on the modern *M&A transactions theories and practices.* Of course, there are many possible avenues to explore the research on the *M&A transaction strategies* and we would like to conclude this innovative article by making the following optimistic statement in *UNCTAD (2000)*: "*M&As* became an important source of finance at the time of the crisis, thereby contributing to the speedy recovery of the economies. True, under normal circumstances it is possible for an acquirer to raise capital in the local financial markets without bringing in new capital from outside. But in the crisis- hit countries which experience a credit crunch and depressed domestic financial markets, cross-border *M&As* necessarily entail fresh capital inflows from the outside."

The *second author* had a wonderful opportunity to discuss the *M&A transactions completion in New York in the USA* and *related research topics* with *Charles K. Whitehead, Professor of Business Law, Cornell University Law School, New York, USA* during the *six informative invited lectures* on the *M&A corporate deals, dynamics of the boards of directors,* and *legal aspects of corporate governance in New York in the USA*, which have been delivered at

*V. N. Karazin Kharkiv National University* in the *City of Kharkiv* in the *State of Ukraine in December, 2014*. The *second author* expresses his personal thanks to *Charles K. Whitehead* for the presentation of the business cases studies with a particular accent on the *poison pills technique application* in the *M&A business deals* in the *USA*. The comprehensive discussions on the *M&A business deals strategies* are also acknowledged.

Finally, it may be interesting to note that the *authors* faced the following situations frequently: Looking at the *Vacheron Constantin Patrimony Traditionnelle World Time Swiss mechanical watch,* we understand that it is time to board for the next international business flight to attend the *business meetings, conferences and symposiums* to deliver our invited presentations; hence, we would like to make a note that the work on the complete version of our research paper is still in progress.

*E-mail: dimitri.ledenyov@my.jcu.edu.au ,

ledenyov@univer.kharkov.ua .

### Firm Theory Science, Business Administration Science:

***Merger and Acquisition Science, Business Administration Science:***

259. Shleifer A, Vishny R W 1991 Takeovers in the' 60s and the '80s: Evidence and implications *Strategic Management Journal* vol **12** pp 51 – 59.

260. Shleifer A 2001 Inefficient markets *Oxford University Press* New York USA.

261. Shleifer A, Vishny R W 2003 Stock market driven acquisitions *Journal of Financial Economics* **70** (3) pp 295 – 311.

262. Shrivastava P 1986 Post-merger integration *Journal of Business Strategy* vol **7** no 1 pp 65 – 76.

263. De S, Duplichan D 1987 Effects of interstate bank mergers on shareholder wealth: Theory and evidence *Geld Banken und Versicherungen* pp 751 – 773.

264. Hansen R G 1987 A theory for the choice of exchange medium in mergers and acquisitions *Journal of Business* **60** pp 75 – 95.

265. Haspeslagh P C, Farquhar A B 1987 The acquisition integration process: A contingent Framework *The 7th Annual International Conference of the Strategic Management Society* Boston USA.

266. Huang Y-S, Walkling R A 1987 Target abnormal returns associated with acquisition announcements: Payment, acquisition form, and managerial resistance *Journal of Financial Economics* **19** (2) pp 329 – 349.

267. James Ch M, Wier P 1987 Returns to acquirers and competition in the acquisition market: The case of banking *The Journal of Political Economy* **95** pp 355 – 370.

268. Neely W P 1987 Banking acquisitions: Acquirer and target shareholder returns *Financial Management* **16** pp 66 – 74.

269. Ravenscraft D J, Scherer F M 1987a Life after takeover *Journal of Industrial Economics* **36** (2) pp 147 – 156.

270. Ravenscraft D J, Scherer R M 1987b Mergers, selloffs, and economy efficiency *Brookings Institution* Washington DC USA.

271. Sicherman N, Pettway R 1987 Acquisition of divested assets and shareholder wealth *Journal of Finance* **42** pp 1261 – 1273.

272. Singh H, Montgomery C A 1987 Corporate acquisition strategies and economic performance. *Strategic Management Journal* **8** (4) pp 377 – 386.

273. Travlos N G 1987 Corporate takeover bids, methods of payment, and bidding firms' stock returns *Journal of Finance* **42** pp 943 – 963.

274. Travlos N G, Waegelein J F 1992 Executive compensation, method of payment and abnormal returns to bidding firms at takeover announcements *Managerial & Decision Economics* **13** (6) pp 493 – 501.

***Business Strategy Science, Strategic Governance Science, Management Science:***